\documentclass[titlepage,oneside]{tcibook}
\usepackage{fancyhea}
\usepackage{work}
\usepackage{bm}       %    enables bold math symbols  e.g.  \bm{\gamma}
\usepackage{graphicx}
\usepackage[usenames]{xcolor} %used for font color
\definecolor{xlinkcolor}{cmyk}{1,1,0,0}
\usepackage[
     colorlinks=true,    % false: boxed links; true: colored links
     linkcolor=xlinkcolor,     % color of internal links
     citecolor=xlinkcolor,     % color of links to bibliography
     filecolor=xlinkcolor,  % color of file links
     urlcolor=xlinkcolor,      % color of external links
     final=true
]{hyperref}
\usepackage{amssymb}
\usepackage{amsmath}
\usepackage{chngcntr}
\usepackage{booktabs}
\usepackage{multirow}
\usepackage{multicol}
\counterwithout{figure}{chapter}
\usepackage{titlesec}

\usepackage{newtxtext,newtxmath} % this is causing the etoolbox warning
\usepackage[USenglish]{babel}
\usepackage[utf8]{inputenc}
\usepackage[T1]{fontenc}
\usepackage{comment}
\usepackage{natbib}
\usepackage{float}

\DeclareUrlCommand\email{\urlstyle{rm}}
\graphicspath{{./}{./figures/}}
\usepackage{enumitem}
\setlist{noitemsep}

\usepackage{lineno}
\usepackage{macros}
\usepackage{aas_macros}

\newcommand{\FIXME}[1]{}
\newcommand{\CHECK}[1]{{#1}}

\newcommand{\Contributors}[1]{ \large \textit{#1} }
\newcommand{\Contact}[1]{}
\newcommand{\Comment}[3]{}

\newcommand{\ADW}[1]{\Comment{blue}{ADW}{#1}} % Alex Drlica-Wagner
\newcommand{\WAD}[1]{\Comment{blue}{WAD}{#1}} % Will Dawson
\newcommand{\AHGP}[1]{\Comment{cyan}{AHGP}{#1}} %Annika Peter
\begin{document}

\raggedbottom

\pagenumbering{roman}

\parindent=0pt
\parskip=8pt
\setlength{\evensidemargin}{0pt}
\setlength{\oddsidemargin}{0pt}
\setlength{\marginparsep}{0.0in}
\setlength{\marginparwidth}{0.0in}
\marginparpush=0pt

% The content begins here

\renewcommand{\chapname}{chap:intro_}
\renewcommand{\chapterdir}{.}
\renewcommand{\arraystretch}{1.25}
\addtolength{\arraycolsep}{-3pt}
\renewcommand{\paragraph}[1]{\emph{#1}\xspace}

\pagenumbering{roman} 
\chapter*{Probing the Fundamental Nature of Dark Matter with the Large Synoptic Survey Telescope}
\vskip -9.5pt
\hbox to\headwidth{%
       \leaders\hrule height1.5pt\hfil}
\vskip-6.5pt
\hbox to\headwidth{%
       \leaders\hrule height3.5pt\hfil}

\begin{center}
  {\Large\bf
    LSST Dark Matter Group\\
    \bigskip
    \today\\
    \bigskip
     v1.1
  }
\end{center}
\eject

\setcounter{page}{1}

%%%%%%%%%%%%%%%%%%%%%%%%%%%%%%%%%%%%%%%%%%%%%%%%%%%%%%%%%%%%%%%%%%%%%%%%%%%%%%%%
% Executive Summary
%%%%%%%%%%%%%%%%%%%%%%%%%%%%%%%%%%%%%%%%%%%%%%%%%%%%%%%%%%%%%%%%%%%%%%%%%%%%%%%%
\begin{comment}

\begin{center}
  {\Large \bf Abstract}
\end{center}
Astrophysical and cosmological observations currently provide the only robust, empirical measurements of dark matter. Future observations with Large Synoptic Survey Telescope (LSST) will provide necessary guidance for the experimental dark matter program. This white paper represents a community effort to summarize the science case for studying the fundamental physics of dark matter with LSST. We discuss how LSST will inform our understanding of the fundamental properties of dark matter, such as particle mass, self-interaction strength, non-gravitational couplings to the Standard Model, and compact object abundances. Additionally, we discuss the ways that LSST will complement other experiments to strengthen our understanding of the fundamental characteristics of dark matter. More information on the LSST dark matter effort can be found at https://lsstdarkmatter.github.io/ .

\clearpage
\end{comment}

\begin{center}
  {\Large \bf Executive Summary}
\end{center}

More than 85 years after its astrophysical discovery, the fundamental nature of dark matter remains one of the foremost open questions in physics.
Over the last several decades, an extensive experimental program has sought to determine the cosmological origin, fundamental constituents, and interaction mechanisms of dark matter. 
While the existing experimental program has largely focused on weakly-interacting massive particles, there is strong theoretical motivation to explore a broader set of dark matter candidates.
As the high-energy physics program expands to ``search for dark matter along every feasible avenue'' \citep{P5Report}, it is essential to keep in mind that the only direct, empirical measurements of dark matter properties to date come from astrophysical and cosmological observations.

The Large Synoptic Survey Telescope (LSST), a major joint experimental effort between NSF and DOE, provides a unique and impressive platform to study dark sector physics.
LSST was originally envisioned as the ``Dark Matter Telescope'' \citep{Tyson:2001}, though in recent years, studies of fundamental physics with LSST have been more focused on dark energy.
Dark matter is an essential component of the standard \LCDM model, and a detailed understanding of dark energy cannot be achieved without a detailed understanding of dark matter.
In the precision era of LSST, studies of dark matter and dark energy are \emph{extremely complementary} from both a technical and scientific standpoint.
In addition, cosmology has consistently shown that it is impossible to separate the \emph{macroscopic distribution} of dark matter from the \emph{microscopic physics} governing dark matter.
In this document, we reaffirm LSST's ability to test well-motivated theoretical models of dark matter: \ie, self-interacting dark matter, warm dark matter, dark matter-baryon scattering, ultra-light dark matter, axion-like particles, and primordial black holes. 
Several of these dark matter models can \emph{only} be tested with astronomical observations.

LSST will observe Milky Way satellite galaxies, stellar streams, and strong lens systems to detect and characterize the smallest dark matter halos, thereby probing the minimum mass of ultra-light dark matter and thermal warm dark matter.
Precise measurements of the density and shapes of dark matter halos in dwarf galaxies and galaxy clusters will be sensitive to dark matter self-interactions probing hidden sector and dark photon models.
Microlensing measurements will directly probe primordial black holes and the compact object fraction of dark matter at the sub-percent level over a wide range of masses.
Precise measurements of stellar populations will be sensitive to anomalous energy loss mechanisms and will constrain the coupling of axion-like particles to photons and electrons.
Unprecedented measurements of large-scale structure will spatially resolve the influence of both dark matter and dark energy, enabling searches for correlations between the only empirically confirmed components of the dark sector.
In addition, the complementarity between LSST, direct detection, and other indirect detection dark matter experiments will help constrain dark matter-baryon scattering, dark matter self-annihilation, and dark matter decay.

The study of dark matter with LSST presents a small experimental program with a short timescale and low cost that is guaranteed to provide critical information about the fundamental nature of dark matter over the next decade.
LSST will rapidly produce high-impact science on fundamental dark matter physics by exploiting a soon-to-exist facility. 
The study of dark matter with LSST will explore parameter space beyond the high-energy physics program's current sensitivity, while being highly complementary to other experimental searches. % BRN text.
This has been recognized in Astro2010 \citep{Astro2010}, during the Snowmass Cosmic Frontier planning process \citep[\eg,][]{1310.8642, 1310.5662, 1305.1605}, in the P5 Report \citep[]{P5Report}, and in a series of more recent Cosmic Visions reports \citep[\eg,][]{1604.07626,1802.07216}, including the ``New Ideas in Dark Matter 2017:\ Community Report'' \citep{Battaglieri:2017aum}.
It is worth remembering that astrophysical probes provide the only constraints on the minimum and maximum mass scale of dark matter, and astrophysical observations will likely continue to guide the experimental particle physics program for years to come.

\clearpage

\begin{center}
  {\Large \bf Preface}
\end{center}

This white paper is the product of a large community of scientists who are united in support of probing the fundamental nature of dark matter with LSST.
%Although LSST was originally proposed as the ``Dark Matter Telescope'' \citep{Tyson:2001}, none of the eight existing LSST Science Collaborations is specifically focused on exploring the microscopic identity of dark matter.
The study of dark matter is currently distributed across several of the LSST Science Collaborations, making it difficult to combine results and build a cohesive physical picture of dark matter.
It was recognized that the existing situation could hamper research on dark matter physics with LSST, and the current effort was started to coordinate dark matter studies among various LSST Science Collaborations, to enlarge the dark matter community, and to strengthen connections between theory and experiment.
The concept for this white paper emerged from a series of meetings and regular telecons organized in 2017--2018 around the topic of astrophysical probes of dark matter in the era of LSST.
Sessions were held at the LSST Project and Community Workshop in 2017 and 2018, multiple LSST Dark Energy Science Collaboration (DESC) meetings, and two dedicated multi-day workshops at the University of Pittsburgh and Lawrence Livermore National Laboratory.
Funding was provided by individual institutions and through a grant from the LSST Corporation (LSSTC) Enabling Science Program.
Through participation in the workshops, numerous telecons, sensitivity analyses, writing, editing, and reviewing, roughly \CHECK{100} scientists have directly contributed to this white paper.
We encourage interested scientists to learn more about this effort at: \url{https://lsstdarkmatter.github.io/}.

\clearpage
% Author list file generated with: mkauthlist 1.2.3+7.g04b131e 
% mkauthlist -f -s -j emulateapj authors.csv tmp.tex 
% python code/mkauthlist.py -f -s -sb -j emulateapj -a data/order.csv data/authors.csv tmp.tex 

\begin{center}
  {\Large \bf List of Contributors and Endorsers}
\end{center}
\bigskip

The following people have contributed to or endorsed the LSST dark matter science case as presented here:

\def\altaffilmark#1{\textsuperscript{#1}}
\def\affil#1{\noindent #1 \\}

\normalsize
\begin{raggedright}
\textbf{Contributors:}
Alex~Drlica-Wagner\altaffilmark{1,2,3,\textdagger},
Yao-Yuan~Mao\altaffilmark{4,*},
Susmita~Adhikari\altaffilmark{5},
Robert~Armstrong\altaffilmark{6},
Arka~Banerjee\altaffilmark{5,7},
Nilanjan~Banik\altaffilmark{8,9},
Keith~Bechtol\altaffilmark{10},
Simeon~Bird\altaffilmark{11},
Jonathan~Blazek\altaffilmark{12},
Kimberly~K.~Boddy\altaffilmark{13}
% ...

%\begin{center}
%(LSST Dark Matter Group)
%\end{center}

$^\dagger$ \email{kadrlica@fnal.gov} \\
$^*$ \email{yymao.astro@gmail.com}

\clearpage
\begin{multicols}{2}
\scriptsize
\parskip=4pt

\affil{$^{1}$ Fermi National Accelerator Laboratory}
\affil{$^{2}$ Kavli Institute of Cosmological Physics, University of Chicago}
\affil{$^{3}$ Department of Astronomy \& Astrophysics, University of Chicago}
\affil{$^{4}$ Department of Physics and Astronomy and Pittsburgh Particle Physics, Astrophysics and Cosmology Center (PITT PACC), University of Pittsburgh}
\affil{$^{5}$ Kavli Institute for Particle Astrophysics and Cosmology, Stanford University}
\affil{$^{6}$ Lawrence Livermore National Laboratory}
\affil{$^{7}$ SLAC National Accelerator Laboratory}
\affil{$^{8}$ GRAPPA Institute, Institute for Theoretical Physics Amsterdam and Delta Institute for Theoretical Physics, University of Amsterdam, Netherlands}
\affil{$^{9}$ Lorentz Institute, Leiden University, Netherlands}
\affil{$^{10}$ Physics Department, University of Wisconsin-Madison}
\affil{$^{11}$ Department of Physics and Astronomy, University of California, Riverside}
\affil{$^{12}$ Institute of Physics, Laboratory of Astrophysics, École Polytechnique Fédérale de Lausanne (EPFL), Observatoire de Sauverny, 1290 Versoix, Switzerland}
\affil{$^{13}$ Department of Physics and Astronomy, Johns Hopkins University}
% ...

\normalsize
\end{multicols}
\parskip=8pt

\end{raggedright}

\tableofcontents 

\eject
\pagenumbering{arabic} 
\setcounter{page}{1}

%%%%%%%%%%%%%%%%%%%%%%%%%%%%%%%%%%%%%%%%%%%%%%%%%%%%%%%%%%%%%%%%%%%%%%%%%%%%%%%%
% Introduction
%%%%%%%%%%%%%%%%%%%%%%%%%%%%%%%%%%%%%%%%%%%%%%%%%%%%%%%%%%%%%%%%%%%%%%%%%%%%%%%%
\chapter{Introduction} \Contact{Alex}
\label{sec:intro}
\bigskip
\Contributors{Alex Drlica-Wagner, Keith Bechtol, Annika H.\ G.\ Peter, Yao-Yuan Mao}

\begin{figure}[t]
\centering
\includegraphics[width=0.85\textwidth]{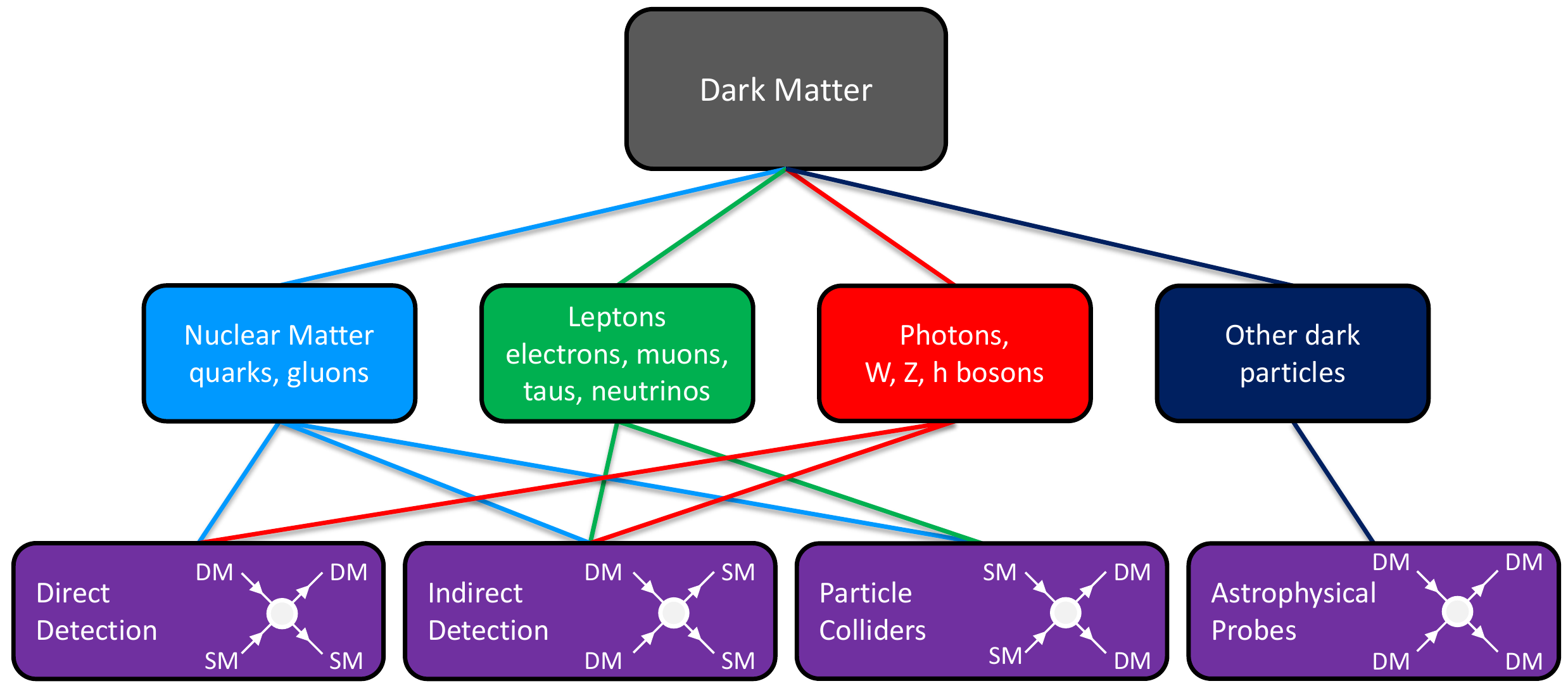}
\vspace{0.5em}
\caption{
\label{fig:interactions}
Dark matter may have non-gravitational interactions, which can be probed by four complementary approaches: 
direct detection, indirect detection, particle colliders, and astrophysical probes.
The lines connect the experimental approaches with the categories of particles that they most stringently probe (additional lines can be drawn in specific model scenarios). 
Figure taken from the Snowmass CF4 Report \citep{1305.1605}.
}
\end{figure}

The fundamental nature of dark matter, which constitutes $\roughly 85\%$ of the matter density and $\roughly 26\%$ of the energy density of the universe, represents a critical gap in our understanding of fundamental physics.
Over the past several decades, experimental searches for non-baryonic particle dark matter have proceeded along several complementary avenues (\figref{interactions}).
Collider experiments (\eg, ATLAS and CMS at the LHC) attempt to produce and detect dark matter particles, while  %\citep{Boveia:2018yeb,Ehret:2010mh,Battaglieri:2017aum}
direct detection experiments (\eg, LUX, LZ, XENON1T, SuperCDMS, ADMX, PICO, DAMiC, SENSEI, CRESST) attempt to directly detect energy deposition from very rare scattering between dark matter and Standard Model particles.
%\citep{1509:02910, 1804.10697, Du:2018uak} 
In parallel, indirect dark matter searches (\eg, {\it Fermi}-LAT, AMS-02, HESS, CTA, HAWC, {\it Chandra}, XMM) seek to detect the energetic Standard Model products from the annihilation or decay of dark matter particles {\it in situ} in astrophysical systems. %\citep[\eg][]{Ackermann:2015,Bulbul:2014,1605.01043,1603.06978}. 
Despite these extensive efforts, the only robust, positive empirical measurements of dark matter to date come from astrophysical and cosmological observations. 

Astrophysics and cosmology offer a complementary technique to study the fundamental properties of dark matter. 
They probe dark matter directly through gravity, the only force to which dark matter is known to couple. 
On large scales, observational data is well described by a simple model of stable, non-relativistic, collisionless, cold dark matter (CDM).
However, many viable theoretical models of dark matter predict observable deviations from CDM, which are testable with current and future experimental programs.
Fundamental properties of dark matter---e.g., particle mass, self-interaction cross section, coupling to the Standard Model, and time-evolution---can imprint themselves on the macroscopic distribution of dark matter in a detectable manner.

In addition, astrophysical observations complement particle physics searches by providing input to direct and indirect dark matter experiments, and by enabling alternative tests of dark matter's non-gravitational coupling to the Standard Model.  
For example, astrophysical observations can be used to i) measure the local distribution of dark matter, an important variable for direct searches, ii) highlight regions of high dark matter density for targeting indirect searches, and iii) identify astronomical objects that can lead to tight constraints on the range of dark matter particle mass and electric charge for a specific dark matter model.  
As the most widely studied CDM particle model, the weakly interacting massive particle (WIMP), becomes more and more tightly constrained, astrophysical observations will provide critical information to help direct future particle physics searches.  
In many cases, observations with telescopes provide \emph{the only} robust, empirical constraints on the viable range of dark matter models.

At the same time, there is immense dark matter discovery potential at the intersection of particle physics and astrophysics.
Detecting a deviation from the gravitational predictions of CDM would provide much-needed experimental guidance on parameters that are not easily measured in particle physics experiments (\eg, dark matter self-interaction cross sections). 
%If, on the other hand, all astrophysical studies of dark matter are found to agree with the CDM predictions, the improved knowledge of dark matter distributions will reduce major sources of theoretical uncertainties in the particle physics experiments. 
%ADW: I don't know what is being referred to here.
Likewise, results from particle experiments can suggest specific deviations from the CDM paradigm that can be tested with astrophysical observations.
The expanding landscape of theoretical models for dark matter provides strong motivation to explore dark matter parameter space beyond the current sensitivity of the high-energy physics program.

The Large Synoptic Survey Telescope (LSST) is a next-generation wide-area optical survey instrument that will probe the fundamental physics of dark matter and dark energy with precise cosmological measurements \citep{0805.2366,2018RPPh...81f6901Z}. 
Following on predecessors such as the Sloan Digital Sky Survey (SDSS) and the Dark Energy Survey (DES), LSST promises to greatly enhance our knowledge of the dark sector of the universe. 
LSST will measure the properties of dark matter over a wide range of astrophysical scales, thereby testing a wide variety of particle physics models (\tabref{models}).
At the largest scales, LSST will use gravitational weak lensing and the large-scale clustering of galaxies to trace the distribution of dark matter.
The profiles of dark matter halos associated with galaxies and clusters of galaxies can be used to test self-interacting dark matter models.
Measurements of the small-scale clustering of dark matter, traced by the faintest galaxies and via gravitational perturbations in strong lenses and stellar tracers, will enable constraints on warm and ultra-light dark matter models.
In addition, the temporal component of the LSST ``wide, fast, deep'' survey will open a new window on the search for compact dark matter, such as primordial black holes.
LSST will provide a rich scientific data set that can be used to develop novel and unanticipated constraints on dark matter properties through precise measurements of physical processes, such as anomalous energy loss in stars that could be produced by axion-like particles.

In this white paper, we present several techniques that LSST will employ to probe the fundamental properties of dark matter. 
In \secref{theory} we discuss several theoretical models of dark matter that can be constrained by LSST.
In \secref{probes} we present several observational probes of novel dark matter physics, and the  measurements that LSST will make to access these probes.
Many astrophysical measurements require collaborative observations between several instruments, and the study of dark matter with LSST is no exception. 
Thus, in \secref{complementarity}, we discuss situations where LSST will complement other astrophysical and particle investigations of dark matter.
Finally, in \secref{discovery} we present two scenarios of dark matter discovery with LSST.
Rather than presenting a comprehensive review of astrophysical probes of dark matter  \citep[e.g.,][]{BuckleyPeter:2017} or an extensive discussion of any particular dark matter model \citep[e.g.,][]{Tulin:2017ara}, we choose to focus on what we believe are some of the most exciting opportunities to study dark matter physics with LSST. 
Our goal is to demonstrate that LSST will not only provide exciting results on the nature of dark matter, but that observations from LSST are {\it critical} to guide future particle physics searches.

\begin{table}[t]
\begin{center}
\begin{tabular}{l c c c}
\hline 
Model & Probe & Parameter & Value \\
\hline 
\hline
Warm Dark Matter  & Halo Mass & Particle Mass & \CHECK{$m \sim 18 \keV$} \\
Self-Interacting Dark Matter & Halo Profile & Cross Section & \CHECK{$\sigmam \sim 0.1\text{--}10\cm^2/\g$} \\
Baryon-Scattering Dark Matter & Halo Mass & Cross Section & \CHECK{$\sigma \sim 10^{-30} \cm^2$} \\
Axion-Like Particles & Energy Loss & Coupling Strength & \CHECK{$g_{\phi e} \sim 10^{-13} $} \\
Fuzzy Dark Matter & Halo Mass & Particle Mass & \CHECK{$m \sim 10^{-20} \eV$}  \\
Primordial Black Holes  & Compact Objects & Object Mass & \CHECK{$M > 10^{-4} \Msun$} \\
Weakly Interacting Massive Particles & Indirect Detection & Cross Section & \CHECK{$\sigmav \sim 10^{-27} \cm^3/\second$} \\
Light Relics & Large-Scale Structure & Relativistic Species & \CHECK{$N_{\rm eff} \sim 0.1$} \\[+0.5em]
\hline
\end{tabular}
\end{center}
\caption{\label{tab:models} Probes of fundamental dark matter physics with LSST. Classes of dark matter models are listed in Column 1, and the primary observational probe that is sensitive to each model is listed in Column 2. The corresponding dark matter parameters are listed in Column 3, and estimates of LSST's senstivity to each parameter are listed in Column 4.}
\end{table}

%%%%%%%%%%%%%%%%%%%%%%%%%%%%%%%%%%%%%%%%%%%%%%%%%%%%%%%%%%%%%%%%%%%%%%%%%%%%%%%%
% Theory and Dark Matter Models
%%%%%%%%%%%%%%%%%%%%%%%%%%%%%%%%%%%%%%%%%%%%%%%%%%%%%%%%%%%%%%%%%%%%%%%%%%%%%%%%
\chapter{Dark Matter Models}
\label{sec:theory}
\bigskip
\Contributors{Simeon Bird, Kimberly K.\ Boddy, Matthew Buckley, George Chapline, Francis-Yan Cyr-Racine, William A.\ Dawson, Alex Drlica-Wagner, Juan Garc\'ia-Bellido, Maurizio Giannotti, Vera Gluscevic, Nathan Golovich, Manoj Kaplinghat, Samuel D.\ McDermott, Michael Medford, Manuel Meyer, Annika H.\ G.\ Peter, Chanda Prescod-Weinstein, Oscar Straniero, Hai-Bo Yu}

In this section, we provide a brief review of several theoretical models of dark matter with a specific focus on the properties of these models that can be explored by LSST. 
We divide the domain of models into three different categories. 
We first discuss reasonably minimal extensions of the popular cold, collisionless particle dark matter paradigm (\secref{particles}). 
We then extend our discussion to much lighter axion-like particle and wave-like dark matter (\secref{axions}). 
Finally, we discuss the potential to constrain alternative compact dark matter models, with a focus on primordial black holes (\secref{machos}).
We stress that exploring a broad theoretical landscape with LSST is strongly motivated by the lack of an experimental discovery of a conventional CDM particle candidate. 

\section{Particle Dark Matter \Contact{Haibo}}
\label{sec:particles}

\Contributors{Hai-Bo Yu, Matthew Buckley, Vera Gluscevic, Kimberly K.\ Boddy, Francis-Yan Cyr-Racine, Annika H.\ G.\ Peter, Manoj Kaplinghat, Alex Drlica-Wagner}

The standard \LCDM cosmological model assumes that dark matter is fully nonrelativistic and interacts purely via gravitational interactions during the process of structure formation. However, a significant dark matter thermal velocity dispersion or the presence of large non-gravitational interactions in the dark sector, such as dark matter self-interactions, couplings to other dark sector particles, or couplings to Standard Model particles, can alter the distribution of dark matter in ways that are observable with LSST. Here we focus on three representive minimal extensions to CDM -- warm dark matter (WDM), self-interacting dark matter (SIDM), and baryon-scattering dark matter (BSDM) --  to demonstrate how measurements of the distribution of dark matter can be used to constrain its micro-physical particle properties. We leave an in-depth discussion of the particle physics responsible for producing these models to the literature; however, we do attempt to connect astrophysical observables to specific terms in the interaction Lagrangian for each model.

\subsection{Warm Dark Matter}
\label{sec:wdm}

In the standard thermal dark matter paradigm, primordial inhomogeneities in the matter density field are washed out by collisional damping and free streaming of particle dark matter \citep{Hofmann:2001,Green:2003un, Bertschinger:2006nq, Loeb:2005pm}.  
For a canonical 100\GeV thermal relic dark matter particle \citep[\eg, the WIMP;][]{steigman1985,Jungman:1995df}, these processes erase cosmological perturbations with $M \leq 10^{-6} \Msun$ \citep[i.e., Earth mass;][]{Green:2003un, 2005Natur.433..389D}. 
Lighter particles continue to free stream until later times, thus suppressing the formation of structure at higher mass scales (\eg, structure formation occurs bottom-up for scales larger than the free-streaming scale and top-down for scales smaller than the free-streaming scale). Because these particles are created while they are semi-relativistic, they are conventionally referred to as warm dark matter (WDM) \citep{Bond:1983hb,Bode:2000gq,Dalcanton:2000hn}. 
WDM constitutes a subclass of sub-GeV dark matter candidates.

One well-motivated WDM candidate is a sterile neutrino, $\nu_{\rm s}$, with a mass in the keV range \citep[\eg][]{Abazajian:2017tcc,Adhikari:2017}. The most relevant Lagrangian term in this case is simply the Majorana mass term,
\begin{equation}
    \mathcal{L} \supset -\frac{1}{2}M_{\rm s}\bar{\nu}_{\rm s} \nu_{\rm s}.
\end{equation}
Interestingly, such a sterile neutrino typically mixes with active Standard Model neutrinos \citep[\eg,][]{Asaka:2005an}, allowing the former to decay and leading to a potentially observable X-ray signal \citep[\eg,][]{Abazajian:2001vt}. A possible hint of such a signal has been found in deep X-ray data in the form of a narrow 3.5 keV line \citep{Boyarsky:2014, Bulbul:2014, Boyarsky:2015, Iakubovskyi:2015}, which has prompted renewed interest in understanding structure formation in WDM cosmologies \citep[\eg,][]{Lovell:2013ola,Bose:2016irl,Bozek:2018ekc}. Generally speaking, both the sterile neutrino mass and its thermal history play an important role in determining the small-scale dark matter distribution within any given particle model. For instance, at a fixed particle mass, a species created in the early universe with a velocity distribution that is skewed towards low-momentum particles \citep[\eg,][]{Shi:1998km,Venumadhav:2015pla} will display less free-streaming damping of cosmological structure than a species with a thermal (i.e., Fermi-Dirac) distribution. To avoid ambiguity, it is customary to quote WDM constraints simply in terms of a particle mass $\mWDM$, assuming that the DM followed a thermal distribution at early times.

The free-streaming scale can be approximated by the (comoving) size of the horizon when the WDM particles become nonrelativistic.
%The comoving horizon size at $z = 10^7$ corresponds to $m = 2.5 \keV$, and is approximately $50 \kpc$, which is significantly smaller than the scale derived above for $\Lstar$ galaxies \citep{Adhikari:2017}
Astrophysical constraints on WDM are generally placed by observing the smallest gravitationally bound dark matter halos.  
In particular, the half-mode mass (the scale at which the dark matter transfer function is reduced by half) represents a characteristic halo mass scale below which halo abundances are suppressed sufficiently to yield observable consequences. 
Assuming the Planck 2016 cosmology \citep{Ade:2015xua}, the half-mode halo mass, \Mhm, is related to the WDM thermal relic particle mass, \mWDM, by \citep[\eg][]{schneider2012,Bullock:2017}
\begin{equation} \label{eqn:Mhm}
    M_{\rm hm} = 5.5 \times 10^{10} \left( \frac{\mWDM}{1 {\rm keV}} \right)^{-3.33} \Msun.
\end{equation}
Thus, an observed suppression in the abundance of dark matter halos smaller than \Mhm could signify the existence of a thermal dark matter particle with mass
\begin{equation} \label{eqn:mWDM}
    \mWDM =  3.33 \left(\frac{M_{hm}}{10^{9} {\rm M_\odot}} \right)^{-0.3} \keV.
\end{equation}
It is important to remember that $\mWDM$ is the thermal-relic-equivalent particle mass. Translating measurements of the halo mass function to constraints on the particle mass for a specific WDM model depends on the specific mapping between particle mass and the early-time momentum distribution.

Measurements of the Lyman-$\alpha$ forest \citep[\eg][]{Viel:2013,2017PhRvD..96b3522I} and ultra-faint satellite galaxies \citep[\eg][]{Jethwa:2018,Kim:2017iwr} place lower bounds on the mass of thermally produced WDM particles at $\roughly 3$--$5\keV$, corresponding to a halo mass scale of $\roughly 10^8-10^9 \Msun$.
The sensitivity and wide-area coverage of LSST has the potential to extend measurements of the dark matter halo mass function by three orders of magnitude, down to $\roughly 10^6 \Msun$ (\secref{halo_mass}). 
These observations have the potential to constrain WDM particle masses \CHECK{$\mWDM \gtrsim 18\keV$}, thus effectively testing putative signatures of keV-mass sterile neutrinos.

\subsection{Self-Interacting Dark Matter}
%\Contributors{Haibo, Francis-Yan, Manoj?,...}
\label{sec:sidm}

The self-interacting dark matter (SIDM) paradigm posits additional interactions in the dark sector \citep[\eg,][]{1992ApJ...398...43C,Spergel:1999mh,Dave:2000ar,Firmani:2000ce}, which allow energy and momentum exchange between particles within dark matter halos \citep[see][for a recent review]{Tulin:2017ara}. The figure of merit for dark matter halo structure is the cross section per dark matter particle mass, $\sigmam$.
Dark matter self-interactions with cross sections per mass roughly equivalent to the strong nuclear force ($\sigmam \sim 1 \cmg$) would imply ${\cal O}(1)$ energy exchange in the central regions of halos within the age of the universe \citep{2012MNRAS.423.3740V,2013MNRAS.431L..20Z,Peter:2013,Rocha:2012jg}. This  would thermalize the inner regions of dark matter halos --- where visible baryonic matter resides --- with observational consequences \citep[\eg][]{Kaplinghat:2013xca}. For low-surface brightness galaxies, SIDM thermalization leads to a cored inner density profile, in contrast to the cupsy profiles predicted in CDM. For high-surface brightness galaxies, thermalization leads to a small core and more concentrated SIDM distribution because of the presence of the baryonic potential \citep{Kaplinghat:2015aga}. It has been shown that SIDM can explain both the diversity and uniformity of galaxy rotation curves, for $\sigmam \gtrsim 1\cmg$ on galaxy scales \citep{Kamada:2016euw,Creasey:2016jaq,Ren:2018jpt}. This diversity of properties within SIDM halos also extends to galaxy cluster scales \citep{Robertson:2017mgj}.

%\ADW{This is coming from Section VI.C in 1705.02358.}
Large self-interaction cross sections are required to modify galactic structure, and such cross sections suggest either strongly-coupled systems \citep[\eg,][]{Frandsen:2011kt,Hochberg:2014dra,Hochberg:2014kqa} or a light mediator with perturbative couplings \citep[\eg,][]{Feng:2009mn,Ackerman:2008gi,Kaplan:2009de,Feng:2009hw,Buckley:2009in,Loeb:2010gj,Tulin:2012wi,Tulin:2013teo,Schutz:2014nka,Blennow:2016gde}. An interesting example of the latter type of model would be to charge dark matter under a $U(1)$ gauge symmetry. %which may be either broken or unbroken. 
Exchange of the gauge boson (a ``dark photon'') then mediates self-interactions, analogous to Rutherford scattering. A phenomenologically similar model replaces the vector mediator with a light scalar. The interaction Lagrangian is then described by %\citep{Tulin:2012wi,Tulin:2013teo}:
\begin{equation}
\label{eq:sidm}
{\cal L_{\rm int}}=\bigg\{
\begin{array}{c l}
g_\chi\bar{\chi}\gamma^\mu\chi\phi_\mu & \text{(vector mediator)}\\
g_\chi\bar{\chi}\chi\phi & \text{(scalar mediator)} \, ,
\end{array}
\end{equation}
where $\chi$ is the dark matter particle (which we assume to be a fermion for concreteness), $\phi$ is the mediator, and $g_\chi$ is the coupling constant. In the non-relativistic limit, self-interactions are described by the Yukawa potential
\begin{equation}
V(r)=\pm\frac{\alpha_\chi}{r}e^{-m_\phi r},
\label{eq:yukawa}
\end{equation}
where $\alpha_\chi = g_\chi^2/4\pi$. In order for annihilation through the mediator to not deplete the dark matter relic abundance during the early universe, it may be necessary to assume asymmetric dark matter (that is, dark matter which is composed mainly of $\chi$, with a minimal admixture of $\bar\chi$). In that case, the vector mediator would provide only a repulsive potential (``$+$'' in Eq.~\ref{eq:yukawa}), while the scalar mediator would have an attractive potential (``$-$'').

In these light mediator models, the self-scattering cross section generally depends on the relative velocity of colliding dark matter particles, $v_{\rm rel}$, and scattering is not isotropic. In practice, we often consider the transfer (viscosity) cross section, defined as $\sigma_T = \int d\Omega(1-\cos\theta)d\sigma/d\Omega$ ($\sigma_V = \int d\Omega\sin^2\theta d\sigma/d\Omega$), to regulate small-angle scatterings and use them as a proxy to match to SIDM N-body simulations with a constant cross section for a given halo-mass scale \citep[][]{Tulin:2013teo,Kahlhoefer:2013dca}. The overall feature of the velocity dependence predicted in the models can be summarized as follows. When the momentum transfer is much larger than the mediator mass, the scattering is in the Rutherford limit, i.e., $\sigmam \propto v^{-4}_{\rm rel}$. While in the opposite limit, $m_\chi v_{\rm rel}\ll m_\phi$, $\sigmam$ is nearly a constant. If the scattering is in the quantum resonant regime for $\chi\textup{-}\bar{\chi}$ collisions, $m_\chi v_{\rm rel}\sim m_\phi$, $\sigmam \propto v^{-2}_{\rm rel}$. Since large dark matter halos have much larger dark matter velocities compared to smaller halos, observations from different scales, ranging from dwarf galaxies to galaxy clusters, provide important tests of these models.

There are numerous observations that are sensitive to dark matter self scattering \citep[\eg,  Table 1 in][]{Tulin:2017ara}. Notably, merging galaxy clusters, such as the Bullet cluster  \citep{Randall:2007ph,2017MNRAS.465..569R}, have been used to put an upper bound on the self-interaction cross section at large particle velocities \citep[\eg,][]{Kahlhoefer:2013dca,Kahlhoefer:2015vua,Kim:2016ujt,Harvey:2016bqd,Robertson:2016qef,Wittman:2017gxn}, yielding $\sigmam \lesssim 2 \cmg$ for $v_{\rm rel} \sim 1000\textup{--}4000 \kms$. Moreover, observations from well-relaxed galaxy clusters \citep{Newman++11,Newman:2013,Newman++13b} show $\sigmam \sim 0.1 \cmg$ for $v_{\rm rel} \sim 1500 \kms$ to be consistent with their inferred core sizes~\citep{Kaplinghat:2015aga,Andrade:2019wzn}. The diversity of rotation curves observed in spiral galaxies can be explained by dark matter scattering with $\sigmam \gtrsim 1 \cmg$ in the range of $v_{\rm rel}\sim50\textup{--}200 \kms$. For these spiral galaxies, the large cross section is driven by galaxies with a high density core. In contrast, high surface brightness galaxies are baryon-dominated in their central regions and are thus effectively insensitive to the value of $\sigmam$ \citep{Kamada:2016euw,Ren:2018jpt}. 

\begin{figure}
\centering
\includegraphics[width=0.6\columnwidth]{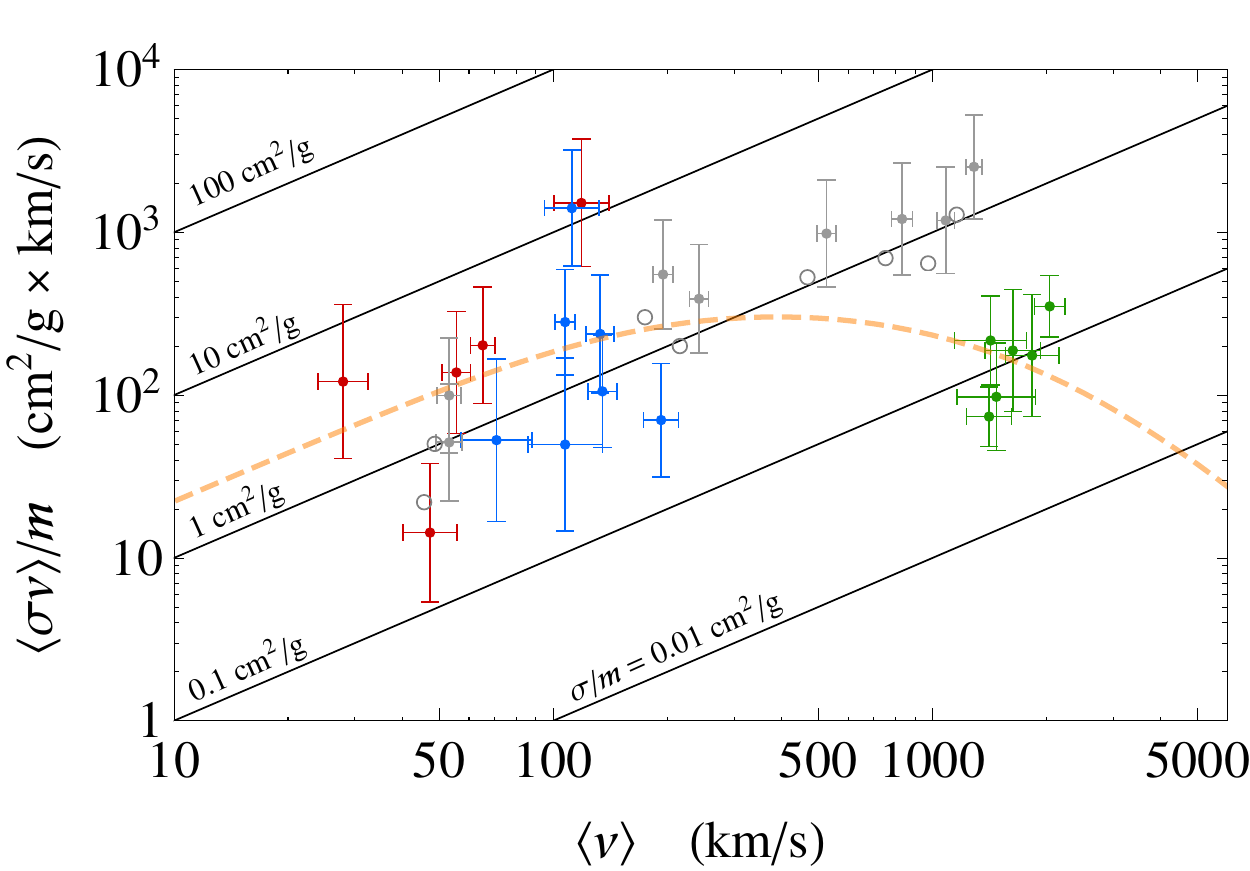}
\caption{Velocity-weighted self-interaction cross section per unit mass as a function of average relative particle velocity in a halo. Data points from astrophysical observations correspond to dwarf galaxies (red), low-surface-brightness galaxies (blue), and galaxy clusters (green). 
Diagonal lines show constant values of $\sigmam$. 
Gray points are fits to mock data from SIDM simulations, with fixed $\sigmam = 1 \cmg$.
Figure taken from \citet{Kaplinghat:2015aga}.
}
\label{fig:sidm_sigma}
\end{figure}

Velocity-dependent SIDM models with $\sigmam \gtrsim 1 \cmg$ in dwarf galaxies and $\sigmam \sim 0.1 \cmg$ in galaxy clusters are able to fit existing observational data (\figref{sidm_sigma}). This result has important implications for the particle properties of SIDM. For instance, consider the dark photon model given in Eq.~(\ref{eq:sidm}) and assume $\alpha_\chi=1/137$ to match the fine structure constant in the visible sector, we can determine $m_\chi\approx 15 \GeV$ and $m_\phi \approx 17 \MeV$ \citep{Kaplinghat:2015aga} and even infer the production mechanism of SIDM in the early universe \citep{Huo:2017vef}. Since LSST will probe scales ranging from the largest galaxy clusters to the smallest dwarf galaxy, it will be able to detect the influence of scattering cross sections at the level of $\sigmam \sim 0.1$--$1 \cmg$ over a wide range of velocities. Thus, LSST will significantly improve our understanding of the self-interacting nature of dark matter. 

It is also natural to expect that SIDM has a modified matter power spectrum compared to CDM. For instance, in SIDM models where the dark matter particle couples to a massless particle in the early universe, either directly or through a light mediator, the tight coupling between dark matter and dark radiation can lead to dark acoustic oscillations \citep{Cyr-Racine:2013ab,Cyr-Racine:2013fsa}, resulting in a suppressed and oscillatory power spectrum \citep[\eg][]{1992ApJ...398...43C,Boehm:2001hm,Boehm:2004th,Feng:2009mn,Aarssen:2012fx}. It has been shown that realistic realizations of SIDM strongly prefer such a scenario \citep{ Huo:2017vef}.

To see the reach of LSST on the SIDM damping effect, we estimate the cut-off scale on the field halo mass due to the dark acoustic oscillations as $M_{\rm cut}\approx0.7\times10^{8}({\rm keV/T_{\rm kd}})^3 \Msun$~\citep{1512.05349}, where $T_{\rm kd}$ is the kinetic decoupling temperature. For SIDM models, where a dark matter particle ($\chi$) couples to a massless fermion ($f$) via a light mediator ($\phi$), $T_{\rm kd}$ is given by~\citep{Aarssen:2012fx,Cyr-Racine:2015ihg}
\begin{equation}
T_{\rm kd}\approx\frac{1.38~{\rm keV}}{\sqrt{g_\chi g_f}}\left(\frac{m_\chi}{100 \GeV}\right)^{\frac{1}{4}}\left(\frac{m_\phi}{10 \MeV}\right)\left(\frac{g_\star}{3.38}\right)^{\frac{1}{8}}\left(\frac{0.5}{\xi}\right)^{\frac{3}{2}},
\label{eq:tkd}
\end{equation}
where $g_f$ is the $\xi\textup{--}f$ coupling constant, $g_\star$ the is the number of massless degrees of freedom at decoupling and $\xi$ parameterizes the ratio of dark-to-visible temperatures. \citet{Huo:2017vef} recasts the Lyman-$\alpha$ bound on WDM to set upper limit on the decoupling temperature $T_{\rm kd}\gtrsim 1 \keV$, corresponding to the minimal halo mass $\roughly 10^8 \Msun$. Since LSST has the potential to extend measurements of the dark matter halo mass function by three orders of magnitude (\secref{halo_mass}), the expected sensitivity on the decoupling temperature is $T_{\rm kd}\sim 10 \keV$. If LSST detects a cutoff on the halo mass function, we can determine the corresponding $T_{\rm kd}$ and further narrow down the particle parameters contained in the Lagrangian via Eq.~(\ref{eq:tkd}) after combining with the measurements of $\sigmam$ discussed above. Moreover, since the damping effect can also suppress the number of subhalos in the Milky Way, we expect LSST to provide another constraint on $T_{\rm kd}$ by providing a more complete census of ultra-faint satellites. In addition, although the acoustic damping effect may look similar to the free-streaming one \citep[\eg][]{1512.05349}, distinct signatures can be imprinted on the halo mass function \citep{Buckley:2014ab,Sameie:2018juk} or the Lyman-$\alpha$ forest spectrum \citep{Krall:2017xcw,Bose:2018juc}. By combining observables, including those from LSST, it might thus be possible to distinguish between WDM and SIDM with a damped matter power spectrum due to early-universe interactions (\secref{combine_probes}).

\subsection{Baryon-Scattering Dark Matter}
%\Contributors{Vera, Kim, ...}
\label{sec:bsdm}

In the standard WIMP scenario, dark matter may be directly observable through its scattering with Standard Model particles.
These models are conventionally probed by direct detection experiments that search for scattering between dark matter particles (from the local Galactic halo) and nuclei in their detectors.
These experiments are placed deep underground to provide shielding from cosmic-ray backgrounds and achieve exquisite sensitivity for low scattering cross sections. 
These experiments are largely insensitive to dark matter with very large scattering cross section because such particles would scatter many times before reaching the experiment, thus losing most of their kinetic energy \citep[\eg][]{Zaharijas:2004jv}.
Thus, it is important to broadly explore parameter space outside the standard WIMP region of interest.

Cosmological and astrophysical observables are unique and complementary probes of baryon-scattering dark matter (BSDM) models.
In particular, they are sensitive to very large (closer to nuclear-scale rather than weak-scale) scattering cross sections and sub-GeV dark matter masses, both of which are inaccessible to direct searches.
Such large cross sections may arise in a number of models.
One such model posits that dark matter is a flavor singlet sexaquark composed of $uuddss$ quarks \citep{Farrar:2017eqq}.
In this case, the scattering cross section with nucleons is expected to be geometric, though velocity-dependent enhancements may exist at very low energies, depending on the form of the sexaquark-nucleon potential.
For the sexaquark to be a viable dark matter candidate, it must be stable or have a sufficiently long lifetime; this criterion sets the sexaquark mass to be below a few GeV.
Alternatively, dark matter may be charged under a dark version of electromagnetism with field strength $\tilde{F}_{\mu\nu}$, which may kinetically mix with ordinary electromagnetism~\citep{Holdom:1985ag}:
\begin{equation}
    \mathcal{L} \supset \frac{\kappa}{2} F^{\mu\nu} \tilde{F}_{\mu\nu} ,
\end{equation}
where $\kappa$ parameterizes the strength of the mixing.
In this scenario, dark matter acquires a fractional amount of electric charge (proportional to $\kappa$ and its dark charge), allowing it to scatter with electrons and protons via Coulomb interactions that have a velocity dependence of $v^{-4}$.
This interaction is significant at late cosmological times, as the universe expands and the momentum of matter redshifts away.

Instead of focusing on particular theories, it is possible to describe the low-energy scattering processes of BSDM models with a nonrelativistic effective field theory~\citep{Fan:2010gt,Fitzpatrick:2012ix,Anand:2013yka}.
The effective Lagrangian has the form
\begin{equation}
    \mathcal{L}_\textrm{eff}(\vec{x})
    = c \Psi_\chi^\ast (\vec{x}) \mathcal{O}_\chi \Psi_\chi (\vec{x})
    \Psi_N^\ast (\vec{x}) \mathcal{O}_N \Psi_N (\vec{x}) ,
\end{equation}
where $\Psi (\vec{x})$ are the nonrelativistic fields for the dark matter, $\chi$, and nucleon, $N$.
Dark matter experiments seek to constrain and measure the coefficient $c$ for a variety of possible operators $\mathcal{O}_\chi$ and $\mathcal{O}_N$ that encode the BSDM physics.
However, regardless of the specific underlying BSDM model, cosmological observables are sensitive only to the magnitude (which scales as $c^2$) and velocity dependence of the cross section.
Thus, while laboratory searches for dark matter typically rely on assumptions about the detailed form of the interaction, cosmology offers very broad and generic probes of dark matter physics.

\begin{figure}
\centering
\includegraphics[width=0.6\columnwidth]{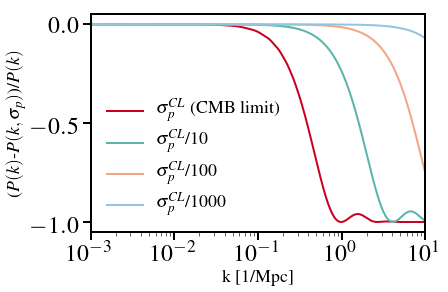}
\caption{Residuals in the linear matter power spectrum between the CDM case and a case where dark matter has a velocity-independent, spin-independent scattering with protons. The dark matter particle mass is set to $1\MeV$, and all other cosmological parameters are set to their best-fit Planck 2015 values \citep{Ade:2015xua}. Different residual curves display cutoffs at different angular scales, controlled by the magnitude of the interaction cross section. The highest cross section shown corresponds to the current 95\% confidence-level upper limit inferred from analyses of CMB data \citep{Gluscevic:2017ywp,Boddy:2018kfv}.
}
\label{fig:dmbaryon_pk}
\end{figure}

In a cosmological setting, scattering results in the exchange of momentum between the dark matter and the baryon fluids.
The momentum transfer induces a drag force, which suppresses structure increasingly at smaller scales. 
The effect of scattering is qualitatively similar to a cutoff in the matter power spectrum arising in the WDM and SIDM scenarios (\figref{dmbaryon_pk}).
This feature can be sought with tracers of matter on all observable scales. 
The best cosmological and astrophysical limits so far come from CMB temperature, polarization, and lensing anisotropy measurements~\citep{Xu:2018efh,Boddy:2018kfv,Gluscevic:2017ywp,Boddy:2018wzy,Slatyer:2018aqg}, cosmic-ray observations \citep{Cappiello:2018hsu}, and Lyman-$\alpha$-forest measurements~\citep{Dvorkin:2013cea,Xu:2018efh}. 
LSST will probe the matter power spectrum on even smaller scales, through substructure measurements from dwarf galaxies in the Local Volume, gaps in stellar streams, galaxy strong lensing, and galaxy-galaxy weak lensing; such observations will substantially extend current experimental sensitivity to BSDM models.

As an example, a measurement of the minimum halo mass translates into an upper limit on the dark matter-proton interaction cross section, based on the corresponding cutoff in the matter power spectrum $P(k)$. \figref{dmbaryon_pk} shows how the position of the cutoff in the linear $P(k)$ varies as a function of the interaction cross section. For example, a lower limit on the cutoff of $k_\text{cutoff} \sim 10/\Mpc$ roughly corresponds to an upper limit on the cross section which is 100 times more stringent than the current limit from CMB searches. Using limits on WDM as a proxy for a suppressed $P(k)$, a minimum halo mass of $\roughly 10^6 \Msun$ would imply an improvement of roughly five orders of magnitude compared to the best current cosmological limits on the interaction cross section.

\section{Field Dark Matter \Contact{Chanda}}
\label{sec:axions}
\Contributors{Chanda Prescod-Weinstein, Samuel D.\  McDermott, Oscar Straniero,  Maurizio Giannotti, Alex Drlica-Wagner, Manuel Meyer}

While current observations of the matter power spectrum constrain the minimum mass of thermally produced dark matter, other mechanisms can  produce dark matter with significantly lower masses. The landscape of light dark matter candidates is vast, and in this section, we specifically focus on the class of axion-like particle (ALP) dark matter candidates.
ALP models span a wide range of viable parameter space (both in coupling strength and mass), and many of the observables described in this section can be generically applied to a broader class of light scalar particles.

The ALP paradigm was inspired by the QCD axion, which arises as a by-product of the most successful solution to the Strong CP Problem in the Standard Model \citep{PecceiQuinn:1977}. 
The cosmological abundance of axions is set by the Peccei-Quinn symmetry breaking scale, $f_\phi$, with a value
\begin{equation}
\Omega_\phi\sim\left(\frac{f_\phi}{10^{11-12}\,{\rm GeV}}\right)^{7/6}.
\end{equation}
This expression may be altered due to the temperature-dependence of the axion mass and ignorance about whether the Peccei-Quinn symmetry breaks before or after inflation. 
QCD theory gives no {\it a priori} prediction for the axion mass; however, in the context of dark matter composed of QCD axions, the axion mass is considered to be $m_\phi< 10^{-3} \eV$. %($10^{-39}$ kg) 
If the initial misalignment angle is order unity, this yields a QCD axion mass of $m_\phi \sim 10^{-5} \eV$. %($10^{-41}$ kg).
The broader category of ALPs possess QCD-axion-like potentials producing light scalar particles that obey a shift symmetry ($\phi \rightarrow \phi + 2\pi n$), but do not obey the same coupling between particle mass and symmetry breaking scale. 
ALPs can be motivated by string theory, where there are many moduli with axion-like potentials, and can produce a range of astrophysical phenomenology.
%ALPs can be sufficiently different from the QCD axion so as to produce notably different astrophysical phenomenology. 
ALPs may be non-thermally produced in the early universe and survive as a cold dark matter population until today \citep[\eg][]{Arias:2012az}.

There has been significant debate in the literature about the astrophysical phenomenology of the QCD axion and ALPs.
\citet{Sikivie:2009} noted that because the axion is a scalar with high abundance in the early universe (circa matter-radiation equality), the axion could potentially settle into a Bose-Einstein condensate (BEC) state, whereby all particles can be described using one coherent ground state wave function. 
Furthermore, \citet{Sikivie:2009} argue that during the radiation-dominated era, axions will rethermalize into BECs with a Hubble-scale correlation length.
This could produce significant observational implications, such as several-kpc-scale caustic structures observable in the stellar distributions of the Milky Way and other low-redshift galaxies \citep[\eg,][]{Natarajan:2006,0805.4556,Rindler-Daller:2013zxa}.

On the other hand, \citet{1412.5930} argue that a particle such as the QCD axion, which has an attractive self-interaction, will not sustain Hubble-scale correlations in an attractive potential.
Instead, \citet{1412.5930} predict that axions will form coherent clumps that have been called ``Axion stars'' or ``Bose stars'' \citep[\eg][]{Kolb:1993}.\footnote{For the QCD axion it would be more appropriate to call these ``axteroids'' (a term coined by Anna Watts) due to their mass of $\roughly 10^{-11} \Msun$ \citep{Tkachev:1991ka,Braaten:2018nag}.} 
Looking beyond the QCD axion, some ALP models suggest that compact BEC ``miniclusters'' could form and grow to $\gtrsim 1\Msun$, at which point they may be detectable by LSST through mergers with other compact objects \citep{1808.04746} or through microlensing \citep{1707.03310}, as discussed in \secref{compact_objects}.

Additional astrophysical constraints on ALPs generally come from proposed couplings with photons and/or electrons. 
For example, the Lagrangian can be expressed as
\begin{equation}
    \mathcal{L} = -\frac{1}{2} \partial_\mu\phi\partial^\mu\phi + \frac{1}{2}m_\phi^2 \phi^2 - \frac{1}{4}g_{\phi\gamma}F_{\mu\nu}\tilde{F}^{\mu\nu}\phi - g_{\phi e}\frac{\partial_\mu\phi}{2m_e}\bar{\psi}_e \gamma^\mu\gamma_5\psi_e,
\end{equation}
where $g_{\phi\gamma}$ is the photon-axion coupling, $g_{\phi e}$ is the axion-electron coupling, $F^{\mu\nu}$ is the electromagnetic field tensor (and $\tilde{F}$ its dual), and $\psi_e$ is the electron field \citep[\eg][]{1302.6283,Redondo:2013wwa}.\footnote{Additional couplings to nucleons are allowed, but are not relevant for the LSST observations discussed here.}
For sufficiently large couplings to photons or electrons, the ALP can manifest as an additional anomalous energy loss mechanism, transporting energy out of the interiors of stars \citep[\eg,][]{Raffelt:1990}.
This energy loss could affect the evolution of stars, for example altering the lifetimes of giant stars \citep{Ayala:2014,Viaux:2013hca,Viaux:2013lha} or the cooling rate of white dwarf stars \citep{Isern:2008nt}.
The precise photometry of LSST will provide sensitive measurements of stellar populations to search for deviations from the predictions of standard stellar evolutionary models (\secref{cooling}).

Astrophysical observations place the only known lower bound on the mass of ALPs and other non-thermally produced ultra-light particles, commonly described as ``fuzzy'' dark matter \citep[FDM; \eg,][]{Hu:2000,Hui:2017}. 
The de Broglie wavelength of these particles is constrained to be smaller than the size of the smallest galaxy, $\mathcal{O}(1\kpc)$, setting a lower limit on particle mass at $m_\phi \gtrsim 10^{-21}\eV$ \citep{1703.04683}. 
In addition, FDM is predicted to produce solitonic cores in the centers of halos, which would measurably affect the velocity profiles of dark-matter dominated galaxies \citep{Robles:2012uy,1807.06018,Schive:2014hza,Du:2016aik,1609.05856}. 
Dark matter substructure is predicted to be less abundant in FDM than its CDM counterpart due to quantum interference effects.
This is similar to the case of WDM, and again dark matter properties can be constrained through mesurements of the least massive dark matter halos.
Combining Equation 8 of \citet[][]{1703.09126} with \eqnref{Mhm} in \secref{wdm}, we can express constraints on the minimum FDM mass, $m_\phi$, as a function of the half-mode halo mass, $M_{\rm hm}$:
\begin{equation}
M_{\rm hm} = 1.2 \times 10^{11} \left( \frac{m_\phi}{10^{-22}\eV} \right)^{-1.4} \Msun,
\end{equation}
or expressed in terms of $m_\phi$, 
\begin{equation}
m_\phi = 3.1 \times 10^{-21} \left( \frac{M_{\rm hm}}{10^{9}\Msun} \right)^{-0.71} \eV.
\end{equation}
LSST will be sensitive to light bosonic dark matter with mass $m_\phi \sim 10^{-20} \eV$ by probing the power spectrum of dark matter halos with half-mode mass of $\Mhm \sim 10^{8} \Msun$. 
Sensitivity to heavier bosonic particles ($m_\phi > 10^{-19} \eV$) would be possible through the detection of even smaller halos ($\roughly 10^{6} \Msun$).

\section{Compact Objects \Contact{Simeon}}
\label{sec:machos}
\Contributors{Simeon Bird, Juan Garc\'ia-Bellido, George Chapline, William A.\ Dawson, Nathan Golovich, Michael Medford}

Compact objects, particularly black holes, represent one of the oldest and most venerable models of dark matter. 
%Primordial black holes could originate from small-scale density fluctuations during the era of inflation. 
Primordial black holes could form at early times from the direct gravitational collapse of large density perturbations that originated during inflation.
The same fluctuations that lay down the seeds of galaxies, if boosted on small scales, can lead to some small areas having a Schwarzschild mass within the horizon, which spontaneously collapse to form black holes \citep{Carr:1974nx,Meszaros:1974,1975Natur.253..251C,Bellido:1996,2016PhRvD..94h3504C}. 
Alternatively, some particle dark matter models may allow dark matter cooling and collapse, providing another mechanism for black hole formation \citep[\eg][]{1705.10341,1707.03419,1707.03829,1802.08206,1812.07000}.
%Because these black holes do not accrete or radiate strongly (at the time of formation there is no gas to form an accretion disc), they are a natural candidate for dark matter 
Such black holes are unlikely to radiate strongly enough from accretion to leave a detectable signature and are thus a natural candidate for dark matter.
The abundance of compact objects tests dark matter through a purely gravitational channel and is thus sensitive to dark matter models that cannot be probed in the laboratory.

Compact object dark matter is fundamentally different from particle models; primordial black holes cannot be studied in an accelerator and can only be detected through their gravitational force. Current constraints suggest that primordial black holes do not make up all of dark matter \citep[\eg][]{Sasaki:2018}. However, these constraints may be evaded if primordial black holes are spatially clustered \citep{Clesse:2015,Clesse:2017}. Moreover, primordial black holes are one possible source of the merging $30 \Msun$ black holes recently detected by LIGO \citep{Bird:2016,Clesse:2016}. This possibility has rekindled interest in these objects, both as a source of dark matter and in their own right.

Limits on the abundance and mass range of primordial black holes are wholly observational. The black hole mass is set by the mass enclosed within the horizon at the time of black hole collapse and thus ranges between $10^{-18} \Msun$ ($10^{15}\g$), below which the black hole would evaporate, and $10^9 \Msun$ ($10^{42}\g$), above which structure formation, Big Bang Nucleosynthesis and the formation of the microwave background would be severely affected \citep{Sasaki:2018}. 
For stellar mass black holes, the gold standard for detecting compact objects is microlensing. Current microlensing constraints set limits on the black hole abundance at the level of $10\%$ for black holes $0.01 - 10 \Msun$ \citep[however, see][]{Calcino:2018}. LSST will revolutionize the astrometric microlensing technique,  constraining the abundance of primordial black holes to a level of $10^{-4}$ of the dark matter over a wide range of masses (\secref{compact_objects}).

As primordial black holes form directly from the primordial density fluctuations, a measurement of their abundance would directly constrain the amplitude of density fluctuations \citep{Carr:1974nx, Meszaros:1974}. %1203.2681 
Although these constraints are several orders of magnitude weaker than, for example, those from the microwave background, they probe small scales between $k = 10^{7} - 10^{19}$ $h$/Mpc, much smaller than those measured by other current and future probes \citep{Bringmann:2012}. Because these scales are highly non-linear in the late-time universe, there is no other possible constraint; the information present at early times has been washed away by gravitational evolution. Primordial black holes are thus a probe of primordial density fluctuations in a range that is inaccessible to other techniques~\citep{Josan:2009,Bellido:2017,Bellido:2018}. These curvature fluctuations are imprinted on space-time hypersurfaces during inflation, at extremely high energies, beyond those currently accessible by terrestrial and cosmic accelerators. 
Our understanding of the universe at these high energies, of order $10^{15} \GeV$ and above, comes predominantly from extrapolations of known physics at the electroweak scale.
Measurements of the primordial density fluctuations via the abundance of primordial black holes would provide unique insights into physics at these ultra-high energies.

In additon, dark matter interactions with the Standard Model may also generate new channels for black hole formation, by triggering collapse of astrophysical objects \citep[\eg][]{1989PhRvD..40.3221G,1004.0629,1012.2039,1405.1031,1804.06740}. LSST could be sensitive to transient events that could be triggered by these formation scenarios \citep[\eg][]{1706.00001} or as part of a muli-messanger campaign to measure sub-solar mass black hole mergers \citep[\eg][]{1808.04771,1808.04772}.

Furthermore, it may be possible for LSST to constrain the existence of ultra-compact mini-halos using correlated microlensing signals \citep{erickcek2011,li2012}. These objects arise from initial overdensities that are too small to collapse into primordial black holes. These overdensities still collapse at high redshift to form low-mass halos; thus, since these objects form early and have few mergers \citep{Bringmann:2012,Delos:2018}, they have a high concentration and a steeper internal density profile than the standard Navarro-Frenk-White shape. In turn, this makes them easier to detect via lensing and harder to disrupt than standard CDM subhalos. Current constraints on these objects are highly model-dependent. In particular, they largely come from counting gamma-ray photons from astrophysical sources under the assumption of a WIMP dark matter annihilation cross-section. LSST will place new constraints on the existence of small halos via micro-lensing and thereby constrain the physics of the inflaton on scales of $k = 10 \textup{--} 10^7 h$/Mpc for the first time in a model-independent way.

%%%%%%%%%%%%%%%%%%%%%%%%%%%%%%%%%%%%%%%%%%%%%%%%%%%%%%%%%%%%%%%%%%%%%%%%%%%%%%%%
% Dark Matter Probes
%%%%%%%%%%%%%%%%%%%%%%%%%%%%%%%%%%%%%%%%%%%%%%%%%%%%%%%%%%%%%%%%%%%%%%%%%%%%%%%%
\chapter{Dark Matter Probes}
\label{sec:probes}
\bigskip
\Contributors{Arka Banerjee, Nilanjan Banik, Keith Bechtol, Kimberly K.\ Boddy, Ana Bonaca,  Jo Bovy, Francis-Yan Cyr-Racine, Alex Drlica-Wagner, Cora Dvorkin, Denis Erkal, Christopher D.\ Fassnacht, David Hendel, Yashar D.\ Hezaveh, Charles R.\ Keeton, Sergey Koposov, Ting S.\ Li, Yao-Yuan Mao, Mitch McNanna, Marc Moniez, Ethan O.\ Nadler, Andrew B.\ Pace, Annika H.\ G.\ Peter, Daniel~A.~Polin, Rogerio Rosenfeld, Nora Shipp, Louis E.\ Strigari, Erik Tollerud, J.\ Anthony Tyson, Mei-Yu Wang, Risa H.\ Wechsler, David Wittman}

Each of the theoretical models described in \secref{theory} produce one or more deviations from the predictions of cold, collisionless, non-interacting dark matter.
These ``probes'' of dark matter physics include: a minimum dark matter halo mass, alterations to halo density profiles, an over-abundance of compact objects, anomalous energy loss, and unexpected correlations in large-scale structure.
In some cases, several distinct physical models of dark matter can be probed by the same (or very similar) observables (\eg, keV-mass thermal dark matter and ultra-light fuzzy dark matter). 
On the other hand, a single probe can manifest itself in a wide range of astrophysical systems (\eg, changes to dark matter profile shape could be observable in the least massive galaxies and the most massive clusters of galaxies).
In this section we do not attempt to provide a comprehensive discussion of all possible astrophysical probes of dark matter physics.
Rather, we focus on specific probes and observables where LSST will have a major impact.
%ADW: should we note something like, "major impact alone or in combination with other observation"?

\section{Minimum Halo Mass}
\label{sec:halo_mass}

\Contributors{Arka Banerjee, Nilanjan Banik, Keith Bechtol, Ana Bonaca, Kimberly K.\ Boddy, Jo Bovy, Francis-Yan Cyr-Racine, Alex Drlica-Wagner, Ana D{\'i}az Rivero, Cora Dvorkin, Denis Erkal, Christopher D.\ Fassnacht, David Hendel, Yashar D.\ Hezaveh, Charles R.\ Keeton, Sergey Koposov, Ting S.\ Li, Yao-Yuan Mao, Mitch McNanna, Ethan O.\ Nadler, Andrew B.\ Pace, Nora Shipp, Erik Tollerud, Mei-Yu Wang, Risa H.\ Wechsler}

The cold, collisionless model of dark matter makes a strong prediction that dark matter halos should exist down to Earth-mass scales (or below) in WIMP and non-thermal axion models \citep{Green:2003un,2005Natur.433..389D,1412.5930}.
Several modifications to the cold, collisionless dark matter paradigm can suppress the formation of dark matter halos on these small scales.
Current observations provide a robust measurement of the dark matter halo mass spectrum for halos with mass $> 10^{10}\Msun$, and the smallest known galaxies provide an existence proof for halos of mass $\roughly 10^8 \Msun - 10^9 \Msun$ \citep{2017MNRAS.467.2019R,behroozi2018,Jethwa:2018,Kim:2017iwr,Nadler:2018,1807.07093}. 
Extending below this halo mass threshold is challenging due to our limited observational sensitivity to the faintest galaxies.
In addition, halos with mass $\lesssim 10^8 \Msun$ are generally expected to host few (if any) stars \citep{1102.4638,1505.06209}, necessitating novel detection techniques that do not rely on the baryonic content of halos.
Here we explore improvements that LSST will make in measuring the faintest galaxies and in probing dark matter halos below the threshold of galaxy formation with stellar streams and strongly lensed systems. We then use these improvements to forecast constraints on specific dark matter models.

\subsection{Milky Way Satellite Galaxies \Contact{Ethan}} 
\label{sec:smallest_galaxies}
\Contributors{Ethan O.\ Nadler, Keith Bechtol, Alex Drlica-Wagner, Mitch McNanna, Andrew B.\ Pace, Yao-Yuan Mao, Erik Tollerud, Risa Wechsler, Francis-Yan Cyr-Racine, Mei-Yu Wang, Kimberly K.\ Boddy, Arka Banerjee}

% Some slides from KITP Dark Matter Program in May 2018 with potential ideas:
% https://drive.google.com/open?id=1BhuwyNE7vClIeVV6FhM99QtictTvpQ0p

\vspace{1em} \noindent \textbf{The Threshold of Galaxy Formation}

Galaxies are born and grow within dark matter halos.
To first approximation, galaxies with the largest stellar masses reside within the highest-mass dark matter halos, while fainter galaxies---which are much more numerous---occupy dark matter halos with progressively smaller typical masses; however, the scatter between stellar mass and halo mass is likely large in the low-mass regime (see \citealt{Wechsler:2018} for a recent review).
Therefore, the smallest and faintest galaxies offer a natural place to search for extremely low-mass dark matter halos, which are in turn sensitive probes of dark matter microphysics. Another advantage of probing low-mass dark matter halos using faint galaxies is that we can study their properties in detail, \eg, via follow-up spectroscopy (\secref{complementarity}).

The challenge in interpreting observations of faint galaxies is the complex relationship between baryons and halos at this extreme mass scale and the effects of baryonic physics both within subhalos and on subhalo populations as a whole \citep[\eg,][]{DOnghia:2009xhq,Brooks:2012ah,errani2017,2017MNRAS.471.1709G,1811.11791,brooks2018}. Nevertheless, probing the extreme faint end of the galaxy luminosity function is valuable both astrophysically and in terms of constraining dark matter models. For example, a driving question for near-field cosmology with LSST is how well we can use the population of Milky Way satellites to constrain the minimum dark matter halo mass necessary for galaxy formation. 
This ``minimum halo mass" depends on the details of reionization and other forms of baryonic feedback that prevent gas from accreting and cooling in low-mass subhalos; however, it might also reflect a cutoff in the subhalo mass function determined by the particle nature of dark matter (\eg, WDM or FDM). In particular, models that produce a cutoff in the matter power spectrum generally suppress the number of subhalos below a characteristic mass threshold (\eqnref{Mhm}). Thus, the existence, abundance, and properties of the smallest galaxies generically lead to constraints on dark matter models that reduce small-scale power.

\vspace{1em} \noindent {\bf Minimum Subhalo Mass Inferred from Milky Way Satellites}

\begin{figure}[t]
\centering
\includegraphics[width=0.775\textwidth]{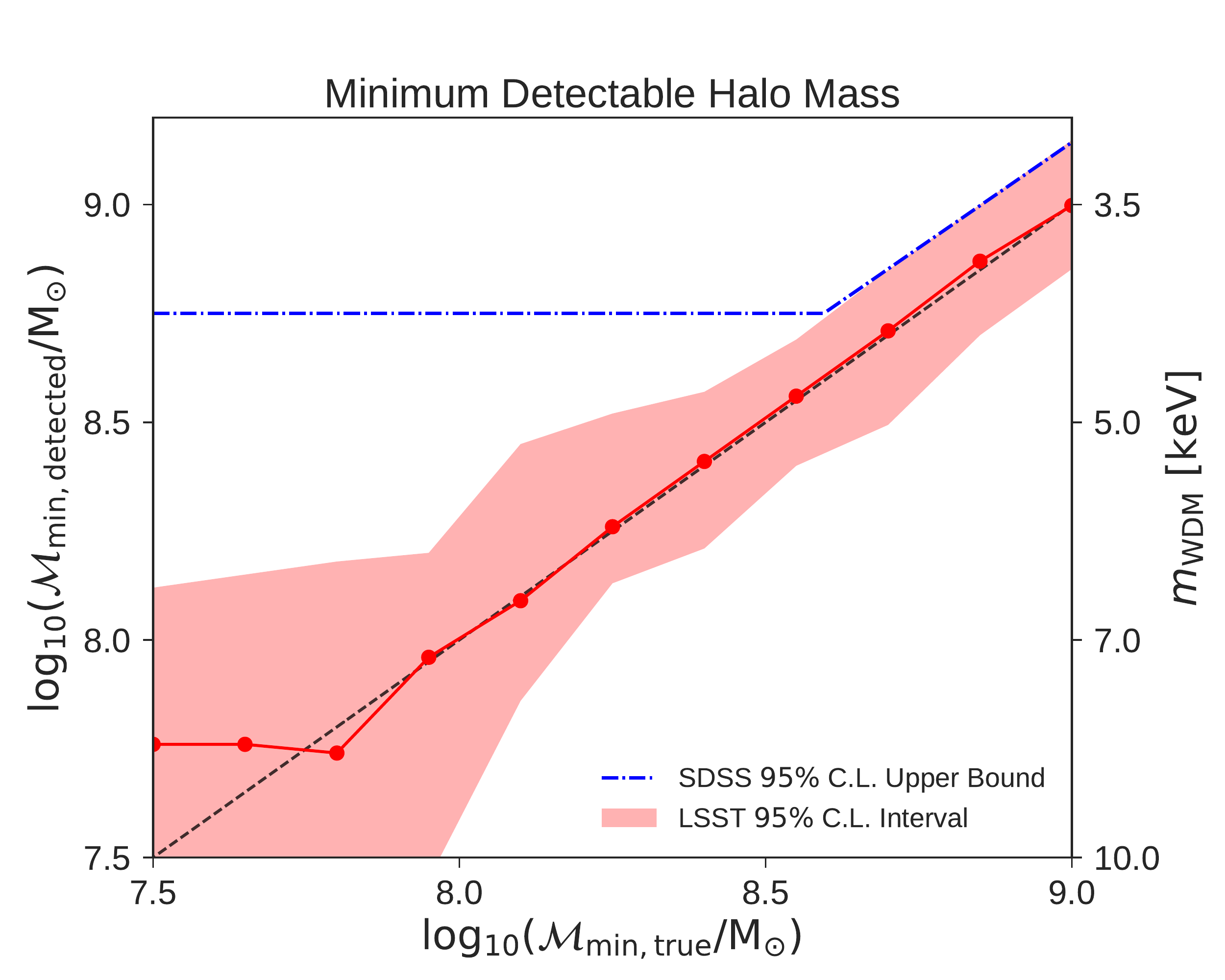}
\caption{Forecast for the minimum dark matter subhalo mass probed by LSST via observations of Milky Way satellites. The red band shows the $95\%$ confidence interval from our MCMC fits to mock satellite populations as a function of the true peak subhalo mass necessary for galaxy formation. Note that we marginalize over the relevant nuisance parameters associated with the galaxy--halo connection---including the effects of baryons using a model calibrated on subhalo disruption in hydrodynamic simulations \citep{2018ApJ...859..129N}---in our sampling. We indicate the corresponding constraints on the warm dark matter mass assuming $M_{\rm hm} = \mathcal{M}_{\rm{min}}$ (see \secref{wdm})}\label{fig:satellite_mmin}
\end{figure}

The least luminous galaxies currently known contain only a few hundred stars and have been found exclusively in the inner regions of the Milky Way due to observational selection effects. Although the census of Milky Way dwarf galaxies has grown from $\roughly 25$ to more than 50 in recent years \citep[\eg, with DES;][]{Bechtol:2015, Koposov:2015, Drlica-Wagner:2015}, our current census is certainly incomplete.
For example, the HSC-SSP Collaboration has detected two ultra-faint galaxy candidates in the first $300 \deg^2$ of the survey \citep{1609.04346,1704.05977}; these galaxies are faint and distant enough to have been undetectable in previous optical imaging surveys. HSC is representative of the depth that will be achieved by LSST over half the sky---an area 60 times larger than the current HSC-SSP footprint. Thus, based on the results of SDSS, HSC, DES, etc., several groups have predicted that LSST could detect %at least 20 --- and as many as 50 \EON{check numbers} \ADW{Seems a bit small. I thought Hargis predicted $\roughly250$.} \AHGP{Check out the numbers from Table 1 of Kim, Peter, \& Hargis: we predict 50-600 depending mostly on the radial distribution of satellites in the halo.  The lower number is the most conservative choice, for a population of UFDs concentrated at the center of the halo, and is consistent with Newton et al.  The largest number is what we predict if the hydro sims of satellite disruption are correct.}--- 
tens to hundreds of new low-luminosity Milky Way satellites, mainly at larger distances and fainter luminosities than those accessible with current-generation surveys \citep{Koposov:2008,Tollerud:2008,Hargis:2014,Newton:2018,Jethwa:2018,Nadler:2018,Kim:2017iwr}. 
In addition, novel techniques, such as the use of the correlated phase space motions of stars \citep{1507.04353,1805.02588} or clustering of variable stars \citep{1507.00734} could further expand the sample of ultra-faint galaxies.
LSST observations of Milky Way satellites therefore offer an exciting testing ground for dark matter models; for example, the measured abundance, luminosity function, and radial distribution of Milky Way satellites \emph{already} place competitive constraints on warm dark matter particle mass at the level of 3--4\keV \citep[\eg,][]{Jethwa:2018,Kim:2017iwr}.%, and these constraints will improve as observations continue to probe the low-mass end of the subhalo mass function.

To relate these questions to LSST observations, we have analyzed simulated ultra-faint galaxies as they would appear in LSST WFD coadd object catalogs to quantify LSST's ability to detect nearby satellite galaxies. %, and we have developed a theoretical framework to connect observations of the Milky Way satellite population to the underlying dark matter subhalo population. 
We detect ultra-faint galaxies as arcminute-scale statistical overdensities of individually resolved stars; in ground-based optical imaging surveys, it is often challenging to classify low signal-to-noise catalog objects near the detection threshold as either foreground stars or unresolved background galaxies. LSST will reach depths at which the galaxy counts far outnumber stellar counts, so the search sensitivity for ultra-faint galaxies will largely be determined by our ability to accurately perform star-galaxy separation at magnitudes $24 < r < 27.5$; importantly, our sensitivity analyses include these effects. 
%In detail, we inject many simulated stellar populations into the center of the LSST DESC DC2 simulated data set and 
We find that Milky Way satellites within $300\kpc$ are well-detected with a surface brightness detection threshold of $\mu = 32\ \rm{mag\ arcsec}^{-2}$ \citep{0912.0201} and an absolute magnitude cutoff of $M_V = 0\magn$.

\figref{satellite_mmin} shows the minimum subhalo mass that LSST can probe via observations of Milky Way satellites, obtained by folding our search sensitivity estimates through a cosmological model of the Milky Wway satellite population that predicts satellite luminosity functions, radial distributions, and size distributions that agree well with current observations. In particular, we generate many mock Milky Way satellite populations using the model presented in \cite{Nadler:2018} given a ``true" value of the minimum peak subhalo virial mass necessary for galaxy formation, $\mathcal{M}_{\rm{min,true}}$, marginalizing over the remaining galaxy--halo connection parameters. We then perform mock observations of these generated satellite populations using the LSST selection function, and we compare these to the true satellite populations by MCMC sampling $\mathcal{M}_{\rm{min}}$ and the remaining galaxy--halo connection parameters assuming that satellite number counts are Poisson distributed in bins of absolute magnitude (see \citealt{Nadler:2018} for details on the fitting procedure). For each value of $\mathcal{M}_{\rm{min,true}}$, this procedure yields a posterior distribution over the minimum halo mass inferred by LSST observations. The red band in \figref{satellite_mmin} illustrates the recovered $95\%$ confidence interval as a function of $\mathcal{M}_{\rm{min,true}}$, and the blue dot-dashed line indicates the minimum halo mass inferred from known classical and SDSS-detected Milky Way satellites. 
%LSST observations recover the true minimum halo mass at large values of $\mathcal{M}_{\rm{min,true}}$, since all of the predicted satellites are observable in this regime, while smaller values of $\mathcal{M}_{\rm{min,true}}$ yield satellites that do not pass our detection criteria, which prevents the lowest-mass subhalos that host satellites to be detected. 
For small $\mathcal{M}_{\rm{min,true}}$, the $95\%$ confidence level upper bound on the lowest detectable subhalo mass improves by a factor of $\sim 5$ with LSST, from $\sim 5 \times 10^{8}\ \Msun$ to $\sim 10^{8}\ \Msun$; this translates to a lower bound of $\sim 7\ \rm{keV}$ on WDM particle mass (see \secref{combine_probes} for details).

Although we have presented a ``population-based'' forecast for dark matter constraints from LSST-detected ultra-faint satellites, we note that kinematic data obtained by follow-up spectroscopy of newly discovered satellites also offers a powerful probe of dark matter microphysics. We estimate the number of LSST-detected Milky Way satellites that can be spectroscopically confirmed in \secref{spectroscopy}, and we forecast the constraints offered by these stellar velocity dispersion measurements for WDM and SIDM in \secref{combine_probes}.

Further extending the sensitivity of LSST to a power spectrum cut-off on scales smaller than the mass threshold for galaxy formation requires techniques that are independent of satellite luminosity and that can detect subhalos purely through their gravitational signatures. Two examples of such probes are described in the following subsections.

\subsection{Stellar Stream Gaps \Contact{David H.}}
\label{sec:stream_gaps}
\Contributors{David Hendel, Nora Shipp, Ting S.\ Li, Ana Bonaca, Jo Bovy, Sergey Koposov, Denis Erkal, Nilanjan Banik, Andrew B.\ Pace}

Stellar streams, in particular the tidally disrupting remnants of globular clusters, are fragile, dynamically cold systems and are sensitive tracers of gravitational perturbations \citep[][]{2002MNRAS.332..915I,2002ApJ...570..656J,2011ApJ...731...58Y,Carlberg:2012}.
The main track of a stream in 6D phase space is shaped primarily by the Milky Way's global matter distribution while the detailed structure of the stream contains information about small-scale perturbations. 
In particular, a dark matter subhalo passing by the stream will provide a net velocity kick, altering the orbits of the closest stream stars.
The main observable consequence of this interaction is the formation of a gap in the density of stars along the stream; the relative depth and size of the underdensity can be used to infer the time since the encounter and the properties of the perturber \citep{Carlberg:2012, Erkal:2015}. The mass required to produce an observable gap \citep[$10^5$--$10^6 \Msun$,][]{erkal2016,bovy:2017} is well below the limit where dark matter subhalos are expected to host galaxies. Thus, stellar streams provide one of the most exciting near-field tests of the minimum subhalo mass.

Current constraints on the minimum subhalo mass from stream gaps are limited by the small number of streams that are bright enough that observations can detect density variations at a useful signal-to-noise ratio. Deep and precise LSST photometry is expected to increase the contrast between streams and the contaminating Milky Way field stars, to have improved star-galaxy separation, and to extend much farther down the color-magnitude diagram for known streams, dramatically increasing our ability to detect density variations and thus leading to the identification of less prominent gaps created by low-mass perturbers. Critically, with LSST we move from examining individual gaps into the regime where we can ask questions about subhalo population statistics and their (in)consistency with cold dark matter.
Here we estimate the least massive subhalo that can be detected with LSST observations of gaps in stellar streams.

We consider a mock-stream observed at a Galactic latitude of $b=-60^\circ$ in the $g$- and $r$-band. We assume the stream is old (12\,Gyr), metal-poor ($Z = 0.0002$), thin (1$\sigma$ stream width of 20\,pc), and cold (velocity dispersion of 1\,km\,s$^{-1}$) We generate synthetic photometry of the stream at a given mean surface brightness (within the $1\sigma$ width) and over a range of heliocentric distances from 10 to $40\kpc$.  Simulated stream stars are drawn from a Chabrier IMF \citep{2003PASP..115..763C}, while a synthetic background of Milky Way stars is generated from the \code{Galaxia} model \citep{sharma2011}. 
We add noise to the simulated photometric measurements for both the stream and Milky Way stars in accordance with expectations for LSST \citep{0805.2366}. 
We then select stars in the color-magnitude diagram that are within $2\sigma$ of the theoretical isochrone of the stream's age and metallicity, where $\sigma$ is the magnitude-dependent photometric uncertainty using the same error model. 
% we assume $\sigma$ is no less then 0.02 mag (i.e. set to 0.02 if it's smaller).
We also assume a limiting magnitude to set the depth of the survey, choosing the point where the photometric uncertainty in either band exceeds 0.1 mag. We apply this color-magnitude selection to determine the density of stream stars and background stars.
The depth of the gap from a given subhalo mass is calculated using the theoretical relation derived by \citet{erkal2016}, assuming that the subhalo impact occurred within the past $0.5\Gyr$, moving at $150\kms$, with an impact parameter equal to the perturber's scale radius. Finally, we define a detection as a gap depth that is $5\,\sigma$ above the noise background (the effects of star-galaxy separation are not considered in this calculation).

\begin{figure}[t]
\centering
\includegraphics[width=0.85\textwidth]{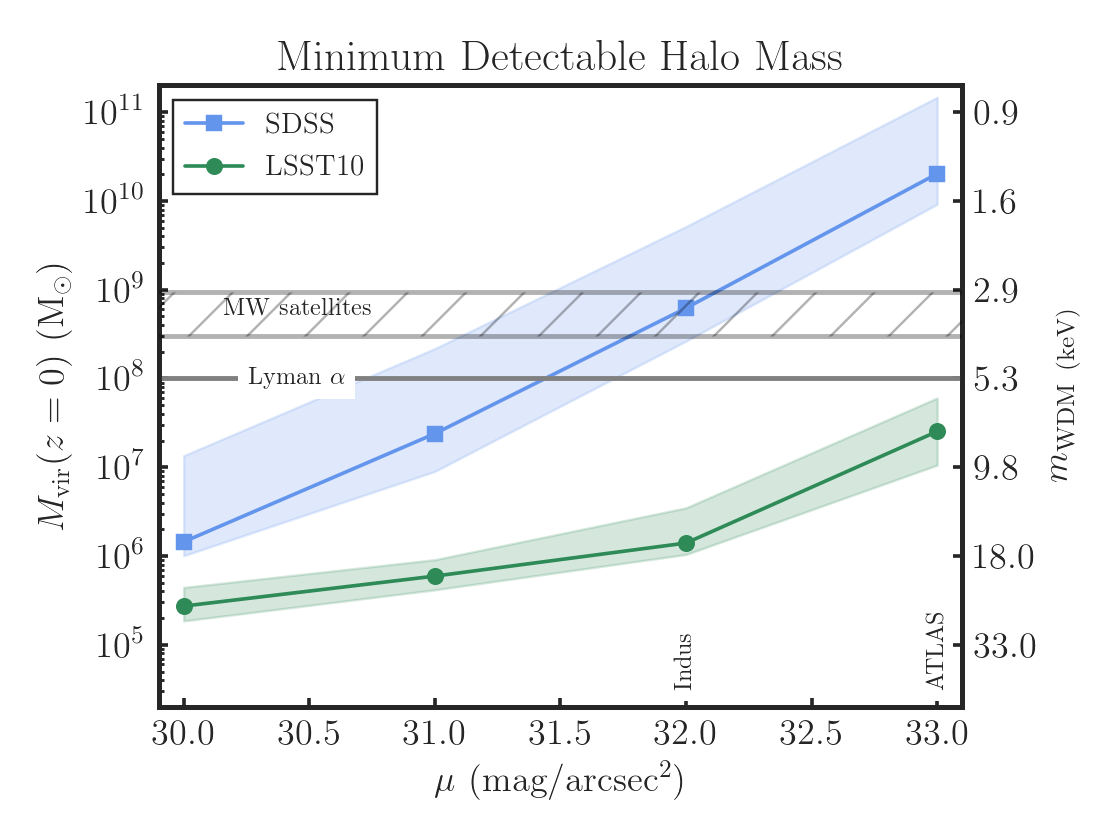}
\caption{\label{fig:streamsurveys} Detection limits for gaps formed from subhalos of different masses using photometry from SDSS (blue) or the 10-year LSST stack (green) as a function of the stream surface brightness.
Shaded regions correspond to a 10-40 kpc distance range, with the lines representing 20 kpc. For streams with surface brightnesses similar to those found in DES, 32--$33 \magn \asec^{-2}$, LSST is expected to probe halo masses two to three orders of magnitude smaller than SDSS and substantially improve the current constraints from Milky Way satellites \citep{Nadler:2018, Jethwa:2018,Kim:2017iwr} and the Lyman-$\alpha$ forest \citep{2017PhRvD..96b3522I}. 
We connect the detected halos to the mass of the warm dark matter particle that would produce a minimum halo of that mass using the relationship determined by \cite{Bullock:2017}. Note that the halo mass definition used here is the $z=0$ virial mass; to relate this quantity to the peak subhalo mass used in our warm dark matter constraints, we have assumed the best-case scenario of no tidal mass loss.
}
\end{figure}

\figref{streamsurveys} shows the lowest-mass dark matter subhalo detectable using the 10-year LSST data as a function of stream surface brightness and heliocentric distance. For a stream with a surface brightness of 33 (31.5) $\mathrm{mag}\,\mathrm{arcsec}^{-2}$, LSST is able to detect subhalos with $M_{\rm vir}(z=0) \sim 2 \times 10^7 \Msun$ ($1 \times 10^6 \Msun$) at 20 kpc.
%\footnote{Note that these correspond to $z=0$ virial masses; to convert these into the peak mass that enters our half-mode mass estimate, we measure the mean $M_{\rm{peak}}$--$M_{\rm{vir}}$ relation from the highest-resolution simulation presented in \cite{Mao2015}.} 
As a comparison, we used the same model to calculate the gap detectability using SDSS DR9 photometry. LSST provides $\roughly 3$ orders of magnitude improvement at low surface brightnesses, where most known (and anticipated) streams lie. Crucially, this pushes the minimum detectable halo mass below current constraints from Milky Way satellites \citep[\eg,][]{Nadler:2018,Jethwa:2018,Kim:2017iwr} or the Lyman-$\alpha$ forest \citep[\eg,][]{2017PhRvD..96b3522I}.
To set constraints on \Mhm and \mWDM, we convert $M_{\rm vir}(z=0)$ to $M_{\rm peak}$ using a relation derived from the highest-resolution simulation presented in \citet{Mao2015}.\footnote{We find a mean relationship of: $\log_{10}(M_{\rm peak}) = 0.88 \log_{10}(M_{\rm vir}) + 1.28$.}
We then follow the same formalism as in \secref{smallest_galaxies} to convert from $M_{\rm peak}$ to $\Mhm$ and calculate \mWDM from \eqnref{mWDM}.

Given a gap density detection threshold and a subhalo population, this formalism can be used to predict the number of gaps in a given stream \citep{erkal2016}. Typically, this predicts $\roughly 1$ gap per stream, making it difficult to interpret well-studied individual streams (i.e. Palomar 5 and GD-1). LSST is expected to measure dozens of streams as precisely as Palomar 5 and GD-1 have currently been mapped and will therefore provide a much stronger constraint: at the 10-year LSST depth, \LCDM predicts that we should observe 17 gaps total in the 13 DES streams reported by \cite{2018ApJ...862..114S}. Observing fewer than 6 gaps would be inconsistent with \LCDM at a 99.9\% level. This prediction assumes that all of the streams have a dynamical age of 8\Gyr and that subhalo disruption follows the model of \citet{erkal2016}.

A single stellar stream is expected to experience several subhalo encounters over its dynamical lifetime. 
Recent strong impacts will result in the observable gaps described above, while weaker encounters will cause less prominent density and track variations.
The effects of ancient impacts will be gradually erased due to the internal velocity dispersion of the stream member stars; however, it is possible to extract statistical information about the impact history of the stream by studying the linear density and track power spectra, both of which are sensitive to the subhalo mass functions.
Impacts from higher-mass subhalos introduce power on large scales, while lower-mass subhalo impacts introduce power on small scales. 
Statistical analyses of the stream density power spectrum have been used to constrain the number of subhalos within the stream radius and the properties of dark matter \citep{bovy:2017, 2018JCAP...07..061B}. 
LSST will allow us to measure the stream density and stream track power spectra at small angular scales that were previously dominated by noise, and \citet{bovy:2017} project that the power spectrum method will be sensitive to subhalos down to mass $10^5 \Msun$. 
In addition, precise measurement of the densities and tracks of multiple streams can be combine to increase statistical power and mitigate systematics from any individual stream.

Depending on their orbits, stellar streams can also be perturbed by the baryonic structures such as the Galactic bar \citep[\eg,][]{erkal2017,pearson2017}, spiral arms \citep{Banik2018}, or giant molecular clouds \citep{amorisco2016}. The resulting gaps may be difficult to distinguish from gaps induced by dark matter subhalos and can bias measurements toward overestimating the number of subhalo impacts on a stellar stream. The only recourse is to carefully examine the streams' orbits to assess these possible confounding factors. Streams with pericenters of $\gtrsim 14\kpc$ should be relatively unaffected by these baryonic factors, and streams on retrograde orbits even less so. In addition, subhalos may also experience extra tidal shocks from the disk, which can alter the number of expected impacts in a given cosmological model \citep[\eg,][]{DOnghia:2009xhq,2017MNRAS.471.1709G}. LSST will mitigate both of these issues by examining streams farther from the center of the Galaxy where these effects are lessened.

In this summary we have only considered the density structure of the stream. However, the perturbation that creates the gap necessarily affects the other phase space dimensions as well. The inclusion of these phase space dimensions allows for an almost unique determination of both the subhalo's internal and impact properties for each gap \citep{erkal2015b}. Furthermore, the perturber's effect produces a correlated signal across observables, improving the precision with which the statistical properties of the stream (\eg, power spectrum and cross-correlation of observables) can be used to measure subhalo properties \citep{bovy:2017}. This provides an exciting opportunity for synergy with current and future spectroscopic and astrometric surveys in addition to precise photometric distances and proper motions from LSST itself. Such efforts will greatly aid in the removal of foreground and background contamination, and they will tighten constraints on the stream progenitor's orbit and provide better measurements of the perturber's mass and size. See \secref{complementarity} for a discussion of some complementary science programs.

\subsection{Strong Lensing \Contact{Chris F.} }
\label{sec:stronglens} 
\Contributors{Christopher D.\ Fassnacht, Cora Dvorkin, Francis-Yan Cyr-Racine, Charles R.\ Keeton, Yashar D.\ Hezaveh, Ana~D\'{i}az Rivero}

Strong gravitational lensing is one of the most powerful probes of dark matter halos beyond the Local Goup. 
Gravitational lensing directly probes the total mass distribution that a light ray encounters and does not require that mass to be luminous or baryonic.
Therefore, an analysis of lensing signals can be used to measure the presence, quantity, and mass of subhalos in massive galaxies and small isolated halos along the line-of-sight.  
The discovery of low-mass dark matter halos is possible even at cosmological distances, where the flux of any luminous material associated with the halos would fall below the detection limits of typical observations.  
Thus, the gravitational lensing approach is highly complementary to Local Group observations.

\begin{figure}
    \centering
    \includegraphics[width=0.4\textwidth]{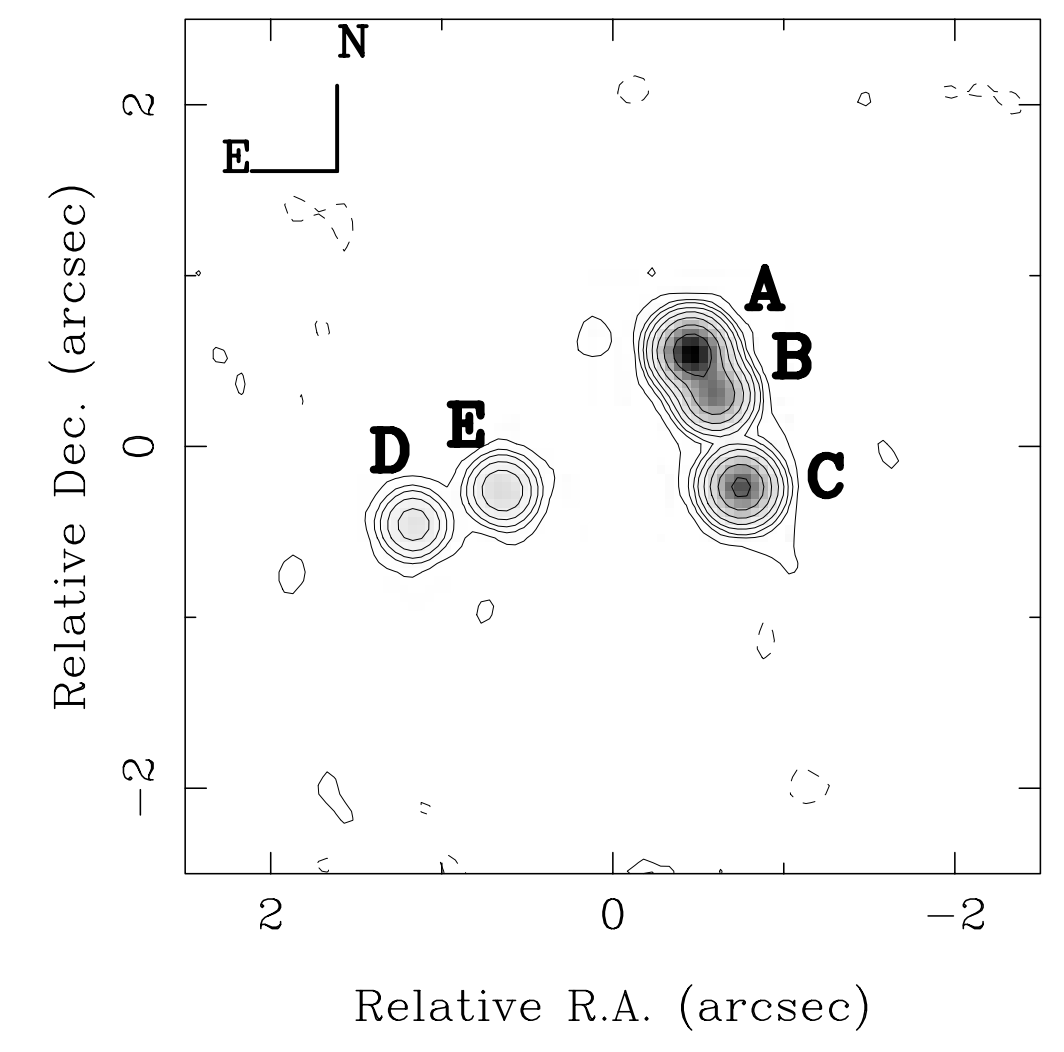}    
    \includegraphics[width=0.44\textwidth]{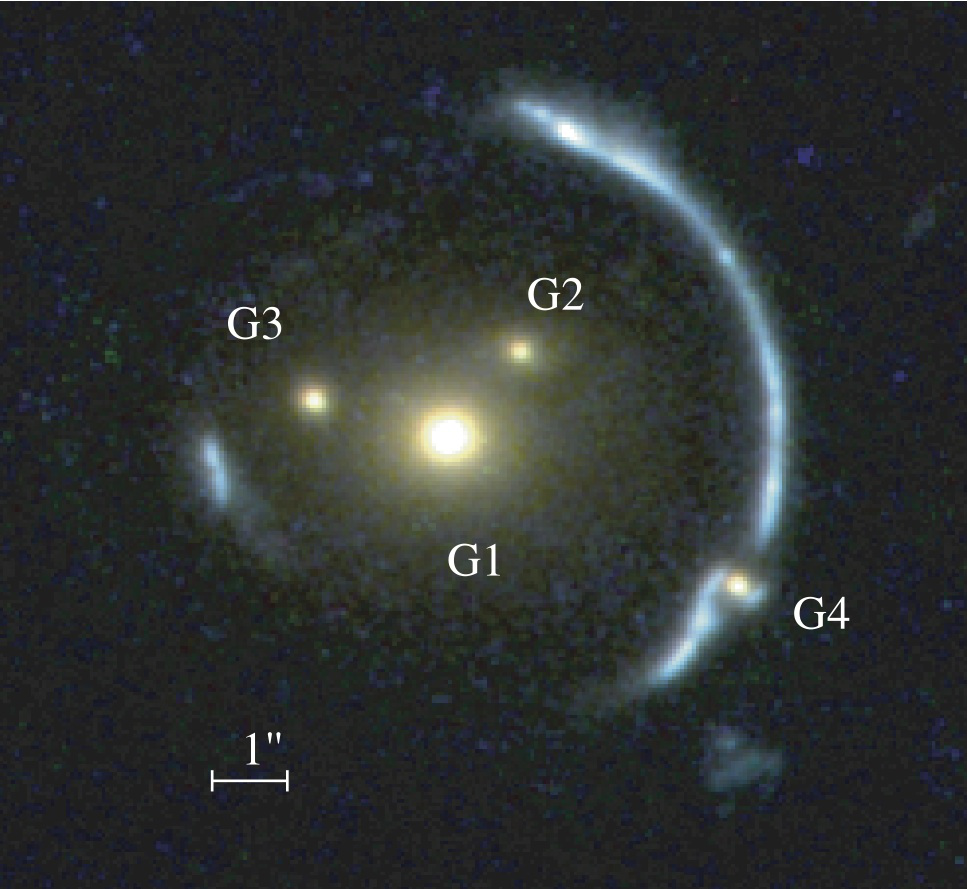}
    \caption{Examples of two gravitational lens systems that exhibit perturbations due to (potentially unseen) halos.  \emph{Left:} Radio-wavelength imaging of a quasar lens system, B2045, that has one of the strongest flux-ratio anomalies known.  Component B should be the brightest of the three close images and instead it is the faintest. Figure from \citet{Fassnacht++99}
    \emph{Right:} HST imaging of the ``Clone'' \citep{2009ApJ...699.1242L}, showing that the long lensed arc is split by the presence of a perturber, in this case galaxy G4.  Note that the location and mass of G4 could have been determined {\em even if G4 were purely dark}.  Figure from \citet{Vegetti_2010_1}.}
    \label{fig:stronglens_examples}
\end{figure}

The (sub)halo-detection techniques described below utilize strong gravitational lensing, in which a massive foreground object bends the light from a background galaxy to produce multiple images of the background object.  
If the emission from the background object is dominated by a single point-like component, such as a quasar or other AGN, the lens system will contain multiple images of that component (\eg, left panel of \figref{stronglens_examples}).
Typically these quasar lens systems consist of two or four images, creating ``doubles'' and ''quads'' respectively. 
If, on the other hand, the background object is dominated by stellar emission, then the lensed emission is in the form of tangentially stretched arcs or a full Einstein ring that surrounds the lensing galaxy (\eg, right panel of \figref{stronglens_examples}).  
In both cases, substructure in the main lensing galaxy and small line-of-sight halos create small perturbations to the lensed images.

\begin{figure}
    \centering
    \includegraphics[width=0.70\textwidth]{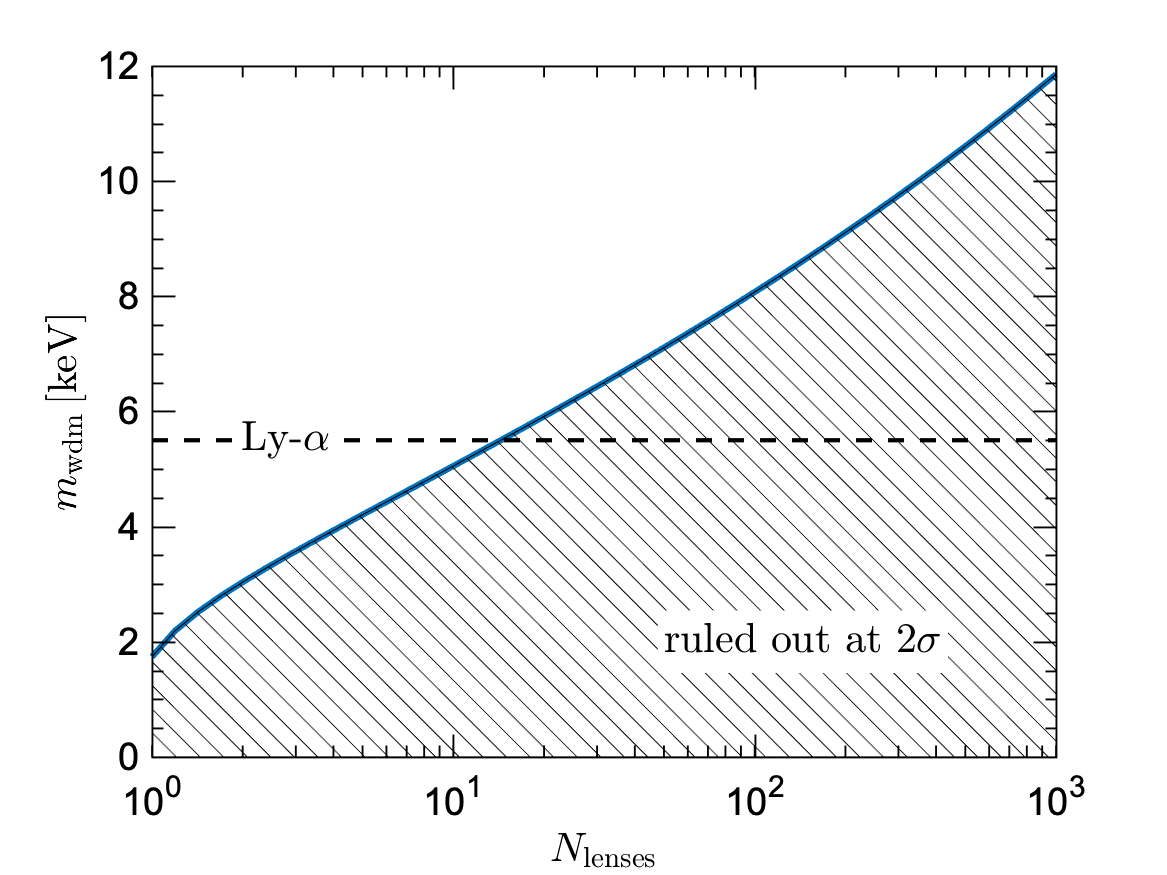}
    \caption{ \label{fig:lensing_wdmlim_vs_nlens} Projected $2\sigma$ constraints on WDM particle mass as a function of the number of strong lens systems that achieve a given (sub)halo mass detection threshold, under the assumption that CDM is correct. These constraints include only the contribution from halo substructure, and do not include the line-of-sight contribution.
Exisiting Lyman-$\alpha$ forest constraints are shown with a dashed horizontal line \citep{2017PhRvD..96b3522I}. Figure based on \citet{Hezaveh_2016ltk}.
}
\end{figure}

As will be described in detail below, there are three main techniques for detecting the presence of dark (sub)halos using strongly lensed systems: analysis of flux-ratio anomalies in lensed quasar systems,
gravitational imaging for lensed galaxy systems, and power spectrum approaches. 
Improved constraints on dark matter properties via these measurements will require: (1) a much larger samples of lens systems, and (2) follow-up observations with high-resolution imaging and spectroscopy.
LSST will play a critical role by increasing the number of lensed systems from the current sample of hundreds to an expected samples of thousands of lensed quasars \citep{O+M10} and tens of thousands of lensed galaxies \citep{Collett2015}.
The vast increases in sample sizes will provide much stronger statistical constraints on dark matter models than are currently possible (\eg, \figref{lensing_wdmlim_vs_nlens}).
The study of lensed systems will also require coordination with other facilities, namely space-based observatories, large ground-based telescopes with adaptive optics systems, ALMA, and very-long-baseline radio interferometry (see \secref{SLcomplement}). These facilities provide the milliarcsecond-scale angular resolution that is required to push the (sub)halo detection sensitivity into unexplored mass regimes.

\vspace{1em} \noindent {\bf Flux-ratio Anomalies}

The presence of clumpy (dark) matter, whether within the main halo of the primary lens or along the line of sight, will perturb the gravitational potential of a strong lens system.
One of the effects of these perturbations is to change the magnification of the lensed images of a background AGN.
The angular scales over which the perturbations are important depend on the mass the perturber, so the presence of a small (sub)halo will typically affect only one of the lensed images and, thus, will change the relative fluxes of the images.
Furthermore, because the image magnification depends on the second derivatives of the gravitational potential, this method is, in theory, sensitive to smaller-mass structures than the gravitational imaging approach described below.\footnote{Indeed, flux-ratio anomalies can be produced by stars in the lens galaxy, through the microlensing phenomenon discussed below.}

The utility of this effect was first presented in \citet{Mao:1998aa}, which considered the effects of spiral arms in the lensing galaxy on the flux ratios of the lensed images, and for many years this was the only lensing technique used to investigate the presence of substructure in massive galaxies.
The approach is to describe the lensing galaxy with a relatively simple smooth single halo model.
These simple models are nearly always capable of fitting the observed positions of the lensed images to within the observational errors.
At that point, any deviation between the model-predicted image fluxes and the observed fluxes could be ascribed to some type of non-smooth / clumpy mass, either in the lensing galaxy or along the line of sight.
At optical and near-IR wavelengths, there are often significant differences between the predicted and observed image fluxes.
However, these perturbations are most likely to be produced by stars in the lensing galaxies, a process known as microlensing, and thus optical and near-IR fluxes are not informative in terms of the statistics of dark matter halos.
What is required is to observe at wavelengths at which the angular size of the emitting region in the background source is large compared to the micro-arcsecond scales at which stars produce their effects.  
Until recently, this meant observing lensed quasar systems at radio or mid-IR wavelengths, which vastly reduced the available sample sizes.

A seminal paper by \cite{Dalal:2002aa} used the statistics of observed flux-ratios in a sample of seven lens systems to place limits on the substructure fraction in the lensing galaxies, i.e., the percentage of the lens mass that is composed of clumpy structures, in the $10^6 - 10^9 \Msun$ range.
The small sample size was set by the number of radio-loud systems that was known at the time and the one lens system with a usable mid-IR data set.  
Because lensed radio-loud AGN are rare, and ground-based high-resolution mid-IR observations are extremely difficult, several novel ideas were pursued to increase the lensing sample in radio \citep[\eg][]{2015MNRAS.454..287J,2018MNRAS.481L..40S,2019MNRAS.483.2125S}.
None the less, the lens sample size has increased slowly over the last decade.
Forecasts based on forward modeling simulations indicate that $\gtrsim$100 well-constrained flux-ratio systems are needed to provide 2$\sigma$ constraints of $10^{7.2} - 10^{7.5} \Msun$ for the half-mode mass in a WDM scenario, corresponding to a $\sim$5--6~keV thermal relic mass \citep{Gilman++18}.
More recent estimates including line-of-sight structure show that similar constraints may be achievable with $\roughly 50$ lenses \citep{1901.11031}.
In either case, increases in sample sizes are required.
The two most promising paths forward are to obtain large lensed quasar samples with LSST and then follow up with either high-resolution mid-IR imaging with JWST, or IFU spectrographs on ELTs or JWST.  The second technique takes advantage of the fact that in lensed AGN, the narrow-line region surrounding the central AGN is larger than the microlensing scale, even though the broad-line region and the source of the continuum emission are not.  Therefore, with high-resolution IFU observations, the narrow-line emission from each lensed image can be spatially resolved, thus providing the required microlensing-free flux ratios \citep{MoustakasMetcalf03, Nierenberg++14, Nierenberg:2017vlg}.

Deep high-resolution imaging in the optical or infrared is also necessary to address possible systematics in the flux-ratio technique.  
Investigations using Keck adaptive optics imaging of radio loud lenses have shown that, in some cases, the observed flux-ratio anomalies can be explained by baryonic structures in the lensing galaxy, namely edge-on stellar disks rather than dark matter halos \citep{Hsueh++2016, Hsueh++2017}.
These baryonic effects were also seen in simulated data \citep{Gilman++2017, Hsueh++2018}.
These studies indicate that a lack of knowledge about the baryonic structure of the lensing galaxy may lead to an overestimate of the amount of clumpy dark matter in the lens or along the line of sight.
With a sample of thousands of quasar lenses expected from LSST, it will be possible to select systems where baryonic effects are minimized.

\vspace{1em} \noindent {\bf Gravitational Imaging}

The presence of a massive peturber along the line of sight can change the shape of lensed emission. 
This effect can be utilized in strong lens system in which the background object is a galaxy that is lensed into long arcs or a complete Einstein ring.
Small (sub)halos that are close in projection to the lensed emission can distort arc shape to a degree that can be detected by high-resolution imaging observations.
This ``gravitational imaging'' technique was proposed by \cite{Koopmans:aa} and further refined by \citet{Vegetti:2008aa,Vegetti:2009aa}.  The size of this effect depends on the mass of the perturber and its projected distance from the lensed arcs, with more massive and closer perturbers having larger effects.
 
The first application of the gravitational imaging technique to real data was for the ``Clone,'' a system for which the primary lensing halo is a compact galaxy group \citep[\figref{stronglens_examples},][]{2009ApJ...699.1242L, Vegetti_2010_1}. In this system, the long lensed arc is broken and split at the location of the peturber, which in this case is a satellite galaxy in the group with a mass of $\roughly 10^{10} \Msun$ \citep[][]{Vegetti_2010_1}.  This massive galaxy located right on the arc produced an effect that could be seen by eye in high-resolution HST imaging.  Lower-mass detections were subsequently made using HST \citep[$\roughly 10^9 \Msun$;][]{Vegetti_2010_2}, Keck adaptive optics \citep[$\roughly 10^8 \Msun$;][]{Vegetti_2012}, and ALMA mm-wave interferometry \citep[$\roughly 10^8 \Msun$;][]{Hezaveh_2016ltk}.  
Note that the masses reported in these papers usually assume a truncated mass distribution (\eg, a pseudo-Jaffe profile) or are explicitly given as mass contained within radii of, \eg, 600\pc, to better match dwarf galaxy measurements made within the Local Group.  Multiplying these values by a factor of 10 gives roughly the expected virial mass of their host halos.
 
The implications for the nature of dark matter from the gravitational imaging technique come from comparing the number of detected halos to those predicted by various dark-matter models.  
For this reason, one of the strengths of the technique is that {\em non-detections} are as valuable as detections, and can be especially powerful at low masses where CDM models predict a large number of halos.
This analysis relies on an understanding of the lowest mass that can be detected at each location in the lens system \citep[\eg,][]{Vegetti2014, Hezaveh_2016ltk, Ritondale++18}.
 
Nearly all previous inferences on dark matter from gravitational imaging have considered solely the expected and measured effects of subhalos within the main halo of the primary lensing galaxy \citep[\eg,][]{Vegetti:2009aa, Vegetti_2012, Vegetti2014, Hezaveh_2016ltk}.
However, an additional perturbation signal is provided by the presence of halos along the line of sight.
An analysis of simulated data has shown that the signal from line-of-sight structures is significant even for lower redshift lenses and is the dominant contribution to any lensing signal for higher redshifts \citep{Keeton:2002ug,Despali++18}.
The line-of-sight structures may very well be a cleaner probe of dark-matter properties than substructures in the lensing galaxies.
This is because the line-of-sight halos are unlikely to have been tidally stripped and thus their measured masses reflect their true masses.
The techniques for including the line-of-sight signal have been developed and applied to recent analyses \citep{Ritondale++18}. 
 
For the relatively high (sub)halo masses that have been probed so far, $\gtrsim 10^9\Msun$, there is little difference between the predictions of CDM and models with a mass cutoff (\eg, WDM).
Therefore, even analyses of $\roughly10$-lens samples have not achieved the statistical precision to distinguish between dark matter models \citep{Vegetti2014, Ritondale++18}.
What is urgently needed is both to increase the sample sizes and, more importantly, to probe further down the mass function.
The mass-detection limit for gravitational imaging is set by three properties of the observations: (1) the signal-to-noise ratio, (2) the angular resolution of the imaging data, and (3) the surface-brightness structure of the lensed background galaxy.  This last point arises because it is easier to detect small astrometric shifts if there are strong gradients in the surface brightness, as opposed to a smooth light distribution.
These properties lead to the need for sensitive high resolution observations of the large samples of appropriate lenses that LSST will provide.
The high-resolutions observations can come from ELTs, which should provide milliarcsecond-scale angular resolution currently only available from VLBI radio observations \citep[\eg][]{2018MNRAS.478.4816S}.
For the subset of LSST lenses that are radio loud, VLBI and ALMA observations will provide excellent complementarity.

\vspace{1em} \noindent {\bf Small-scale Structure Power Spectrum}
\begin{figure}[t]
\centering
\includegraphics[width=0.6\textwidth]{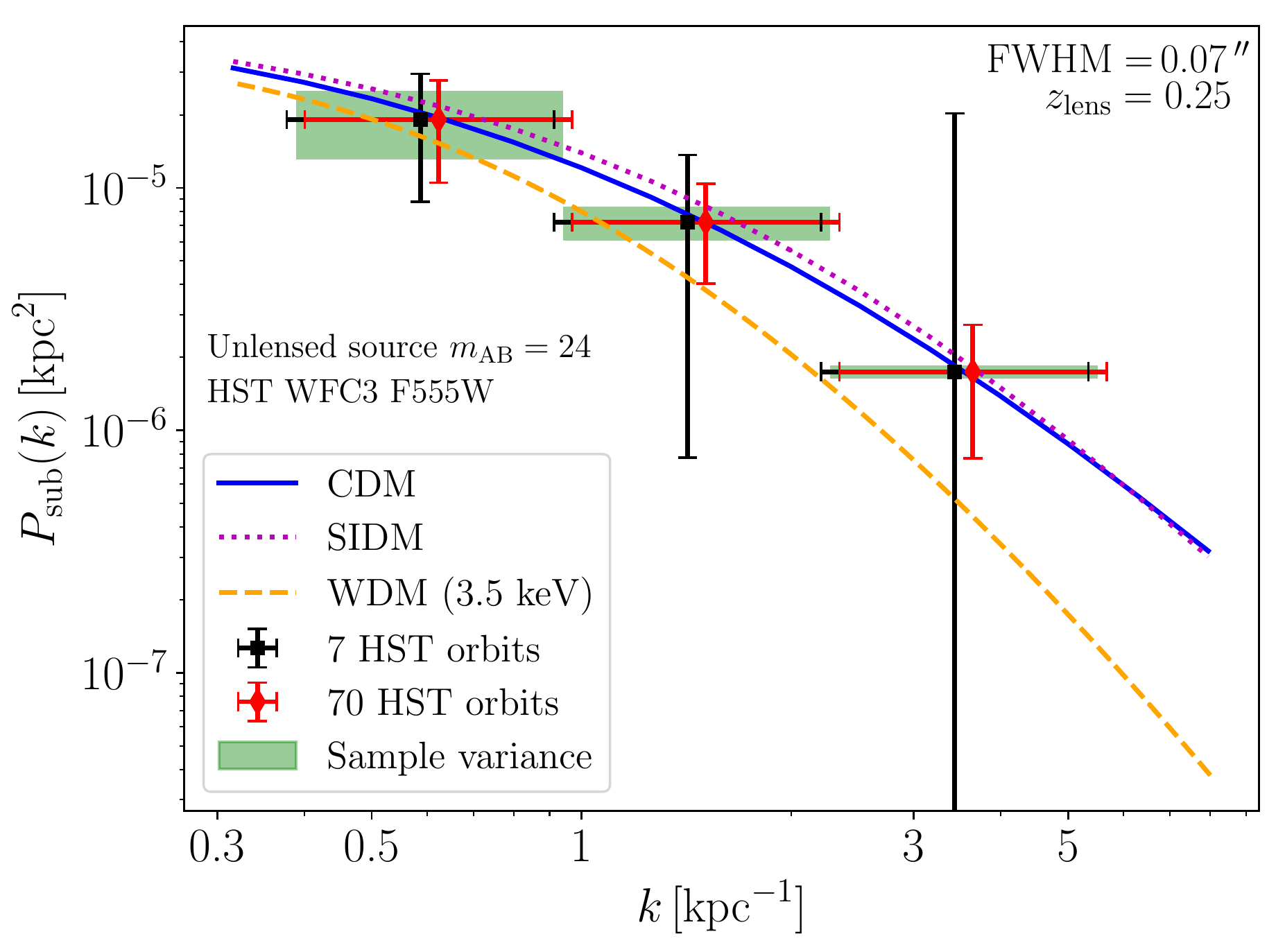}
\caption{Fisher forecast for the substructure convergence power spectrum in three logarithmic wavenumber bins. We consider here observations with the wide-field camera 3 (WFC3) aboard the Hubble Space Telescope (HST) using the F555W filter, resulting in a point-spread function FWHM of $0.07$ arcsec. The source is placed at $z_{\rm src}=0.6$ with an unlensed magnitude $m_{\rm AB}=24$. The error bars show the $1$-$\sigma$ regions, while the green rectangles display the sample variance contribution within each bin. We conservatively assume that only half of each orbit is available for observation. The blue solid line shows the fiducial substructure power spectrum model used in the forecast, which corresponds to a CDM population of subhalos modeled with truncated NFW profiles. The dotted magenta line shows the power spectrum for SIDM, assuming a subhalo core size equals to $70\%$ of the scale radius. For comparison, the orange dashed line shows the substructure power spectrum for a thermal relic warm DM with mass of 3.5 keV \citep{Viel:2013}. Figure adapted from \cite{Cyr-Racine:2018htu}. \label{fig:pksub_fisher}}
\end{figure}

While gravitational imaging can detect highly significant and well-localized perturbers along lensed arcs and Einstein rings, less massive perturbers or those located farther away from lensed images typically lead to observational signatures that are too subtle to be detected individually. However, the large number of such perturbers, both as subhalos within the lens galaxy and as field halos along the line of sight, means that their collective effect might be detectable at the statistical level \citep[\eg,][]{Birrer2017}. The power spectrum of the lensed deflection field is a particularly powerful quantity for capturing the aggregate behavior of lensing perturbers. This approach was proposed in \cite{Hezaveh_2014}, and further expanded in \cite{Rivero:2017mao}, \cite{Chatterjee_2017}, and \cite{Cyr-Racine:2018htu}. 
Furthermore, \cite{Daylan:2017kfh} proposed a technique to constrain the statistical properties of dark matter subhalos in the lens galaxies by studying the joint perturbations of unresolved subhalos.

A key advantage of this power spectrum approach is that it describe the effect of perturbers in terms of a \emph{spatial fluctuation} basis instead of the more traditionally used \emph{mass} basis. The power spectrum directly captures the spatial scales on which perturbers influence the lensed images without having to invoke the notion of (sub)halo density profile, the latter of which is usually required to map from the perturber mass space to the resulting spatial deflection field. As such, the power spectrum is a natural language to describe the collective effects of small lensing perturbers. 

To develop intuition about which dark matter properties could be probed from measurement of this new lensing statistic, \cite{Rivero:2017mao} developed a general formalism to compute from first principles the convergence power spectrum for different populations of subhalos (not yet including line-of-sight perturbers). The authors pointed out that this power spectrum can be mainly described by three quantities: a low-wavenumber amplitude, that depends on the subhalo abundance and on specific statistical moments of the subhalo mass function; on a turnover scale, that probes the truncation radius of the largest subhalos in the system; and on a higher-wavenumber ($k\gtrsim 1 \kpc^{-1}$) slope, that probes a combination of the subhalo inner density profiles and of a possible cutoff in the primordial matter power spectrum. These theoretical findings were then confirmed numerically in \cite{Brennan:2018jhq} using a semi-analytic galaxy formation model, and in \cite{Rivero:2018bcd} using high-resolution $N$-body simulations. Measurements of the power spectrum promise a wealth of information about the behavior of dark matter on small scales.

Several challenges need to be addressed to fully enable the constraining abilities of power spectrum measurements. Most importantly, the degeneracy between the possibly complex brightness profile of the source and the statistical effects of the lensing perturbers needs to be accurately explored. Also, the importance of line-of-sight structure remains to be properly quantified, and the effect of lens galaxy light and other luminous foregrounds on the power spectrum inference needs to be better understood. Finally, since instrumental artifacts such as a mismodeled point-spread function or camera sensitivity could potentially mimic a power spectrum signal, it is likely that these effects would have to be reconstructed at a higher precision than what is typically done for gravitational imaging. 

Thus far, measurement of the lensing power spectrum has been attempted by \cite{Bayer:2018vhy}, and an upper limit on its amplitude was derived using HST archival data. The currently known samples of galaxy-galaxy lenses numbers in the few dozens, and LSST is expected to increase this number several-fold as mentioned above. High-resolution follow up using either space-based or AO-enabled ground-based observatories will be required to measure the power spectrum from these targets and thus probe small-scale structure in a new way (\secref{SLcomplement}). \citet{Cyr-Racine:2018htu} has performed detailed forecasts for the sensitivity of different observational scenarios to the perturber power spectrum for lenses of the type that LSST is expected to discover at optical wavelengths (\figref{pksub_fisher}). It was found that images only a factor of a few deeper than what is currently typically available \citep[\eg, from the SLACS sample][]{Bolton2008} could be sufficient to detect the overall amplitude of the lensing power spectrum. On the other hand, constraining the slope at larger wavenumbers, which could help distinguish between WDM and CDM (\figref{pksub_fisher}), would require much deeper imaging.

\subsection{Satellite Joint Analysis  \Contact{Francis-Yan} }
%Minimum Halo Mass and Density Profile Measurements: 
\label{sec:combine_probes} 
\Contributors{Francis-Yan Cyr-Racine, Ethan O.\ Nadler, Manoj Kaplinghat, Alex Drlica-Wagner, Arka Banerjee, Kimberly K.\ Boddy}
 
As stated above, a cut-off in the matter power spectrum is related to early universe kinematics of dark matter particles and interactions of dark matter particles with a relativistic species. This cut-off directly affects both the subhalo mass function and the luminosity function of Milky Way satellites. Importantly, it also affects the internal properties of subhalos with masses near this threshold. Indeed, the power spectrum cut-off delays halo formation on mass scales near to and smaller than the cut-off, hence lowering the concentration of these objects at fixed halo mass \citep[\eg,][]{Dunstan:2011bq}. Since SIDM (\secref{sidm}) can also have a large impact on the central densities of small halos, combining information from minimum halo mass and density profile measurements (\secref{profiles}) can jointly constrain the presence of a cut-off and of a non-vanishing $\sigma_{\rm SIDM}$, hence simultaneously probing the cold and collisionless tenets of the CDM paradigm. As an illustrative example, we estimate the potential sensitivity of an analysis combining kinematic measurements of LSST-discovered Milky Way satellites with the minimum halo mass forecasts shown in \figref{satellite_mmin}. While not discussed here, we note that both stellar stream gaps and strong lensing could also be used for such joint analysis since they in principle have some sensitivity to internal properties of small halos as well. 

%\vspace{1em} \noindent {\bf Joint impact of a cut-off and self-interaction on the central densities of satellites}
\vspace{1em} \noindent {\bf Joint impacts on the central densities of satellites}

As a demonstration of the power of LSST to probe both a power spectrum cut-off and dark matter self-interaction, we focus here on a simplified summary statistic which captures the essence of LSST's constraining power. The reader should keep in mind that a more detailed analysis using the full complexity of the LSST data set (and its spectroscopic follow-up) could unveil even more information about dark matter physics. Specifically, we consider here the impact of self-interaction or a cut-off on the cumulative number of satellites above a given luminosity threshold that have stellar velocity dispersion within their half-light radius above a minimum value $\sigma_{\star,\rm lim}$, $N_{\rm sat}(L_\star>L_{\rm lim}, \sigma_\star> \sigma_{\star,\rm lim})$. Here, we use the stellar velocity dispersion, $\sigma_\star$, as a probe of the subhalo's central density.

%We model the connection between the central density of subhalos and the self-interaction cross section or cut-off in the power spectrum in the following way. 
For simplicity, we parameterize the cut-off in the power spectrum using the thermal WDM mass, $\mWDM$ \citep[\eg,][]{Bode:2000gq}, but note that other physics such as interactions with a relativistic species \citep[\eg,][]{Boehm:2004th,Cyr-Racine:2015ihg} could also cause a small-scale suppression of power. It is well-known that dark matter self-interaction creates constant density cores in the subhalos \citep{Spergel:1999mh}, which usually lowers the central density as compared to NFW halos. In the limit of large cross sections or significant subhalo mass loss, the self-interactions could also lead to core collapse and an increase in the subhalo central density \citep{Balberg:2002ue,Ahn:2004xt,Nishikawa:2019lsc}. 

In the absence of significant self-interaction, the power spectrum cut-off affects the central density through a modification of the subhalo concentration-mass relation \citep{Dunstan:2011bq,schneider2012,Lovell:2013ola,Bose:2016irl}. We adopt the following form for this modification  
\begin{equation}
c(M_{\rm vir}; \mWDM) = c_{\rm CDM}(M_{\rm vir})\left(\frac{M_{\rm vir}}{10^{12} \Msun} \right)^{\Delta\alpha(\mWDM)}\,,
\end{equation}
where $M_{\rm vir}$ is the subhalo virial mass, $c_{\rm CDM}$ is the subhalo concentration in the standard CDM case \citep{Moline:2016pbm}, and $\Delta\alpha$ is a power-law index that depends on the power spectrum cut-off. In this small self-interaction cross section limit, we assume that the core size is negligible and the density profile is of the NFW form. Note that stellar feedback can change this situation, a manifestation of the well-known degeneracy between feedback and self-interactions \citep[\eg][]{1996MNRAS.283L..72N}. 

In the opposite limit of large cores created by self-interactions, the central density $\rho_0$ of the subhalo can be written as \citep{Nishikawa:2019lsc}
\begin{equation}
\rho_0=\rho_{\rm s} f(t/t_0)\,,
\end{equation}
where $t$ is time elapsed since infall, $t_0 = a(\sigmam)\rho_{\rm s} v_0$ with $v_0^2=4\pi G \rho_{\rm s} r_{\rm s}^2$ and $a=\sqrt{16/\pi}$ for a hard-sphere interaction~\citep{Balberg:2002ue}. Here, $\rho_{\rm s}$ and $r_{\rm s}$ are the NFW density and scale radius parameters, respectively. The function $f$ encodes the evolution in time of the subhalo's central density in the presence of self-interactions, which includes its initial suppression due to core formation, and its subsequent increase due to the onset of the gravo-thermal instability \citep{Ahn:2004xt}. Importantly, the onset of this latter phase depends on the satellite's orbital history, with highly tidally stripped subhalos reaching it on a much shorter timescale than field halos \citep{Nishikawa:2019lsc}. To incorporate these results we make the further assumption that the tidal evolution of subhalos is not highly sensitive to dark matter particle physics. Current simulations tend to support this point of view, although it is likely that the stripped mass fraction will be somewhat larger for subhalos that have lower central densities due to either a cut-off in the power spectrum or self-interactions \citep{Lovell:2013ola,Dooley:2016ajo}. 

We adopt the following simple model for relating the mean stellar dispersion $\bar{\sigma}_\star$ to the central density $\rho_0$ \citep{Wolf:2009tu}
\begin{equation}
\bar{\sigma}_\star = 1 \kms \sqrt{\frac{\rho_0}{0.1 \Msun/{\rm pc}^3}} \left ( \frac{R_{\rm h}}{50 {\rm pc}} \right ) \,,
\end{equation}
which is valid as long as the core radius $r_{\rm c}$ is much larger than the projected half-light radius $R_{\rm h}$. In the opposite case where $r_{\rm c} < R_{\rm h}$, we modify this relation by putting an upper bound on $f(t/t_0)$ at a value given by $1/(x_{\rm h}(1+x_{\rm h})^2)$, where $x_{\rm h} = R_{\rm h}/r_{\rm s}$, and where the core radius is defined by the equation $\rho_{\rm NFW}(r_{\rm c}) = \rho_0$.   

To map the present-day mass of our subhalos to their luminosities, we combine the zoom-in simulations presented in \cite{Mao2015} with the subhalo--satellite galaxy model presented in \cite{Nadler:2018} to obtain the probability that a subhalo of present-day virial mass $M_{\rm{vir}}$ hosts a satellite of luminosity $L_\star$ and stellar dispersion $\sigma_\star$. The mapping from subhalos to satellites includes a prescription for hydrodynamic effects such as enhanced subhalo disruption due to a galactic disk, the galaxy formation threshold due to reionization, and a flexible model for the relationship between luminosity and subhalo peak circular velocity. To characterize $P(L_\star|M_{\rm{vir}})$, we sample from the posterior distribution of model parameters from the fit to classical and SDSS-detected satellites in \cite{Nadler:2018}, generate a large number of satellite population realizations for each Milky Way host halo, and fit the resulting $P(L_\star|M_{\rm{vir}})$ relation with a log-normal distribution. We also use these simulated satellite populations to perform a large number of mock LSST observations to obtain the probability distribution of our summary statistic $N_{\rm sat}(L_\star>L_{\rm lim}, \sigma_\star> \sigma_{\star,\rm lim})$.

\begin{figure}[t]
\centering
\includegraphics[width=0.75\columnwidth]{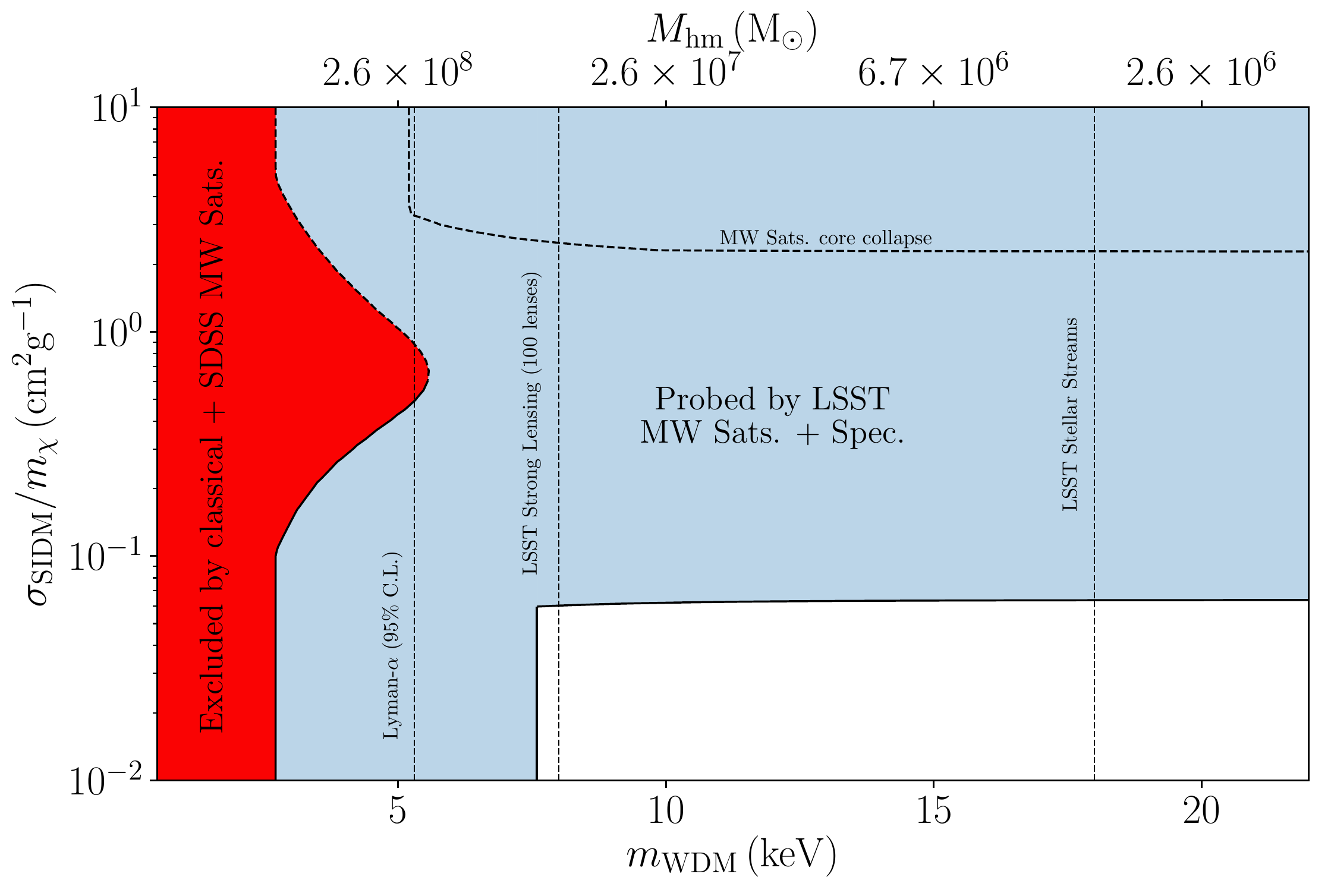}
\caption{\label{fig:sidm_wdm} Projected joint sensitivity to WDM particle mass and SIDM cross section from LSST observations of dark matter substructure. 
The red region is ruled out at 95\% confidence level by current observations of the Milky Way satellite population (the dashed part of the contour is subject to uncertainties due to core collapse, see main text).
The dashed vertical lines correspond to current constraints on the minimum WDM mass from the \Lya forest \citep{2017PhRvD..96b3522I}, and the projected sensitivity of LSST-discovered strong lenses and stellar streams.
%LSST will be sensitive to deviations from the standard CDM scenario through several different probes.
The discovery of additional Milky Way satellites with LSST and their subsequent spectroscopic follow-up will probe the region in blue.
The region with large SIDM cross section delimited by the dashed line labelled ``MW Sats core collapse'' is likely to be probed by Milky Way satellite galaxies, but the simple analysis performed here is insufficient to quantify its sensitivity due to halo core collapse in this regime \citep{Nishikawa:2019lsc}. 
We caution that the exact shape of this region will depend on the amount of tidal disruption that subhalos experience.  
The top axis displays the corresponding half-mode mass as per Eq.~\eqref{eqn:Mhm}. 
Note that $\sigmaSIDM$ stands for the self-interaction cross section evaluated at velocities relevant for Milky Way satellite galaxies ($v_{\rm rel}\sim5$--$50 \kms$). }
\end{figure}

Using the mapping from $L_\star$ and halo mass to $M_{\rm vir}$, our summary statistic can be computed using the expression:
\begin{align}\label{eq:Nsat_above_thresh}
 N_{\rm sat}(L_\star>L_{\rm lim}, \sigma_\star> \sigma_{\star,\rm lim}) &=  \int_{L_{\rm lim}} dL_\star \int_{\sigma_{\star,\rm lim}} d\sigma_\star \int dM_{\rm vir} \frac{dn}{d M_{\rm vir}} P(L_\star|M_{\rm vir}) \\ \nonumber
 &\qquad\qquad \qquad \times \delta\left(\sigma_\star-\bar{\sigma}_\star(\rho_0,c(M_{\rm vir};\mWDM))\right),
\end{align}
where $dn/dM_{\rm vir}$ is the redshift zero subahlo mass function (which depends on $\mWDM$) and $\delta$ is the Dirac delta function. Given values of $\mWDM$ and $\sigmam$, we can use Eq.~\eqref{eq:Nsat_above_thresh} to compute the number of Milky Way satellites observable with LSST that have stellar dispersion above our chosen threshold. As in \secref{smallest_galaxies},
we take $M_V=0$ mag and $\mu=32$ mag/arcsec$^2$ as our detection threshold for LSST, and set $\sigma_{\star,\rm lim}=2.6 \kms$. This latter choice is driven by the minimum stellar dispersion value obtained in our mock observation of satellite galaxies passing the LSST detection threshold as described above, assuming standard CDM. Since measuring stellar dispersions will require spectroscopic follow-up of LSST-discovered satellite galaxies, we fold in our analysis the probability that a given target can be followed up with 30-meter class telescopes given its luminosity and heliocentric distance, as provided in \secref{spectroscopy}.

The resulting projected joint sensitivity to the WDM particle mass and the SIDM cross section are shown in \figref{sidm_wdm}. 
The red region to the left of the figure is already excluded by observations of known classical and SDSS-discovered Milky Way satellites. 
The dashed vertical lines show current constraints from the \Lya forest \citep{2017PhRvD..96b3522I} and the projected sensitivity of strongly lensed systems and stellar streams discovered by LSST.
In blue, we show the region of SIDM-WDM parameter space that would be probed by LSST+spectroscopic measurements of the Milky Way satellite population. 
In the white region at low SIDM cross sections, the central core caused by self-interaction is too small to significantly affect the dynamics of the satellites. In the blue region at high cross sections delimited by the dashed line, tidally-stripped subhalos may undergo core collapse and may thus have similar stellar dispersion to their CDM counterparts, resulting in a loss of sensitivity to the SIDM cross section from the simple statistics given in Eq.~\eqref{eq:Nsat_above_thresh}. However, we note that the expected diversity of the Milky Way satellite population within this region of dark matter parameter space \citep{Nishikawa:2019lsc} is likely to be distinguishable from that predicted by CDM, and we thus include it in the parameter space that LSST can probe. While not shown in \figref{sidm_wdm}, complete gravothermal collapse and subhalo evaporation at very high SIDM cross section ($\gtrsim 10\cmg$) likely imply that these high values are already ruled out.  We thus see that spectroscopic follow-up of LSST-discovered satellites could significantly improve our knowledge of dark matter physics in the prime parameter space corresponding to $\mWDM \sim 5-15 \keV$ and $\sigmam \sim 0.1-10 \cmg$.

\section{Halo Profiles} 
\label{sec:profiles}
\Contributors{Susmita Adhikari, Robert Armstrong, William A.\ Dawson, Alex Drlica-Wagner, Nathan Golovich,  M.\ James Jee, Yao-Yuan Mao, Annika H.\ G.\ Peter, Daniel A.\ Polin,  J.\ Anthony Tyson, David Wittman}

The standard CDM model predicts that dark matter halos should be ``cuspy,'' \ie with inner density profiles asymptoting to high central densities.
This results from the inability of collisionless dark matter to redistribute kinetic energy, and is born out in numerical simulations which give rise to a family of cuspy halo profiles \citep[\eg, the NFW profile;][]{Navarro:1996gj}.
If dark matter is able to interact through scattering or the exchange of some light mediator (\secref{sidm}), then the density of halos could instead flatten out to produce dark matter ``cores'' \citep{Spergel:1999mh}.
These interactions can also lead to an isotropization of dark matter velocity distribution, leading to more spherical halos \citep{Peter:2013}.
Thus, measurements of the radial density profiles and shapes of dark matter halos are sensitive to the microphysics governing dark matter self-interactions.
Due to the natural possiblity that dark matter scattering has a non-trivial velocity dependence (\secref{sidm}), it is important to probe halo profiles over a wide range of mass scales.
Here we explore the contributions that LSST will make towards measuring the profiles of dark matter halos in isolated small galaxies and clusters of galaxies.
We highlight these systems because they reside at opposite extremes of the galaxy mass spectrum where dark matter dominates over baryonic processes that can also alter the shapes of halos.

\subsection{Dwarf Galaxies as Lenses \Contact{Yao}}
\label{sec:halo_profile_group}
\Contributors{Yao-Yuan Mao, M.\ James Jee, Alex Drlica-Wagner, J.\ Anthony Tyson, Annika H.\ G.\ Peter, Chihway Chang, Rachel Mandelbaum, Manoj Kaplinghat}

Dwarf galaxies ($M_\star \lesssim 10^{9} \Msun$) provide the best visible tracers of low-mass dark matter halos. 
The relatively low baryonic content makes dwarf galaxies sensitive probes of  dark matter physics through the shape of their dark matter halo profiles. 
In particular, the ``core-cusp'' problem in dwarf galaxies has been cited as one of the most significant challenges to CDM \citep[\eg,][]{2010AdAst2010E...5D,Bullock:2017}.
The standard CDM model predicts that dark matter halos should have steeply rising (``cuspy'') central densities in contrast to the shallower (``cored'') mass profiles that are observationally inferred for many dwarf galaxies.  
Evidence for cored profiles exists for Milky Way satellite galaxies from kinematic and theoretical studies \citep[\eg,][]{Walker:2009, 2012ApJ...759L..42P}, and is stronger when one studies the inner density profiles of dwarf galaxies based on high-resolution neutral hydrogen surveys \citep[\eg,][]{Begum:2008,Hunter:2012,Cannon:2011,Oh:2015}. 
Many of these observations show inferred central slopes of the dark matter density profile, $\rho(r) \sim r^{-\gamma}$, that are significantly shallower ($\gamma \approx 0$--$0.5$) than the CDM prediction $\gamma \approx 0.8$--1.4 \citep{Navarro:2010}.

A wide range of solutions to the core-cusp problem have been proposed including observational, astrophysical, and dark matter explanations.
From a dark matter perspective, SIDM can significantly suppress the the central density of halos.
A self-interaction cross-section of $\sigma / m_\chi \sim 1 \cmg$ can explain the diversity of rotation curves seen in low-mass spiral galaxies \citep[\eg,][]{1504.01437,Kamada:2016euw,Tulin:2017ara}.
In addition, ultra-light or fuzzy dark matter has also been suggested as a possible solution to the core-cusp problem through the formation of uniform density solitonic cores \citep[\eg,][]{1502.03456,Hui:2017}. 
However, baryonic feedback remains a major complication for interpreting central density profile measurements in a dark matter context \citep{1996MNRAS.283L..72N,2005MNRAS.356..107R,2008Sci...319..174M,2012MNRAS.421.3464P,Madau:2014,Read:2016}. 
If dwarf galaxies form enough stars, energy from SN explosions can flatten the profiles of dark matter and baryons; however, if too many stars are formed, the excess baryonic mass can have the opposite effect of steepening the slope of the central density profile \citep{Bullock:2017}.
Technical challenges in implementing multi-phase gas and baryonic physics make it difficult to directly address and calibrate baryonic predictions based on hydrodynamical simulations \citep{Tollet:2016,1611.02281,Sawala:2016}.
However, one key prediction is that the creation of cores will be sensitive to the exact star formation history \citep[\eg,][]{governato2012,dicintio2014,onorbe2015,Read:2016,1811.11768,2019MNRAS.tmp....3R}.
Thus, robust measurements of both the stellar and dark matter mass of dwarf galaxies is essential to investigate the effect of baryonic feedback on the central dark matter density.
In addition, it has been argued that significant observational and astrophysical systematics, such as beam smearing, center offsets, inclinations, and non-circular motions can bias central density measurements toward flatter profiles \citep[\eg,][]{astro-ph/0006048,2004ApJ...617.1059R,2008AJ....136.2761O,2016MNRAS.462.3628R}. 
Thus, accurate independent measurements of dwarf galaxy density profiles are critical.

LSST can provide joint statistical measurements of both the central density and stellar content of dwarf galaxies. 
The stacked gravitational weak lensing signal from a large sample of dwarf galaxies will provide the most direct measurement of the amount and distribution of dark matter.  
In this section we predict the sensitivity of LSST to a stacked weak lensing signal from dwarf galaxies.

\vspace{1em} \noindent {\bf Dwarf galaxy lenses}

\begin{figure}
\centering
\includegraphics[width=0.6\columnwidth]{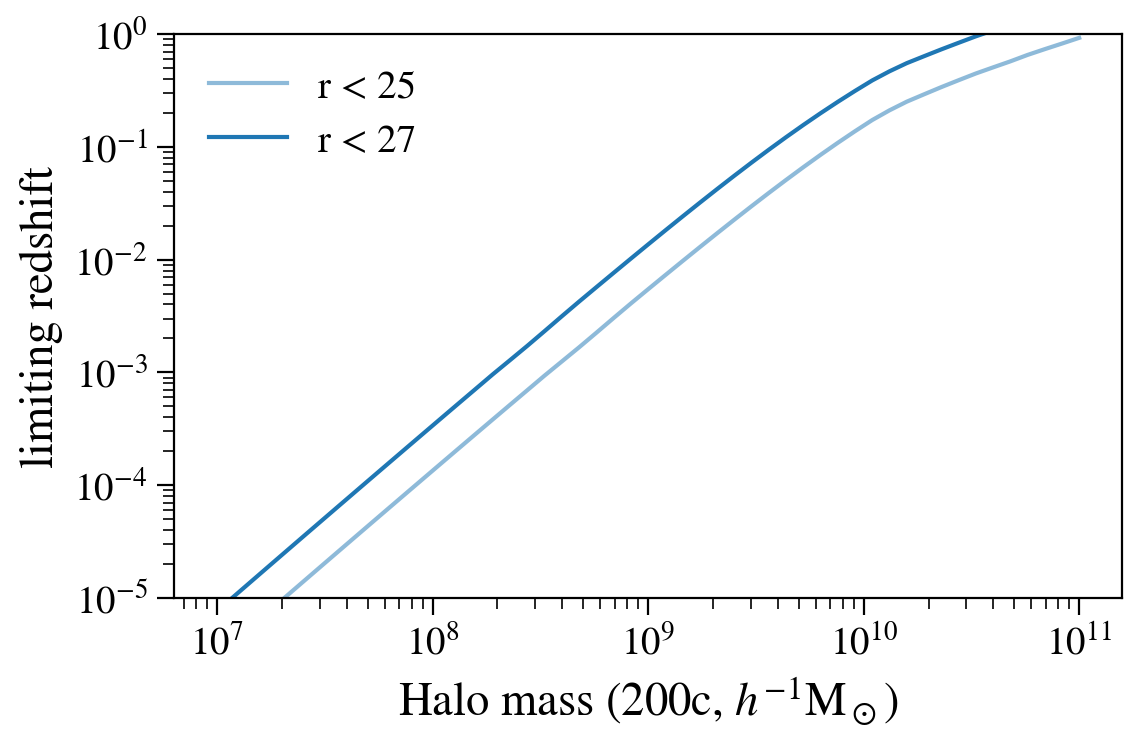}
\caption{\label{fig:dwarf_redshift} Limiting redshift for detecting a dwarf galaxy that lives in a dark matter halo of certain masses, assuming a luminosity--halo mass relation derived from the  subhalo abundance matching technique, which matches galaxy luminosity from the GAMA luminosity function to present-day halo mass ($M_{200c}$) by their respective abundance.}
\end{figure}

We are interested in estimating the number of isolated dwarf galaxies accessible to LSST as a function of dark matter halo mass.
To predict the abundance of the dwarf galaxy sample, we assume the mass-to-light ratio derived from the subhalo abundance matching technique, which links the global galaxy luminosity function with (sub)halo mass function by their respective abundance \citep[\eg,][]{2004ApJ...609...35K,2013ApJ...771...30R}. We use \code{colossus} \citep{2018ApJS..239...35D} to obtain the halo mass function and adopt the global galaxy luminosity function measured by GAMA \citep{2015MNRAS.451.1540L}. We match galaxy luminosity to current halo mass with the definition of $M_{200c}$. We also assume the mass-to-light ratio does not evolve significantly in this low-redshfit regime. 
We use this predicted galaxy luminosity to estimate the limiting redshift for dwarf galaxy detection as a function of galaxy halo mass for two LSST limiting magnitudes: $r \sim 25$ and $r \sim 27$. 
\figref{dwarf_redshift} shows that to probe dark matter halos with mass $\lesssim 10^9 \Msun$, it will be necessary to select galaxies at $z < 0.01$. 
While selecting very low-$z$ galaxies with photometric data is challenging, current projects like the SAGA Survey \citep{Geha:2017} have shown that it is possible using data from SDSS. 
Future large, multi-object spectrographs will greatly expand the spectroscopic data for training these selections. 
It will also be possible to use morphological information to select nearby dwarf galaxies.
LSST will be able to distinguish a dwarf galaxy with $M_V=-14$ from background galaxies of the same apparent magnitude out to a distance of $\roughly 100 \Mpc$ \citep[Section 9 of][]{0912.0201}.

\vspace{1em} \noindent {\bf Source galaxies}

The conservative LSST 10-year ``gold'' sample for cosmic shear measurements of dark energy is expected to have a source galaxy density of $\roughly 27 \amin^{-2}$ \citep{Chang:2013,1809.01669}. 
However, we expect that the dwarf lensing analysis can retain significantly more source galaxies for the following reasons.
(1) Our measurement uncertainty is dominated by the low number of dwarf galaxy lenses, rather than the  multiplicative shear measurement bias that must be strictly controlled for dark energy measurements. This allows us to include fainter, smaller, and more blended sources.
(2) Unlike the lenses used for cosmic shear measurements, the dwarf galaxy lenses are at very low redshift. This means that most detected sources are background galaxies.
(3) We expect to be able to combine shape measurements from multiple filters, which could increase the source density by $\roughly 80\%$. 
Combining these factors, we estimate a source galaxy density of $50 \amin^{-2}$, which is consistent with the fiducial, multi-band estimate of \citet{Chang:2013}.
The primary focus of the source galaxy selection will be to avoid catastrophic \photoz outliers (low-$z$ galaxies reported at high-$z$), which typically occur for less than a few percent of galaxies in current surveys \citep{1406.4407}. 

\vspace{1em} \noindent {\bf Sensitivity}

\begin{figure}
\centering
\includegraphics[width=\columnwidth]{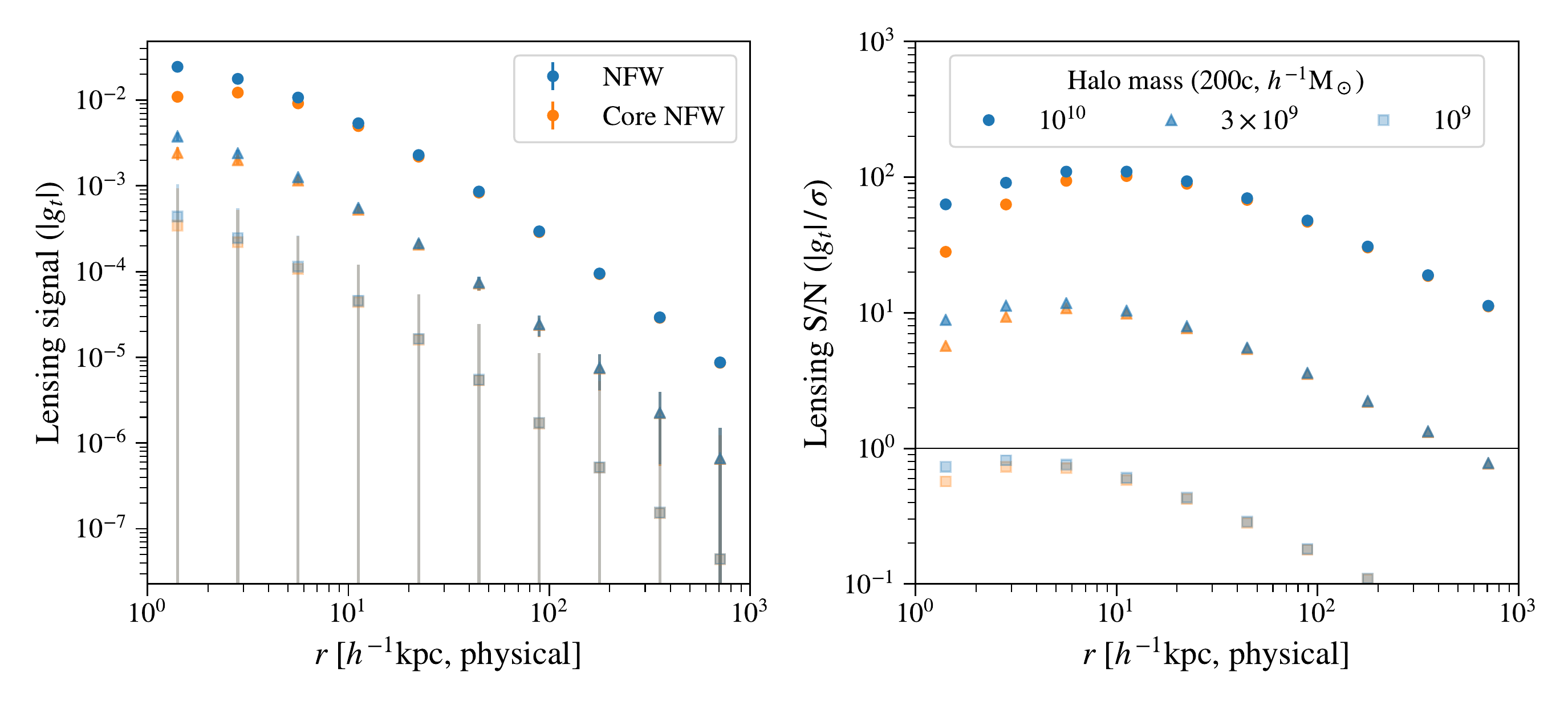}
\caption{
\label{fig:dwarf_sn} Lensing signal (reduced tangential shear; \textit{left}) and signal-to-noise (\textit{right}) for stacked samples of dwarf galaxies in three different mass bins (shown by different shapes of markers), each with width of 0.5 dex in mass. Two different density profiles are used for this calculation: the NFW profile (blue) and a NFW profile with a core (orange). 
The calculation assumes perfect selection of dwarf galaxies within the redshift range over which they are detectable by LSST. 
Source galaxies are assumed to be at $z=1.2$, with a surface number density of $50\,\amin^{-2}$, and a shear uncertainty of $\sigma_\gamma = 0.25$ per component.}
\end{figure}

We calculate the expected strength of a lensing signal for three different bins in halo mass,  $M_{200c} = \{10^{10},\, 3\times10^9,\, 10^{9}\}\,h^{-1}\Msun$, each with a width of $0.5$\,dex in mass. 
These samples correspond to $N = \{1.2\times10^8,\, 7.8\times10^6,\, 1.6\times10^5\}$ dwarf galaxies out to a redshift of $z = \{0.35,\, 0.07,\, 0.014\}$, respectively.
Source galaxies are placed at $z = 1.2$ with a density of $50 \hbox{ arcmin}^{-2}$ and a shear uncertainty of $\sigma_\gamma = 0.25$.
We model the mass distribution in each dwarf galaxy with an NFW halo assuming the concentration--mass relation from \citet{1809.07326}.
We calculate the shear from the stacked dwarf galaxy lens sample using \code{colossus} \citep{2018ApJS..239...35D}, assuming that each lens is placed at the limiting detectable redshift.
The results are shown in \figref{dwarf_sn}, where we find that LSST has the potential to measure the lensing shear with ${\rm S/N} \gtrsim 10$ for halos with $M \gtrsim 3 \times 10^9\,h^{-1}\Msun$.
Note that some of our assumptions are clearly optimistic. In particular, the number density of the source galaxies we assumed is high, and the assumption of perfect lens galaxy selection is also unlikely to hold. Nevertheless, since the S/N ratio goes $\sim 1/\sqrt{N_\text{lens} N_\text{src}}$, and thus lowering these numbers by a factor of $\sim 2$ would still yield a very high S/N ratio.

As mentioned earlier, a cored density profile is a signature of SIDM, hence we also calculate the shear signal for a modified NFW profile with a central core,
\begin{equation}
\rho_\text{core}(r) = \rho_\text{NFW}(r) \times (1 -  e^{-3r/r_s})\,,
\end{equation}
where $\rho_\text{NFW}(r)$ is a standard NFW profile and $r_s$ is the scale radius of the NFW profile. 
We show the predicted shear signal from the cored profile in \figref{dwarf_sn} to be compared with the signal from NFW profile. 
We see that the overall signal-to-noise does not change much with the profile.  
However, to statistically distinguish the different profiles, one needs to measure the shear at very small angular scales ($< 10\,h^{-1}\kpc$, which corresponds to 2.9\,arcsec at $z=0.35$ and 10\,arcsec at $z=0.07$). This small-scale regime is where the systematics due to PSF modeling and blending would dominate. 
In other words, while the numbers of source and lens galaxies that LSST can find will be high enough to distinguish the difference between the two profiles, shear measurement systematics may present the major obstacle. 

The median seeing of LSST is about 0.7\,arcsec \citep[LSST SRD,][]{LPM-17}. Since the dwarf galaxy lenses are at very low redshift ($z=0.07$ for the $M_\text{halo}=3\times10^9\,h^{-1}\Msun$ sample), the angular scale ($\sim$10\,arcsec) that we would use to distinguish the cored profile is still well above a few times the median seeing. However, the uncertainty in PSF models can affect the shape measurement up to the scale of 3\,arcmin \citep{2012MNRAS.427.2572C}. We believe that, with improved PSF models and marginalization over model uncertainty, it will still be feasible to  utilize dwarf galaxy lenses to distinguish different halo profiles at small scales. 

\subsection{Galaxy Clusters \Contact{Susmita}}
\label{sec:halo_profile_clusters}
\Contributors{Susmita Adhikari, William A.\ Dawson, Nathan Golovich, David Wittman,  M.\ James Jee, Annika H.\ G.\ Peter, Daniel A.\ Polin, Robert Armstrong}

Galaxy clusters are the most massive gravitationally bound structures in the universe. The high matter density and high velocity dispersions of clusters make them ideal laboratories for testing dark matter self-interaction models in a very different regime from individual galaxies.
In the following section we discuss several probes that use galaxy clusters to constrain the nature of dark matter.  We show that current constraints from many different cluster-scale probes are of the order of $0.1$--$1\cmg$.  To understand why this is so, it is important to note that the average column density of a cluster-scale halo is of the order of $1 \g \cm^{-2}$.  Improved cross section constraints will come from a combination of the large statistical data sets that will be collected by LSST and other telescopes in the LSST era, and more sophisticated theoretical predictions for observables for specific SIDM models.

\vspace{1em} \noindent {\bf Distribution of matter and substructure}

As we describe below, the current best cluster-scale SIDM constraints come from the radial dark matter profiles of halos.  However, cluster-scale halos that consist of SIDM and CDM exhibit other differences, which may prove to be highly constraining given the vastly detailed LSST cluster data sets.  Significantly more theoretical work is required to project robust constraints in the LSST era for those probes.

\paragraph{Radial profile:} Interactions among dark matter particles allow for the exchange of energy between different parts of the halo. The high number of interactions near the dense central region of a dark matter halo increases the temperature, or the velocity dispersion, near the central region. This process can be thought of as a transfer of heat from the outer (hotter) parts of the halo to the inner (colder) region. The excess dispersion due to self interaction leads to flattening of the inner density of the halo, leading to the formation of a cored density profile. For cluster-scale halos, the high densities near the center make the timescales for thermalization shorter at a given cross section than they are for lower-mass objects (although it must be noted that low-mass halos are generally older and have a longer time to thermalize).  The short thermalization time is important because dark matter thus behaves as a fluid in the innermost part of cluster-scale halos, and can relax to a hydrostatic equilibrium configuration at the center of the halo, where baryons dominate the potential \citep{Kaplinghat:2015aga}.  Depending on the merger history, cluster-scale halos can be as cuspy as those in CDM-only simulations (for recent mergers), or relax to a hydrostatic equilibrium (for highly relaxed systems) in which the dark matter halo has a small but relatively dense core \citep{Robertson:2017mgj}.

Density profiles of massive galaxy clusters therefore serve as probes for SIDM. Clusters tend to be dark matter dominated outside the very central regions, and they are the only known systems where the matter distribution can be individually mapped to the virial radius using weak lensing. Strong lensing also provides a measure of cluster mass independent of the dynamical state. And stellar kinematics of the central galaxy can be used to measure the dark matter density profile in the innermost regions. LSST will produce an unrivaled catalog of strong and weak lensing measurement of cluster density profiles. This, in concert with X-ray mass estimates and stellar kinematics, will provide a strong test of the NFW dark matter density profile predicted by cold, collisionless dark matter \citep{Newman:2013,Kaplinghat:2015aga,Robertson:2018anx,Andrade:2019wzn}. Moreover, the strong lensing cross section is an additional probe of the density profile \citep{Robertson:2018anx}.  For hard-sphere scattering, cross section constraints are of the order of $0.1$--$1\cmg$, but without fully quantified systematic uncertainties.

\paragraph{Halo shape:} Apart from the density profile itself, in SIDM models, dark matter velocity distributions become more isotropic than in the CDM model, especially at the center of the halo.  Correspondingly,  the halo density profile becomes more spherical.  Historically, constraints from cluster and galaxy ellipticies \citep{Miralde-Escuda:2000} provided strong constraints on the cross section of SIDM; however, later investigations found these constraints to be somewhat optimistic \citep{Peter:2013}. 
Recent measurements of the shapes of cluster-scale dark matter halos include studies with: cluster members \citep{2018MNRAS.475.2421S},  X-rays \citep{Hashimoto:2007},  lensing \citep{Mandelbaum:2006, Evans:2009, Oguri:2010}, and combinations of observables \citep{Clampitt:2016, Sereno:2018}.  
Current constraints on the cross section are sensitive to the order of $\sigmam \sim 1 \cmg$.
Several groups have shown in $N$-body simulations that the effects of SIDM with a cross section of roughly $\text{(a few)}\times 0.1$--$1 \cmg$ are potentially observable, although baryons can alter the probability distribution function of halo shapes by an amount that is not yet robustly quantified \citep[\eg][]{Peter:2013, Robertson:2017mgj, Brinckmann:2018}.

\paragraph{Substructure:} Structures form hierarchically in the standard CDM scenario: small objects form first and merge to form larger mass structures such as galaxy clusters. These clusters continue to accrete smaller halos and some of these small structures survive as subhalos within the cluster. It is therefore interesting to study the distribution of substructures within larger halos, to understand how the distribution is affected by self interactions among dark matter particles. 

Subhalos can be affected by SIDM models in three different ways within a cluster. First, dark matter particles in subhalos can evaporate due to interactions with the particles in the host cluster. Subhalos lose mass when they enter a cluster. In the CDM scenario particles that are at larger radii and are loosely bound get stripped as the subhalo orbits within a cluster. In SIDM models, evaporation due to self-scattering leads to additional mass loss. Unlike tidal stripping, self-interactions can also affect the inner regions of the subhalos. Simulations show that evaporation is inefficient at increasing the subhalo disruption rate unless hard-sphere cross sections are of order $\sigmam \sim 10\cmg$, or subhalos are on nearly radial orbits through the cluster center \citep{2012MNRAS.423.3740V,Rocha:2012jg,Dooley:2016ajo}. 

While this generally means that the total subhalo mass function within the virial volume is largely unaffected relative to CDM, other effects of evaporation may be detectable. Measuring the mass and the profile around cluster satellites (especially as a function of orbit eccentricity) using galaxy--galaxy lensing to measure the mass-to-light ratio of subhalos can be a promising probe for dark matter physics \citep{Natarajan:2017sbo}. The lensing signal around subhalos is weak and will be contaminated by the cluster mass profile, so methods like subtracting the lensing signal from diametrically opposite points within the cluster can be used to extract the signal. Given the statistics of cluster galaxies in LSST, it is ideally suited for a study of the weak lensing signal of subhalos.  
Second, as subhalos are also tracers of the dark matter density field within the cluster, their orbits will be affected by the change in the potential of the cluster near the core relative to CDM.  This effect can lead to an imprint in the radial distribution of subhalos in clusters, generally by making the subhalos less concentrated toward the halo centers.

Third, non-expulsive interactions can lead to a drag force on subhalos.  This has several potentially interesting observable consequences.  The location of the splashback radius is sensitive to dynamics of subhalos within the cluster. The splashback radius is the boundary of the multistreaming region of a halo and is the largest apocenter of recently accreted objects \citep{Diemer:2014xya,Adhikari:2014lna}. The slope of the density profile of a halo falls off rapidly in a narrow localized region around this radius, and the splashback radius is observed as a minimum in the slope of the projected number density profile of galaxies \citep{More:2016vgs,Baxter:2017csy,Chang:2017hjt}.
The apocenter of the orbits of subhalos can change if there is extra drag beyond dynamical friction \citep{Kummer2018}.  
Therefore measuring the location of the splashback radius can help distinguish between different models of dark matter, although the difference between splashback locations in the CDM and SIDM scenarios has not yet been well quantified.

Similar to the situations discussed above and in merging clusters (Section~\ref{sec:merging_clusters}), the drag force due to non-expulsive interactions may also lead to offsets between the light distribution and dark matter distribution of individual satellites with respect to their subhalos. Small offsets between the subhalo and the galaxy within it may be detectable by indirect means: the potential gradient established by the dark matter at the position of the stellar centroid would induce a U-shaped warp in the stellar disk facing the direction of infall, and a longer-lasting disk thickening. Numerical simulations show these to be observable by current and next-generation photometric surveys under SIDM models with $0.5 \cmg \lesssim \sigmam \lesssim 1 \cmg$~\citep{Secco}. While S-shaped disks formed by tidal distortions of the stellar light profile are abundantly observed in cluster environments, indicating that they are readily induced by ``baryonic effects,'' these effects are not likely to generate prominent U-shaped warps. Such warps are only formed by a differential force on the disk and its halo, due to, for example, the SIDM drag.  The offset between a satellite galaxy and its subhalo may also be observed directly or statistically with strong lensing \citep{Massey2011,Massey:2017cwf}, but the magnitude of the effect is highly model-dependent (depending strongly on the angular and velocity dependence of the cross section).  Current limits are $\mathcal{O}(1\cmg)$ for specific non-hard-sphere models \citep{Harvey:2015hha}.  

\vspace{1em} \noindent {\bf Merging Galaxy Clusters \Contact{Nate?}}
\label{sec:merging_clusters}

% intro para
In the previous section, we considered subhalos to be minor merger events onto the main cluster.  Major cluster mergers can probe the nature of dark matter by serving as the biggest ``dark matter colliders'' on account of their high mass and large collision velocities. Dense halos falling together at thousands of km\,s$^{-1}$ provide an environment where the scattering of dark matter particles off each other would have observable effects.  The observable effects vary depending on the dark matter model and the configuration of the merger \citep{Kim:2016ujt}. 
Cluster mergers may also be able to distinguish between particle models that yield frequent scattering with low momentum transfer (as in a long-range force) and those that yield infrequent scattering with high momentum transfer (as with hard sphere or contact scattering) due to their differing phenomenology in the merger environment.  This is in contrast to the halo radial profile and shape constraints discussed in the previous section, for which the energy and momentum transfer rate matters most and for which there is no preferred direction in the problem.

% why LSST discovery is important
The best known example of a colliding cluster system is the Bullet Cluster, which has been frequently studied as a laboratory for SIDM \citep{Randall:2007ph,2017MNRAS.465..569R}. 
However, since a cluster merger is an eons-long process of which we have only a single snapshot, the measurement uncertainty is dominated by our very limited knowledge of the merger history. While it will remain critical to investigate individual clusters in great detail, the power of LSST lies in systematically analyzing a population of merging clusters with a consistent method, thereby constraining the properties of dark matter.
LSST will contribute to better and more robust constraints not only through the study of already known systems, but also by enabling the discovery of many more merging systems. Because mergers displace plasma from galaxies, they are best discovered by cross-correlation of LSST optically-detected clusters with radio and X-ray surveys \citep{Golovich:2018,Wilber2018}.

%offsets

The first SIDM constraints based on a merging galaxy cluster came from the Bullet Cluster, which was originally identified as an extremely hot X-ray cluster with two galaxy peaks. Higher resolution optical and X-ray imaging revealed a spectacular post-merger system with a clear X-ray cold front and shock. The spatial agreement of the galaxies and mass centroids obtained by weak lensing, and the disassociation of the intra-cluster medium (ICM) led to the constraint $\sigmam \lesssim 2 \cmg$ for hard-sphere scattering \citep{Markevitch2004,Randall:2007ph,2017MNRAS.465..569R,Robertson:2016qef}. Many other dissociative mergers have been found and studied, with roughly similar cross section limits \citep[but with greater systematic uncertainty, \eg,][]{bradac2008}. 

After several ``dissociative'' mergers had been discovered, ensemble studies of the offsets between dark matter, galaxies, and gas were utilized to drive down the Poisson noise from inference on individual systems. \citet{Harvey:2015hha} modeled 72 subclusters within 30 merging systems to place the strongest constraint on SIDM ($\sigmam<0.47 \cmg$).
The study assumes a simplified drag force model where dark matter behaves similar to the ICM. However, \citet{Wittman:2017gxn} reanalyzed the sample including more comprehensive data. They identified several substantial errors that were driving the result and obtained a revised limit of $\sigmam \lesssim 2\cmg$.

The drag force model applies best to particle models with frequent interaction and low momentum transfer per interaction. In models with infrequent, high momentum transfer interactions (including hard-sphere scattering), dark matter particles may be scattered out of the cluster entirely. (Evaporation also occurs for small-angle scattering, though---see \citealt{Kahlhoefer:2013dca}.) This mass loss may be detected by comparing the mass-to-light ratio of merging clusters with those of non-merging clusters, on the assumption that the merger does not affect the galaxy light. This argument leads to a constraint of $\sigmam \lesssim 1 \cmg$, similar to current constraints on the drag model. However, the assumption that the galaxy light is unaffected is a source of uncertainty here. The LSST discovery of many more merging clusters, with six-band LSST photometry, will help us quantify this source of uncertainty. 

Several billion years post-pericenter, after a merging cluster has coalesced into a single cluster, SIDM will still create a cored dark matter distribution in the center of the cluster. For $\sigmam \sim 1 \cm^{2} \g^{-1}$, this core is $\roughly 100 \kpc$ (although the baryonic potential can alter the dark matter distribution). \citet{Kim:2016ujt} presented the effect of this on the brightest cluster galaxies (BCGs) up to 10 Gyr post-pericenter. They demonstrated a wobbling in the BCG as it is able to oscillate about the shallow potential for many oscillations. \citet{1703.07365} analyzed a small set of massive clusters and compared the BCG location with a strong lensing based estimate of the gravitational potential centroid. They compared these observations with hydro-CDM simulations to show that the observations suggest a cored dark matter halo in these clusters of $\roughly 10\kpc$. \citet{Harvey:2018uwf} recently studied cluster-scale halos in hydrodynamic simulations, and saw offsets that grew with cross section and halo mass, although with a smaller amplitude than the dark-matter-only simulations of \citet{Kim:2016ujt} implied. LSST will characterize thousands of relaxed clusters that invariably will have undergone a merger in their history. With deep and relatively high resolution imaging, LSST will allow for single snapshots of the BCG alignment in every massive cluster, and also for detection of faint strong lensing streaks in many of these systems.

% Esra added the text below
SIDM properties are sensitive to the separation between the centroid of the X-ray emitting hot plasma, i.e., intra-cluster medium (ICM), galaxies, and dark matter. Accurate measurements of the X-ray centroid of the X-ray emitting gas in clusters of galaxies requires sub-arcsec imaging with X-ray telescopes. The {\it Chandra} X-ray observatory with 0.5~arcsec FWHM PSF currently provides the most precise location of the ICM. Next-generation high spatial resolution X-ray observatories, \eg, {\textit Lynx} and {\textit AXIS} with much higher throughput, will provide accurate measurements of centers of high-redshift clusters ($z > 1$) in the 2030's and will enable tests of SIDM models over a much larger redshift range.

\section{Compact Object Abundance \Contact{Will}}
\label{sec:compact_objects}
\Contributors{William A.\ Dawson, Nathan Golovich, Simeon Bird, Yacine Ali-Ha\"imoud, Juan Garc\'ia-Bellido, Marc Moniez, Michael Medford, Robert Armstrong, Jessica Lu, Casey Lam}

\begin{figure}[t]
\centering
\includegraphics[width=0.75\textwidth]{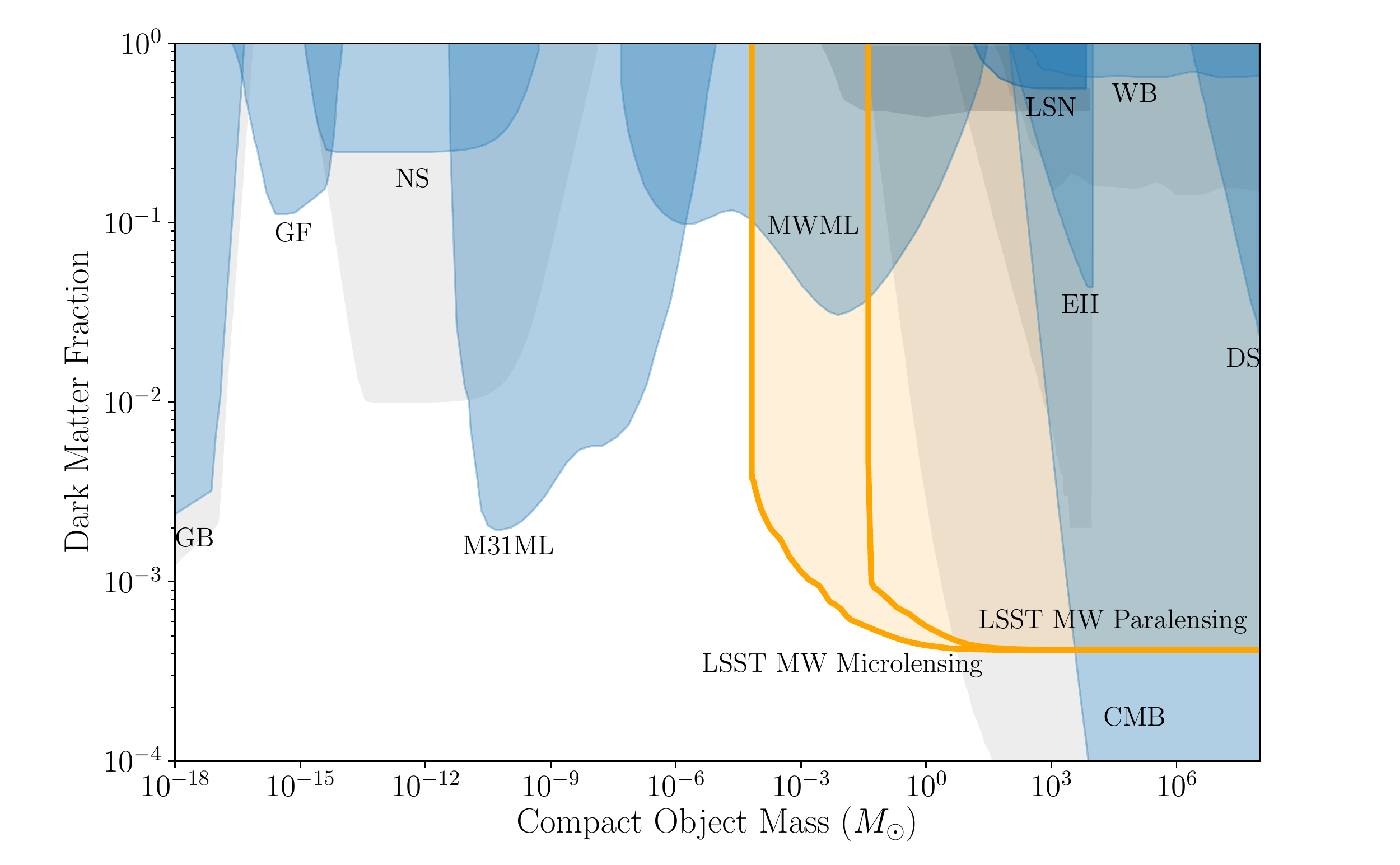}
\caption{\label{fig:macho_constraints}
    Constraints on the maximal fraction of dark matter in compact objects from existing probes (blue and gray) and projections for LSST (gold).
    Existing constraints include: lack of extragalactic gamma-rays from PBH evaporation \citep[EGR;][]{0912.5297, 1604.05349}, gamma-ray femtolensing \citep[GF;][]{1204.2056}, neutron star capture \citep[NS][]{1301.4984}, M31 microlensing \citep[M31ML][]{1701.02151}, Milky Way microlensing \citep[MWML;][]{2007A&A...469..387T, 2001ApJ...550L.169A, 2009MNRAS.397.1228W}, lensing of supernovae \citep[LSN;][]{1712.02240,1712.06574}, Eridanus II and other dwarf-galaxy constraints \citep[EII;][]{2016ApJ...824L..31B, 1611.05052}, wide binary stars \citep[WB;][]{2009MNRAS.396L..11Q, 2004ApJ...601..311Y}, cosmic microwave background \citep[CMB;][]{2017PhRvD..95d3534A, 2008ApJ...680..829R}, and disk stability \citep[DS;][]{1985ApJ...299..633L, 1994ApJ...437..184X}.
    %To improve figure clarity we have not shown some astrophysical constraints where they are less sensitive than a presented constraint; see \citet{2016PhRvD..94h3504C} for a more complete review.
    There are a range of constraints for most astrophysical probes in the literature due to varying assumptions within a single work (EGR, NS, and EII) and reanalysis/disagreements between groups (WB, CMB).
    We present more conservative constraints in blue and more aggressive constraints in gray.
    %The LSST M31 microlensing projection is based on extrapolating HSC constraints \citep{1701.02151} assuming a 10-day mini-survey of M31 with a $12\second$ cadence between exposures.
    %% Such a survey is approximately 10 times longer with an order of magnitude faster cadence than the existing HSC survey.
    The projected LSST Milky Way (MW) microlensing and paralensing constraints come from a Monte Carlo analysis where lenses were injected into light curves based on LSST OpSim cadence simulations. 
    %% (see \url{https://github.com/lsstdarkmatter/dark-matter-paper/issues/8} for details).
    The paralensing constraint assumes that the secondary microlensing parallax signal is used for discovery without incorporating the primary heliocentric microlensing signal.
    The projected LSST sensitivity does not include contamination from conventional astrophysical microlensing events.
}
\end{figure}

MAssive Compact Halo Objects \citep[MACHOs;][]{1991ApJ...366..412G} have a long history as dark matter candidates \citep{1974ApJ...193L...1O, 1980ApJS...44...73B, 1981ApJ...243..140G, 1986ApJ...304....1P, Bellido:1996, Clesse:2015, Bird:2016, Clesse:2016}. 
Cosmological observations of the CMB, BAO, and deuterium abundances have shown that compact objects must be non-baryonic if they are to make up a majority of dark matter \citep[\eg][]{Ade:2015xua}. 
As described in \secref{machos}, this has led to the identification of primordial black holes (PBHs) as the most popular candidate for MACHO dark matter \citep{Bellido:1996}.
There are a number of astrophysical probes that constrain the PBH dark matter abundance over mass scales ranging from $10^{-17}-10^{15}\,M_\odot$ (\figref{macho_constraints}).
At the lowest masses ($M < 10^{-9}\Msun$), PBHs are constrained through the non-detection of PBH evaporation in the extragalactic gamma-ray background \citep[\eg,][]{0912.5297, 1604.05349}, non-detection of femto-lensing of gamma-ray bursts \citep[\eg,][]{1204.2056}, the rate of SN Type 1a \citep{1805.07381}, and neutron star capture \citep[\eg,][]{1301.4984}.
The landscape of intermediate-mass MACHOS ($10^{-11} \Msun < M < 10 \Msun$) is predominantly constrained by microlensing observations, which limit the monochromatic compact dark matter fraction to be $\lesssim 10\%$ over this mass range \citep[\eg,][]{2001ApJ...550L.169A, 2007A&A...469..387T, 2009MNRAS.397.1228W, 1509.04899, 1701.02151, Calcino:2018}.
At the high-mass end ($M \gtrsim 10^3\Msun$), PBH dark matter is subject to constraints from dynamical stability of wide binary stars \citep[\eg,][]{2009MNRAS.396L..11Q, 2004ApJ...601..311Y}, star clusters \citep[\eg,][]{2016ApJ...824L..31B, 1611.05052}, dwarf galaxies \citep{1704.01668}, and the Galactic disk \citep[\eg,][]{1985ApJ...299..633L, 1994ApJ...437..184X}.
Lyman-$\alpha$ observations disfavor PBHs with $M > 10^4\Msun$ based on an observed plateau in the Poisson term of the matter power spectrum \citep{astro-ph/0302035}.
Strong constraints have also been placed on the abundance of PBHs with mass $\gtrsim 1 \Msun$ using CMB anisotropies \citep{2008ApJ...680..829R}.
However, these constraints have been shown to be extremely model dependent and were relaxed substantially in subsequent studies \citep{2017PhRvD..95d3534A}.
%Several indirect constraints have been published that rule out most of the mass scales above the sensitivity of these microlensing surveys; however, these constraints rely on complex astrophysical assumptions.
This, in addition to recent direct observations by LIGO of mergers of $10-50 \Msun$ black holes, potentially with less spin than expected \citep{1602.03837, LIGOScientific:2018b, LIGOScientific:2018a},
has motivated a renewed interest in PBH dark matter with larger masses than have been so far constrained directly by microlensing.

In this section, we focus on the ability of LSST to directly detect signals of compact halo objects through precise, short ($\sim30\,$s) and long-duration ($\sim$ years) gravitational microlensing observations.
If scheduled optimally, the wide field-of-view, high cadence, and precise photometry of LSST will provide sensitivity to microlensing event rates corresponding to $\roughly 0.03\%$ of the dark matter density in compact objects with masses $>0.1\Msun$ (\figref{macho_constraints}).
We briefly mention that LSST will also probe PBHs by determining the rate of type 1a SN, identifying candidate wide-binary star systems at greater distance than is possible with \Gaia, and through dedicated mini-surveys of high stellar density fields (similar to that performed with HSC by \citealt{1701.02151}).

\subsection{Microlensing}
\label{sec:microlensing}

\begin{figure}
\centering
\includegraphics[width=0.7\columnwidth]{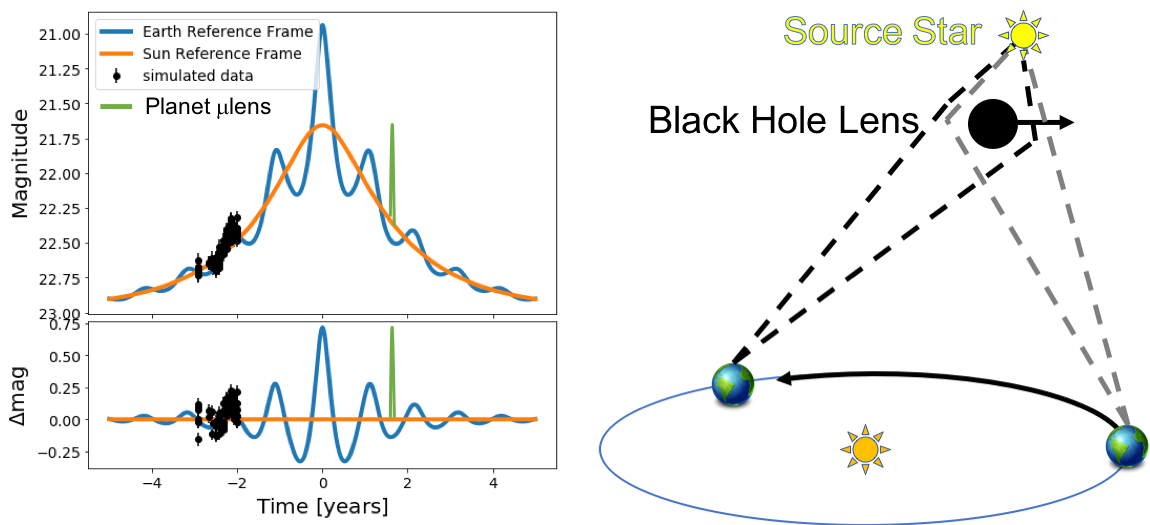}
\vspace{1em}
\caption{\label{fig:microlensing_cartoon}
    \emph{Left:} 
        A descriptive example of the microlensing and paralensing signals for a $23 \magn$ source being lensed by a $50 \Msun$ black hole. 
        For events with an Einstein crossing time much less than a year ($\lesssim 1 \Msun$), the microlensing magnification will appear symmetric in time (orange curve).
        For microlensing events lasting on the order of a year or more ($\gtrsim 1 \Msun$), the lensing geometry changes due to parallax as Earth orbits the Sun.
        This paralensing signal has a period of a year, with the phase determined by the coordinates of the source star, making it robust to astrophysical systematics.
        %It is also possible to detect binary dark matter, and extend the mass range to planet mass compact dark matter, via the source passing through one of the gravitational lensing caustic curves formed by the binary lens.
        %The green curve on top of the heliocentric orange curve is representative of a typical planetary microlensing event caused by a caustic crossing.
        %While LSST can measure these events if lucky, we will rely on LSST to detect the heliocentric microlensing event and trigger targeted follow-up higher cadence observations to measure the planetary microlensing event.
        The black data points are representative of extending the LSST wide-fast-deep cadence into the Galactic plane. \WAD{Need to update this figure with the LSST WFD cadence.}
        \emph{Right:} 
        A cartoon diagram of paralensing. 
        For microlensing events lasting on the order of a year or more, the lensing geometry changes as Earth orbits the Sun, leading to a parallax effect.
        %\Contributors{Will D., PALS Collaboration}
    }
\end{figure}

Gravitational microlensing, the achromatic brightening and dimming of background stars due to the transit of a massive compact foreground object, can be used to directly detect and measure the properties of PBHs.
The idea of employing microlensing to search for compact objects in the Galactic halo was proposed by \citet{1986ApJ...304....1P}, and several photometric surveys commenced in the 1990's including MACHO \citep{1992ASPC...34..193A}, OGLE \citep{1992AcA....42..253U}, and EROS \citep{1993Msngr..72...20A}.
These collaborations provided the first \emph{direct} constraints on the compact nature of dark matter; however, they were limited by image quality, analysis techniques, and computational resources.
These limitations, combined with the $\roughly10$-year duration of these surveys, led to a loss of sensitivity at $M \gtrsim 1 \Msun$.
LSST can surpass this limitation by directly detecting events based purely on the parallactic component of the lensing signal (\figref{microlensing_cartoon}).
Furthermore, by supplementing the LSST survey with astrometric microlensing surveys using other telescopes (HST, Keck AO), LSST can break the lensing mass-geometry degeneracies and make precise measurements of individual black hole masses, thereby measuring the black hole mass spectrum in the Milky Way halo.
If compact objects make up a significant fraction of dark matter, LSST will provide insight into the primordial perturbations and early universe equation of state, in the case of PBHs, or provide evidence that dark matter particle physics is complex enough to allow significant cooling channels. 
Dark matter with sufficient self-interactions to cool will also affect halo profiles, and LSST will be well-placed to distinguish different models for the formation of novel compact objects.
%Thus, if PBHs make up a significant fraction of dark matter, LSST will effectively measure their ``particle'' properties. 
%A precise measurement of the PBH mass spectrum will provide insight into the fundamental physics of the early universe.

\noindent \textbf{The Microlensing Signal}
%Gravitational microlensing occurs when a massive lens passes between a background source and an observer, causing the light from that source to pass through a warped space-time acting as a lens. \WAD{cite Wambsganss for a detailed review}

Gravitational microlensing results in two potentially observable features: (1) photometric microlensing, a temporary amplification of the brightness of a background source, which is achromatic in the case of unblended sources, and (2) astrometric microlensing, an apparent shift in the centroid position of the source.
The characteristic photometric signal of a simple point-source, point-lens (PSPL) model as observed from the center of the solar system is symmetric, achromatic, and has both a timescale and maximum amplification that depend on the mass of the lens.
LSST will observe billions of stellar sources in multiple filters over several years to enable the detection of thousands of microlensing events across a wide range of timescales and consequently a wide range of masses.
This simple PSPL model is complicated by astrophysical factors including the velocity distribution of sources and lenses, extinction due to Galactic dust, blending in dense stellar fields, and the shift in perspective resulting from viewing a microlensing event while the Earth revolves around the Sun.
Fortunately, these complications can be addressed and disentangled to arrive at the mass of the gravitational lens and a detection of dark matter via microlensing \citep{1405.3134,1509.04899}.

One particularly powerful feature for long-duration microlensing events results from the change in the geometric configuration of the source-lens-observer system as the Earth orbits the Sun (\figref{microlensing_cartoon}).
The change in viewing angle and distance results in a parallax effect that imposes a 1-year periodicity on top of an otherwise symmetric microlensing light curve.
This additional signal exists irrespective of the mass of the lens, providing an independent measurements of the distance of the lens and breaking the mass-distance degeneracy of a microlensing signal \citep[\eg,][]{1509.04899}.
This enables microlensing to directly constrain compact dark matter at much larger mass scales ($M \gtrsim 1\Msun$), where the duration of the event is $\gtrsim 1$ year. 
Based on Figure 3 from \citet{1509.04899}, we expect that LSST will be able to use paralensing to detect $10\Msun$ lenses out to a distance of $\roughly5 \kpc$.

Moreover, if primordial black holes are significantly clustered in the halo of our galaxy, forming pockets of hundreds of massive PBH, then one should expect to detect a few very-long-duration microlensing events, of order decades~\citep{Bellido:2017}. If one of the constituent PBH in the cluster happen to be directly along the line of sight of the magnified star then one would expect, on top of the smooth Paczynski curve, a sharp caustic associated with the Einstein-ring crossing time of the individual PBH within the cluster~\citep{Bellido:2018}.

%\ADW{What is the maximum distance for which the parallax signal is measureable?}

%\WAD{achromatic}
Gravitational lensing is achromatic, making the multiple filter observations of LSST a key advantage for distinguishing a microlensing signal from other astrophysical transient and variable objects.
The benefit derived from multi-filter observations will depend strongly on the selected LSST cadence. 
Microlensing signals are more easily extracted from frequent observations in fewer filters, as long as sources are observed in at least two colors.

\noindent \textbf{Microlensing Systematics}

As with any empirical observation, microlensing measurements are subject to systematics that must be accounted for.
We briefly summarize these systematics and strategies for mitigating their effect.

\paragraph{Variable and binary stars:} Temporally variable objects are a potential source of false detections at all timescales. However, the microlensing signal can be distinguished because, modulo secondary ambiguous blending effects, it is achromatic. Most astrophysical variable and binary stars, by contrast, are associated with a temperature dependence and thus have chromatic variability. To leverage this fact, it is necessary for LSST to survey high-stellar-density fields in at least two bands. Furthermore, for microlensing events with durations $\gtrsim 1$ year, mimicking the annual parallax signal imprinted on the microlensing signal would require a binary or variable star with a period of exactly 1 year.

\paragraph{Blending:} Given the typical scale of the Einstein radius (the approximate region where a microlensing signal is detectable), the odds of a microlensing event along an arbitrary line of sight are $\roughly 10^{-7}$--$10^{-5.5}$ \citep[\eg,][]{2000ApJ...541..734A,2006ApJ...636..240S}.
Due to the low probability of a microlensing event, most observable microlensing events will be in dense stellar fields (e.g., the Galactic plane, Magellanic Clouds, M31), driving microlensing surveys to these regions.

These dense survey fields, coupled with LSST depth and ground-based PSF with FWHM $\sim 0.7 \arcsec$ \citep{0805.2366}, lead to significant ambiguous blending (i.e., multiple objects within a single PSF).
Towards the Galactic center there are $\roughly 50$ stars within an LSST PSF, similarly there are $\roughly 15$ and $\roughly 5$ stars per PSF in the Galactic bulge and disk, respectively \citep{1806.06372}.
LSST can overcome much of the blending problem to detect variations in the brightness of stars, including the detection of microlensing lightcurves, through the process of difference imaging.
With difference imaging, reference images are built through the coaddition of multiple observation of the same fields, resulting in a deep image of the static sky. Reference images are then scaled and PSF-matched to individual observed science images and subtracted off, resulting in a difference image that only contains the signal that differs from the static sky. 

Difference imaging should reduce many of the systematic effects associated with blending. Follow-up high resolution imaging from space or gronud-based adaptive optics can mitigate most blending issues for detected microlensing events.
%; however, it will still be difficult to characterize the intrinsic source star baseline photometric properties. Crowding can also lead to a secondary chromatic signal with exactly the same duration/shape as the microlensing event, although with different amplitude. 
Blending systematics could also be mitigated through spectroscopic follow-up of microlensing events before and after crossing. 
%Lensing by low-mass stars will alter the spectrum of the source, while PBH lensing will leave the spectrum unchanged.

\paragraph{Cadence:} The temporal cadence of LSST observations will be important for optimizing sensitivity to microlensing events. 
While high-mass black holes should be accessible through relatively sparse observations distributed over the course of the year, smaller black holes require higher cadence observations.
\citep{1812.03137} suggest an observation strategy that includes a survey of the Galactic Bulge, Galactic Plane, and Magellanic Clouds with a reduced filter set at a cadence of 2 – 3 days. 
Observing the same region in at least two filters within the same night will allow tests of achromaticity.
\citet{1812.03139} suggest very high-cadence observations of the Magellanic Clouds (continuous 15s exposures), which should be sensitive to microlensing events from low-mass PBHs and scintillation light from invisible baryons \citep[\eg,][]{2003A&A...412..105M}.
The short readout time of the LSST Camera (2\,s) should allow for high-cadence observations of very short duration, subsolar mass PBHs, similar to the HSC observations of M31 \citep{1701.02151}.

\paragraph{Galaxy Model:} While microlensing can be detected independent of any detailed knowledge of Galactic structure, properly incorporating uncertainty in the Galactic dust, stellar velocity distributions, and dark matter halo model is essential to interpret the microlensing signal in the context of dark matter.
Significant improvements in our understanding of the Milky Way's dark matter halo have been made on this front since the first microlensing surveys \citep[\eg,][]{Calcino:2018}, and LSST will further improve these estimates (\secref{direct}).
In addition, systematic  microlensing measurements, especially when extended to the Galactic spiral arms, have the potential to strongly constrain the baryonic structure of the Galaxy in terms of mass density distribution and kinematical structure \citep[\eg,][]{Moniez:2017}.

%\paragraph{Binary vs isolated stars:} A potential systematic for the paralensing signal is binary star systems with approximately year-long periods. This is unlikely to be a significant systematic due to the low probability of having an achromatic binary system with a year long period. %\WAD{There was a researcher who did some of these studies for our microlensing group. Need to reference his work/arguments.} 

\noindent {\bf Projected Sensitivity}

%We estimate the projected sensitivity of an LSST survey optimized for the detection of microlensing signals from PBHs and present the results in \figref{macho_constraints}.
%The LSST M31 microlensing projection is based on extrapolating the \citet{1701.02151} by assuming a ten-day mini-survey of M31 with a 12 second cadence between exposures.
%Such a survey is approximately 10 times longer than that of \citet{1701.02151}, with an order of magnitude faster cadence.

We project the LSST Milky Way (MW) microlensing and paralensing sensitivity  from a Monte Carlo analysis of injected light curves based on LSST OpSim cadence simulations.
%\footnote{Details available in: \url{https://github.com/lsstdarkmatter/dark-matter-paper/issues/8}.}
We inject simulated microlensing signals onto the OpSim lightcurves for $2\times10^5$ PBH lenses in the mass range from $10^{-2}\Msun$ to $10^{8}\Msun$ distributed using a triangular density distribution, where $p(r = 0 \kpc) = 0$ and $p(r = 8\kpc) = 0.25$.
The sources are assumed to be $23\magn$ stars located at $8\kpc$ with measurement noise generated from the standard LSST photometric error model \citep{0805.2366}.
LSST is expected to detect $\roughly 10^{9}$ such stars, with an optical depth of $4.48 \times 10^{-6}$ \citep{2006ApJ...636..240S}.
The paralensing constraint comes from assuming that only the secondary microlensing parallax signal is used for discovery, and not the primary heliocentric microlensing signal.
These sensitivity projection do not attempt to separate microlensing events produced by PBHs from conventional microlensing events produced by stars and compact objects resulting from stellar evolution. 
Thus, these projections should be interpreted as the limiting sensitivity of LSST to \emph{all} microlensing events, regardless of their origin. 
In \secref{pbh_discovery} we discuss the expected conventional microlensing event rates in more detail.

\section{Anomalous Energy Loss Mechanisms\Contact{Maurizio}}
\label{sec:cooling}
\Contributors{Maurizio Giannotti, Oscar Straniero, Samuel D.\ McDermott, Alex Drlica-Wagner}

Observations of stars provide a mechanism to probe temperatures, particle densities, and time scales that are inaccessible to laboratory experiments.
Since conventional astrophysics allows us to quantitatively model the evolution of stars, the detailed study of stellar populations can provide a powerful technique to probe new physics.
In particular, if new light particles exist and are coupled to Standard Model fields, their emission would provide an additional channel for energy loss. 
Such anomalous energy loss mechanisms would change the time that stars spend in specific stellar evolutionary phases.
Such deviations are a robust predictions of light, weakly coupled particles, and the general agreement between observations and Standard Model predictions has been used to constrain the properties of many types of new particles \citep{hep-ph/0611350, 1210.1271, 1302.3884, 1305.2920, 1611.03864, 1611.05852, 1803.00993}.

\begin{figure}[t]
\centering
\includegraphics[width=0.55\columnwidth]{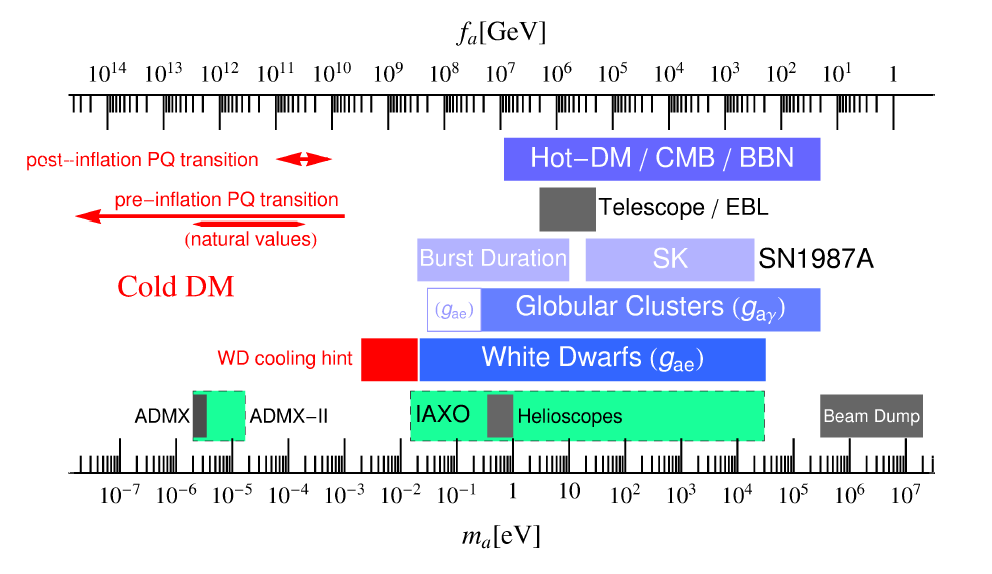}
\includegraphics[width=0.44\columnwidth]{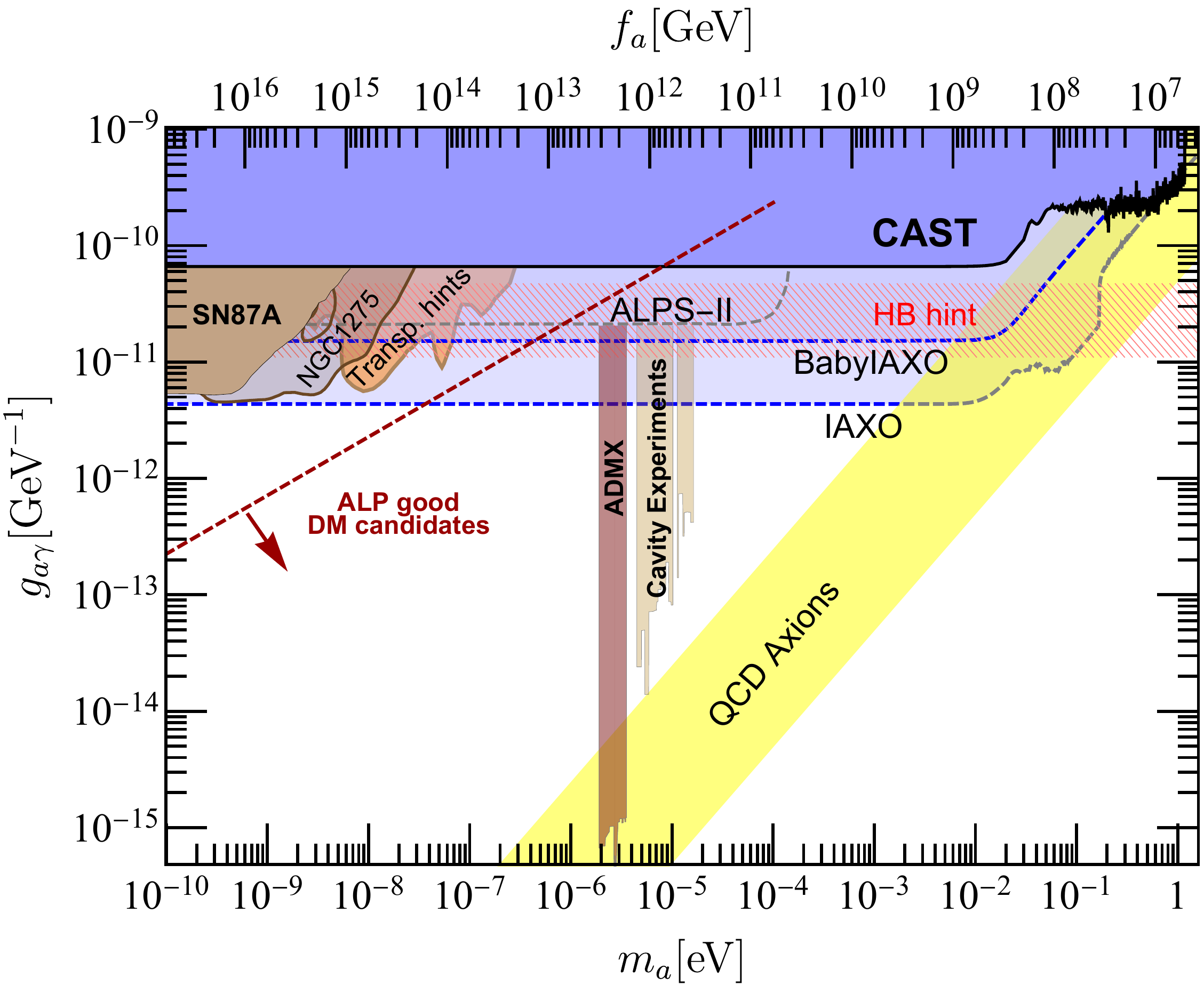}
\caption{Left: Existing experimental and observational constraints on the QCD axion \citep{Redino:2015}.  
Right: Constraints on ALP coupling to photons in the parameter space of coupling strength and particle mass \citep{DiVecchia:2019}.
Astrophysical observations of horizontal branch stars (HB) are interpreted assuming the ALP only interacts with photons.
Note the wide range of mass and coupling scales that are constrained by these observations (see text for more details).
\label{fig:axions}
}
\end{figure}

While the predictions of the Standard Model are broadly consistent with observations of stellar evolution, several independent observations have shown a systematic preference for an additional subdominant energy-loss mechanism (see \citealt{Giannotti:2017hny} for a recent review).
These observations include red giants branch (RGB) stars, in particular the luminosity of the tip of the branch~\citep{Viaux:2013lha,Viaux:2013hca}; 
horizontal branch stars (HB), specifically by comparing the number of HB and RGB stars~\citep{Ayala:2014,Straniero:2015nvc};
variable white dwarf (WD) stars, for which the cooling efficiency was extracted from the rate of the period change~\citep{KeplerEtAl,Isern:1992gia,BischoffKim:2007ve,Corsico:2012ki,Corsico:2012sh,Corsico:2014mpa,Corsico:2016okh,Battich:2016htm}; 
and the WD luminosity function (WDLF), which describes the distribution of WDs as a function of their luminosity~\citep{Isern:2008nt,Bertolami:2014wua,Isern:2018uce}.
Observed discrepancies between these stellar measurements and predictions from conventional models of stellar cooling can be interpreted as the need for additional energy loss (\figref{axions}).
\cite{Giannotti:2015kwo} provide a systematic analysis of the new-physics interpretation of stellar observations  where cooling anomalies have been reported, and they conclude that axions and ALPs are the best candidates to account for the observed discrepancies. 
While these `hints' of anomalous cooling represent subdominant deviations to broadly successful models of stellar evolution, it is imperative to explore possible signatures of new physics when they arise.

LSST will greatly improve our understanding of stellar evolution by providing unprecedented photometry, astrometry, and temporal sampling for a large sample of faint stars \citep{0912.0201}.
These observations will allow us to better assess the significance of claimed anomalies, and will further guide constraints on (or detection of) new physics.
A better understanding of astrophysical energy transport will ultimately help shed light on the physics of light, weakly-coupled particles and will offer an invaluable guide to future experimental searches for axions and ALPs~\citep{Irastorza:2018dyq}.

\subsection{White Dwarf Luminosity Function}

The white dwarf luminosity function (WDLF) plays a particularly significant role in our understanding of stellar cooling and offers a fundamental method to test new physics.
Measurements of the slope of the WDLF can probe additional energy loss mechanisms and the production rate of the novel particle responsible for the nonstandard cooling.
The general agreement between the observed WDLF and predictions from standard astrophysics has been used to place bounds on the axion-electron coupling \citep{Isern:2008nt,Bertolami:2014wua}, on the anomalous neutrino magnetic moment \citep{Bertolami:2014noa}, on the kinematic coupling of dark photons to standard photons \citep{Chang:2016qfl}, and on the variation of the gravitational constant \citep{Althaus:2011ca}.
However, several recent analyses of the WDLF have shown a preference for additional energy loss with respect to the Standard Model predictions.
In particular, \cite{Bertolami:2014wua} used data from the Sloan Digital Sky Survey (SDSS) and the SuperCOSMOS Sky Survey (SCSS) to show a $2 \sigma$ discrepancy from the Standard Model prediction, which could be explained by axions coupling to electrons with $g_{\phi e}\simeq 1.4\times 10^{-13}$.\footnote{The additional energy can also be accounted for by dark photons~\citep{Giannotti:2015kwo,Chang:2016qfl}, but not by anomalous neutrino electromagnetic form factors~\citep{Bertolami:2014noa}.}
These measurements of the WDLF have guided experimental searches for axions and ALPs, particularly the IAXO~\citep{Irastorza:2011gs,Armengaud:2014gea,Giannotti:2016drd} and ALPS II~\citep{Bahre:2013ywa,ALPSII} experiments.

Observations from the \Gaia satellite have already increased the catalog of WDs by an order of magnitude with respect to SDSS \citep{1805.01227,1807.02559,1807.03315}.
The growing sample of WDs with precisely measured distances will enable an improved measurement of the WDLF.  
However, the completeness of the \Gaia sample is limited to WDs within 100 pc~\citep{1807.03315}.
%Oscar. In the GAIA DR2 about 260.000 WDs have been identified. However the sample is complete up to G=20-21, only for those within 100 pc, which are about 11.000 stars. The major problem is that the completeness drops at low Galactic latitudes, and the magnitude limit of the catalogue varies significantly across the sky as a function of Gaia’s scanning law.  A larger and more complete sample will be certainly available with the final data release.
LSST is expected to detect WDs that are 5 to 6 magnitudes fainter than those detected by \Gaia, ultimately increasing the census of WDs to tens of millions~\citep{0912.0201}.
LSST will provide more complete and homogeneous samples of WDs, allowing for a significant reduction in both the statistical and systematic uncertainties in measurements of the WDLF. 
LSST is expected to measure hundreds of thousands of WDs in the Galactic halo, enabling the construction of a reliable luminosity function of halo WDs. 
By deriving independent WDLFs from different Galactic populations it will be possible to reduce uncertainties related to star formation histories, and to ultimately provide a more clear assessment of the physical origin of the cooling anomalies \citep{Isern:2018uce}. 

The growing sample of WDs will similarly increase the known population of variable WDs with pulsation periods of 100s--1500s. 
Measured changes in the pulsation periods of WDs can be used to directly constrain the rate of cooling \citep[\eg][]{1007.2659}.
Indeed, hints of anomalous cooling from axions have been claimed \citep[\eg][]{Corsico:2012ki,Corsico:2012sh}, though more recent analyses set upper limits at the level of $g_{\phi e} < 3.3 \times 10^{-13}$ \citep{Battich:2016htm}. 
LSST will greatly increase the sample of pulsating WDs, enabling high-cadence follow-up observations to precisely measure changes in pulsation period and probe anomalous cooling mechanisms.

\subsection{Globular Cluster Stars}

Massive stars, specifically those close to the helium burning phase, provide another excellent environment to study anomalous energy loss mechanisms. 
Stellar evolutionary codes such as MESA \citep{1009.1622} provide a good model for the evolution of massive stars, allowing constraints to be placed on novel particle production \citep[\eg,][]{1210.1271,1611.05852}.
Several recent analyses of giant branch stars have reported deviations from standard stellar model predictions that can be interpreted as a signature of anomalous energy loss.
For example, studies have shown a brighter-than-expected tip of the RGB (TRGB) in the M5 globular cluster~\citep{Viaux:2013lha,Viaux:2013hca}, indicating somewhat over-efficient cooling during the evolutionary phase preceding the helium flash.
The anomalous brightness, $\Delta M_{I,{\rm TRGB}}\simeq 0.2$ mag in absolute $I$-band magnitude, observed in M5 can be interpreted as an anomalous cooling of a few $10^{33}$ erg/s.
Such cooling could be accounted for by a neutrino magnetic moment or an axion-electron coupling of the order of that predicted from the WDLF~\citep{Viaux:2013lha}. 
These constraints can be improved using multi-band photometry of multiple globular clusters \citep[\eg,][]{Straniero:2018fbv}.

Advances in the analysis of globular cluster RGB stars are currently limited by modeling uncertainties on the stellar evolution of the giant branch.
Fundamental improvements should be expected in the near future. 
In particular, exquisite astrometry from the {\it Gaia} satellite will precisely determine cluster distances, currently the largest sources of observational uncertainty in the determination of the absolute luminosity of the TRGB.\footnote{The {\it Gaia} data relevant for GCs are expected in 2022~\citep{Gaia}.}
Moreover, the angular resolution of the next-generation space-based missions, such as JWST~\citep{Gardner:2006ky}, will enlarge the statistical sample of RGB members near the cores of GCs. 
The brightness of RGB stars in nearby GCs limits the contributions of LSST, which saturates at $g \sim 17$ mag and suffers from crowding near the cores of GCs.
However, LSST will enable independent measurements of GC distances, providing a valuable handle on systematic uncertainties of {\it Gaia} observations.\footnote{The parallaxes of bright ($G<14$ mag) sources can be derived with a median uncertainty of 0.04 mas in {\it Gaia} DR2. However, for fainter stars the parallaxes become sensitive to systematic errors.  Presently, these systematics hamper a precise determination of GC distances \citep{Chen:2018}.}
Moreover, it is likely that the homogeneity and precise photometry of LSST will improve the calibration of the bolometric corrections for RGB and HB stars and ultimately contribute to a more clear assessment of the cooling of GC stars.

\subsection{Massive Stars and Core-Collapse Supernovae}

\begin{figure}[t]
\centering
\includegraphics[width=0.5\columnwidth]{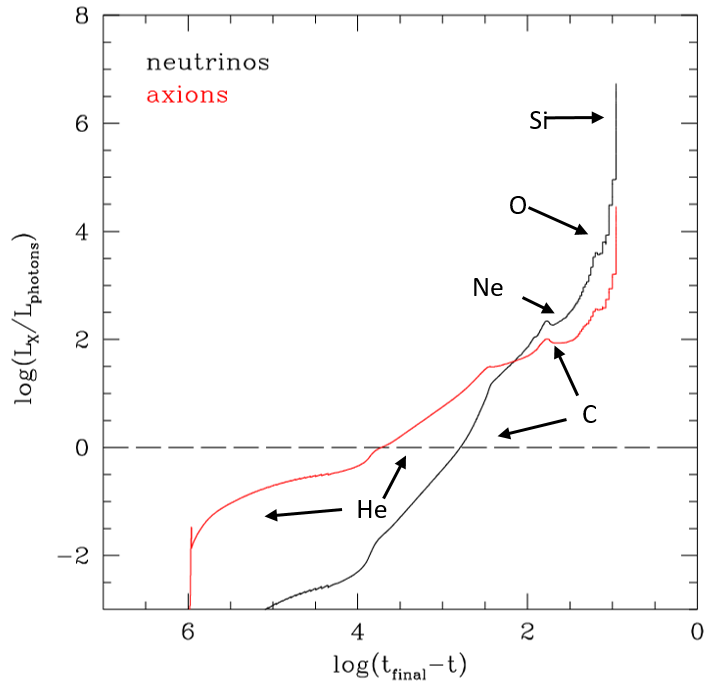}
\caption{Evolution of the luminosities of neutrinos and axions in a $M=18 \Msun$ stellar model ($t$ in years), relative to the photon luminosity. The various evolutionary phases are indicated. 
%In the calculation of the axion rate we have assumed the presently available upper bounds, as obtained from astrophysical constraints, for the axion-photon and axion-electron coupling constants.
In the calculation of the axion rate, we have assumed the current upper bounds on the axion-photon and axion-electron coupling.
\ADW{Do we need a reference for this figure?}
}
\label{fig:massivestar}
\end{figure}

The cores of massive stars are among the most powerful natural laboratories to investigate the possible production of weakly interacting light particles, particularly axions. 
The energy loss rate via axions is quite sensitive to temperature. 
For instance, the rate of the Primakoff process (the photon-axion conversion in the static electric field of ions and electrons) scales as $T^4$. 
\figref{massivestar} shows the evolution of the neutrino and axion luminosities in a $18\Msun$ star. 
After He burning, the central temperature rapidly increases, becoming larger than $10^9$ K. 
In standard stellar models (no axions or other non-standard cooling), the energy loss by neutrinos largely overcomes the energy loss by photons. This rapidly decreases the evolutionary time scale and determines the chemical and physical structure of the star at the onset of core collapse. 
In this context, an additional energy loss mechanism may significantly affect the pre-explosive stellar structure and, in turn, may determine the success or failure of a core collapse supernova (CCSN). 
Such an effect may be revealed by connecting CCSNe to their massive star progenitors, something that will be enabled by the wide area and high temporal cadence of LSST.

Another powerful strategy by which novel particles can be constrained is by considering the evolution of the neutrino cooling phase of nearby SNe  \citep[\ie, SN1987A][]{Burrows:1988, Raffelt:1988}.
The simplest and most robust method by which such constraints can be implemented is the so-called ``Raffelt criterion,'' which limits the luminosity of new particles to be below the luminosity of neutrinos during the neutrino-cooling phase \citep{hep-ph/0611350}.
Neutrino observations of SN1987A have been used to place limits on a wide variety of new particles \citep{hep-ph/0207098, 1611.03864, 1611.05852, 1803.00993, 1808.10136}.
Again, it is possible that a subdominant release of energy into new particles is responsible for resolving some lingering inconsistencies with the Standard Model-only picture of CCSNe explosions \citep{0806.4273, 1805.07381}, though such an effect is difficult to resolve analytically on top of other qualitative uncertainties of the Standard Model-only picture \citep{1809.05106, 1811.11178}.
LSST, in combination with upcoming neutrino experiments (i.e., DUNE), will help reduce Standard Model uncertainties and expand this analysis to future and more distant CCSNe \citep{1807.10334}.

LSST is expected to discover $\roughly 3.5\times10^5$ CCSNe per year \citep{Lien:2009}. 
%\ADW{Better to update this from Goldstein et al. (2018).}\AHGP{I think the Goldstein ref is with respect to lensed SNe}\ADW{Goldstein starts from a full population analysis, which is what we would take.}
For nearby SNe, LSST will be able to resolve massive progenitor stars in pre-explosion imaging. 
So far, a clear identification of progenitor stars has been obtained only for about 20 type II SNe \citep{Smartt:2015}.  
The identification of a much larger number of massive stars before they explode is a mandatory step for understanding the CCSNe process and the possible activation of non-standard cooling processes during the late evolution of massive stars. 
Recent theoretical studies have investigated the conditions for which a massive star successfully bounces after core collapse, giving rise to a SN \citep[\eg,][and references therein]{OConnor:2011,Sukhbold:2016}.   
In particular, it was found that the ability to explode predominantly depends on the structure of the progenitor. 
%Therefore, the impact of LSST in this field is twofold: it will constrain the supernova engine and may provide hints for new physics beyond the standard model. 
Therefore, besides providing more clear insight into the SN engine, LSST will provide a solid framework to test the presence of novel cooling channels efficient during pre-SN evolution, constraining or hinting at the existence of axions or other weakly interacting particles.

\section{Large-Scale Structure \Contact{Tony}}
\Contributors{J.\ Anthony Tyson, Rogerio Rosenfeld, Keith Bechtol, Francis-Yan Cyr-Racine}
\label{sec:lss}

LSST is anticipated to produce the largest and most detailed map of the distribution of matter and the growth of cosmic structure over the past 10 billion years. The large-scale clustering of matter and luminous tracers in the late-time universe is sensitive to the total amount of dark matter, the fraction of dark matter in light relics that behave as radiation at early times, and fundamental interactions in the dark sector. For example, the high densities of matter and energy in the early universe imply that even extremely weakly coupled dark matter particles can leave detectable imprints on the galaxy distribution today. Dark matter that couples to photons, neutrinos, or a light scalar field could produce dark acoustic oscillations analogous to baryon acoustic oscillations \citep{Cyr-Racine:2013fsa}. Large-scale structure probes would be particularly competitive for multi-component dark matter models in which only a fraction of the dark matter couples to dark radiation. Meanwhile, LSST will enable tests of dark matter with non-standard gravitational interactions, such as violations of the equivalence principle \citep{Bonvin:2018} and interactions between dark matter and dark energy. The nature of the dark matter can be broadly described by a so-called Generalized Dark Matter model \citep{Hu:1998}, which has been recently explored in a study of the dark matter equation of state through cosmic history \citep{Kopp:2018}. 
Dark matter probes involving large-scale structure highlight the interconnectedness of dark matter and dark energy research, both in terms of testing the validity of the standard cosmological paradigm, and overlap in the specific analysis methods employed. The studies described in this section are illustrative examples of how the nature of dark matter can be probed employing the same galaxy clustering and weak lensing techniques as used in dark energy constraints. 

As one specific example, measurements of large-scale structure with LSST will enhance constraints on massive neutrinos and other light relics from the early universe that could compose a fraction of the dark matter. The combination of galaxy imaging and redshift surveys together with CMB experiments in the coming decade will make it possible to measure the density of relativistic particles in the early universe at the percent level. The existence of the cosmic neutrino background can already be inferred from temperature fluctuations of the CMB \citep{Planck:2018_cosmo_params} and Big Bang nucleosynthesis \citep{Cooke:2018}, and a major goal of upcoming cosmology experiments is to measure the sum of neutrino masses at the few milli-eV level using weak gravitational lensing and galaxy clustering \citep[e.g.,][]{CMB-S4:2016,DESI:2016,Mishra-Sharma:2018}.\footnote{The LSST Dark Energy Science Collaboration ``Science Roadmap'' is available at \href{http://lsstdesc.org/sites/default/files/DESC_SRM_V1_4.pdf}{this url}.} In the standard cosmological model with three neutrinos, the effective number of relativistic free-streaming species is $N_{\rm eff}$ = 3.046. Several classes of dark matter models, such as axions, axion-like particles, dark photons, and light sterile neutrinos, predict deviations in $N_{\rm eff}$ that could be measurable with LSST and CMB experiments, even if the light species decouple before the QCD phase transition \citep{Font-Ribera:2014,Baumann:2018}. Galaxy surveys contribute to constraints on $N_{\rm eff}$ by independently constraining the Hubble constant, which is partially degenerate with $N_{\rm eff}$, and importantly, by measuring the broadband shape and phase of the galaxy power spectrum, which are sensitive to the gravitational influence of free-streaming light relics. The broadband galaxy power spectrum is less dependent on the primordial helium abundance and can lend confidence to CMB measurements of $N_{\rm eff}$.

\begin{figure}[t]
\centering
\includegraphics[width=0.9\columnwidth]{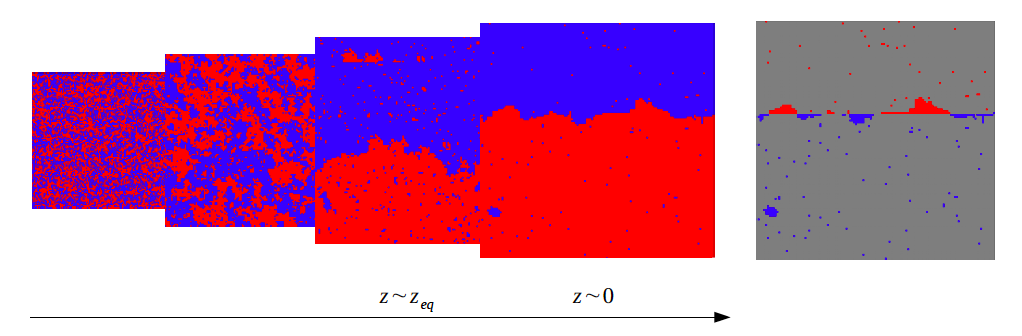}
\caption{A schematic diagram of the emergence of dark energy anisotropy from an Ising model phase transition and a coupling with the anisotropic distribution of dark matter. Figure taken from \citet{1810.11007}.}
\label{fig:DMDEmap}
\end{figure}

Another example demonstrating the potential of large-scale structure to study dark matter involves probing possible couplings between dark matter and dark energy \citep[\eg][]{0801.1565}. %, \eg, if dark energy were an aspect of dark matter at late times.
%The physics of dark matter could be probed via LSST 
The observed inhomogeneities in dark matter on large scales could then be reflected as spatial inhomogeneities in the late-time acceleration. 
This could be observed as spatial variations in cosmic acceleration by LSST correlated with the large-scale weak lensing maps of dark matter. 
Such a spatial correlation is a generic prediction if dark matter and dark energy are causally related or if dark energy is emergent \citep{1801.09658}. 
A spatially complicated potential leads to a small cosmological constant from an energy difference between its global and local minima, and dark energy and dark matter are thereby intertwined. 
If so, and if the universe on sub-horizon scales is not homogeneous, then spatial fluctuations in one component should be correlated with spatial fluctuations in the other, particularly near the epoch of emergence.

There are other models where dark matter and dark energy are intertwined, such as models where they interact 
\citep{Amendola:1999er,Holden:1999hm}.
Interacting models can be described phenomenologically via two fluids that can exchange energy and momentum, 
described by energy-momentum tensors that are not individually conserved. One can parameterize the coupling of
dark matter and dark energy by writing the divergence of the individual energy-momentum tensors as 
\begin{eqnarray}
\nabla_{\mu} T^{(DE)}\,^{\mu}_{\nu} &=& C^{(DE)}_{\nu}, \label{cons_phi} \\
\nabla_{\mu} T^{(DM)}\,^{\mu}_{\nu} &=& C^{(DM)}_{\nu}, \label{cons_dm}
\end{eqnarray}
where the superscript $(DM)$ stands for the dark matter fluid and $(DE)$ for the dark energy.
The conservation of the total dark component energy-momentum tensor 
(we assume the separate conservation of the energy momentum of radiation and baryons),
\begin{equation}
\label{energyconservation}
\nabla_{\mu} \left[ T^{(DM)} \,^{\mu}_{\nu} + T^{(DE)} \,^{\mu}_{\nu} \right]= 0,
\end{equation}
implies that
\begin{equation}
C^{(DM)}_{\nu}=-C^{(DE)}_{\nu}.
\end{equation}

The coupling between dark matter and dark energy is determined by the function $C^{(DM)}_{\nu}$ which is usually
written as
\begin{equation}
C^{(DM)}_{\nu} = (8\pi G)^{1/2} \,\beta\rho_{DM}\nabla_{\nu} \phi,
\end{equation}
where $\beta$ is a constant that expresses the coupling strength. In this model, dark energy must be dynamical and here it is modeled by a scalar field  $\phi$, such as a quintessence field.
In this model, $\beta$ is the only new parameter in addition to the usual description of the dark energy sector.
The standard uncoupled case is recovered for $\beta=0$.

There is a vast literature studying this class of models that can modify both the evolution of the 
background cosmology as well as the evolution of perturbations. For instance, it has been recently claimed
that such a model can ease the tension in the measurements of $\sigma_8$ from CMB and galaxy surveys 
\citep{Barros:2018efl}.

LSST can separately map dark matter and dark energy at a redshift where they have roughly comparable influences on the expansion rate of the universe. 
Studies of SNe and 3$\times$2pt clustering provide two complementary methods of reconstructing dark energy in independent patches within the LSST survey footprint \citep[\eg][]{0902.2590}.
%There are two complementary methods of reconstructing dark energy on the sky: SNe and 3$\times$2pt in independent patches within the survey footprint \citep[\eg][]{0902.2590}.
%\citep[Figure 15.9 in ][]{0912.0201}.
The transition from a dark matter-dominated universe to one with late-time acceleration (dark energy) may hint at some connection between these two components. 
An angular cross correlation between maps of dark energy and tomographic weak lens maps of dark matter could yield a non-zero signal.  
If so, the ratio of the cross correlation to the auto-correlations would be a diagnostic of the underlying physics. 
In this scenario, measurements of dark energy anisotropy become a probe of the nature of dark matter; \figref{DMDEmap}, reproduced from \cite{1810.11007}, illustrates a Ginzburg-Landau phase transition model that results in correlated dark matter-dark energy anisotropy. 
Quadrupole and higher-order correlated anisotropies are generated around redshift $z=0.7$.  
This is accessible in LSST maps of dark energy and dark matter in a broad redshift shell.

\paragraph{Systematics and synergies:}
Systematics in the dark energy and dark matter maps on large angular scales must be reduced below the level of any dark matter-dark energy correlation signal.  
For example, systematics in apparent magnitude and \photoz due to uncorrected extinction from Galactic dust would be one focus. 
Encouragingly, the two measures of dark energy anisotropy depend differently on wavelength-dependent extinction. 
A useful null test will be the cross correlation between dark energy and dark matter maps with dust maps.  
This would set the floor for residual extinction systematics, forming the basis for a forward simulation of the resulting dark energy-dark matter false correlation. 
As in analysis of CMB data, cuts on Galactic latitude can reveal the level of residual systematics. 
Finally, any dependence of the cross-correlations on redshift could discriminate between models as well as detect redshift-dependent systematics.

For detection of low multipole sky correlations, observations in the north as well as the south will be useful.
There is important synergy with WFIRST and EUCLID observations in the north.  
These complementary data could be calibrated and tested by joint null tests in overlap areas with the LSST survey.

%%%%%%%%%%%%%%%%%%%%%%%%%%%%%%%%%%%%%%%%%%%%%%%%%%%%%%%%%%%%%%%%%%%%%%%%%%%%%%%%
% Complementarity
%%%%%%%%%%%%%%%%%%%%%%%%%%%%%%%%%%%%%%%%%%%%%%%%%%%%%%%%%%%%%%%%%%%%%%%%%%%%%%%%
\chapter{Complementarity with Other Programs}
\label{sec:complementarity}
\bigskip

\Contributors{Kimberly K.\ Boddy, Esra Bulbul, Johann Cohen-Tanugi, Alessandro Cuoco, William A.\ Dawson, Alex Drlica-Wagner, Cora Dvorkin, Christopher Eckner, Christopher D.\ Fassnacht, Vera Gluscevic, Shunsaku Horiuchi, Charles R.\ Keeton, Ting S.\ Li, Marc Moniez, Manuel Meyer, Lina Necib, Ethan O.\ Nadler, Jeffrey A.\ Newman, Eric Nuss, Andrew B.\ Pace, Justin I.\ Read, Joshua D.\ Simon, Erik Tollerud, David Wittman, Gabrijela Zaharijas}

LSST will uniquely complement several other experimental studies of dark matter.
Below we summarize some of these complementary probes, with a specific focus on spectroscopic observations, high-resolution imaging, indirect detection experiments, and direct detection experiments.
While LSST can substantially improve our understanding of dark matter in isolation, these experiments are essential to provide a holistic picture of dark matter physics.
This section is not intended to be comprehensive, but rather serves to demonstrate the influence that LSST will have on dark matter studies generally.

% Spectroscopy
\section{Spectroscopy \Contact{Ting}}
\label{sec:spectroscopy}
\Contributors{Joshua D.\ Simon, Ting S.\ Li, William A.\ Dawson, Denis Erkal, David Wittman, Erik Tollerud}

The power of photometric and astrometric measurements from LSST will be significantly augmented by additional spectroscopic observations.
In particular, spectroscopic follow-up studies will provide kinematic and redshift information for many of the objects studied by LSST.
Given the faintness and high density of targets that are expected from LSST, community access to multi-object spectrographs on large-aperture telescopes is essential for these studies \citep{2016arXiv161001661N}. 
Due to LSST's location in the southern hemisphere, southern spectroscopic facilities are best at maximizing observational overlap.

Many next-generation telescopes and instruments are currently under development or construction. 
These facilities can be broadly divided into two categories: wide-field, massively multiplexed spectroscopy on medium- to large-aperture telescopes ($\roughly$8--10-meter class), and giant segmented mirror telescopes (GSMTs, $\roughly30$-meter class) with relatively smaller fields of view. 
%These instruments can be broadly divided into two categories: massively multiplexed spectrographs on 8- to 10-meter telescopes, and giant segmented mirror telescopes (GSMTs, $\roughly 30$-meter class) with smaller fields of view. 
The former category includes facilities on existing and future telescopes including the Prime Focus Spectrograph (PFS) instrument on the Subaru telescope \citep{2014PASJ...66R...1T} that is currently under construction; the Maunakea Spectroscopic Explorer \citep[MSE;][]{MSEbook2018}, a planned 11.25m, wide-field, optical and near-infrared facility completely dedicated to multi-object spectroscopy; the Southern Spectroscopic Survey Instrument (SSSI), a project concept recommended for consideration by the DOE’s Cosmic Visions panel \citep{1604.07626, 1604.07821}, and a possible future ESO wide-field spectroscopic facility. 
The latter category is populated by new facilities such as the Thirty Meter Telescope \citep[TMT;][]{1505.01195}, the Giant Magellan Telescope \citep[GMT;][]{GMT:2018}, and the Extremely Large Telescope \citep[ELT;][]{EELT:2009}.

In this section, we illustrate several examples of how complementary spectroscopy will improve the measurement of dark matter properties with LSST.

\subsection{Milky Way Satellite Galaxies \Contact{Josh}}\label{sec:MW_sats_spec}
%\Contributors{Josh, Ting, Erik, ...}
In \secref{smallest_galaxies} we discussed the derivation of an upper limit on the minimum dark matter halo mass based on the observed luminosity function of satellites discovered by LSST. 
%\ET{Should we mention the fact that confirmation of the LSST discoveries *also* may require spectroscopy at least for some of the ambiguous cases?}  
An alternative approach is to obtain spectroscopy of individual stars in each satellite to measure its velocity dispersion, from which the central mass and density can be inferred.  Then one can compare either the densities or the circular velocity function directly with theoretical predictions without assumptions about the subhalo mass function or the stellar mass-halo mass relation.

\begin{figure}
  \centering
  \includegraphics[width=0.49\textwidth]{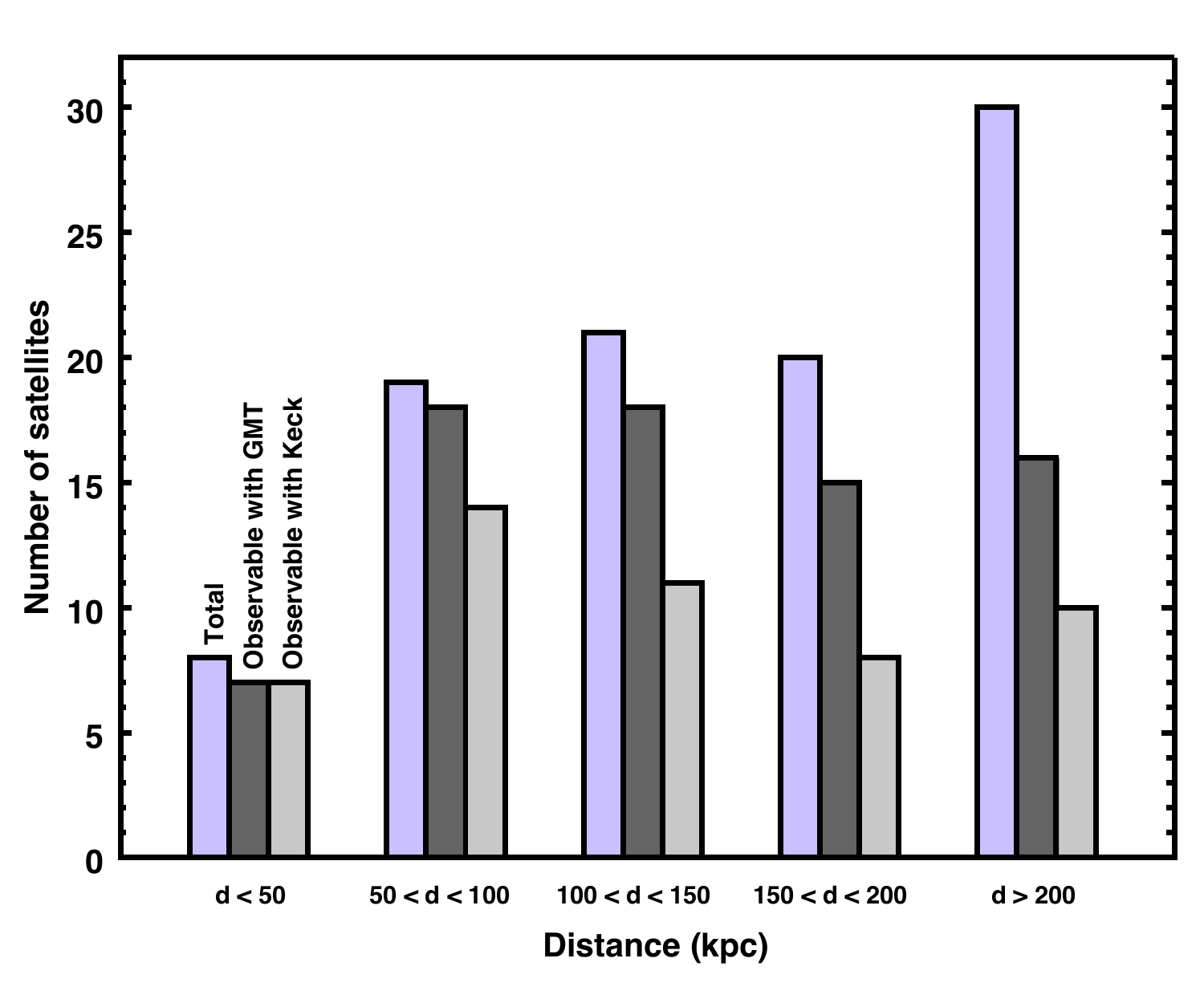}
  \includegraphics[width=0.50\textwidth]{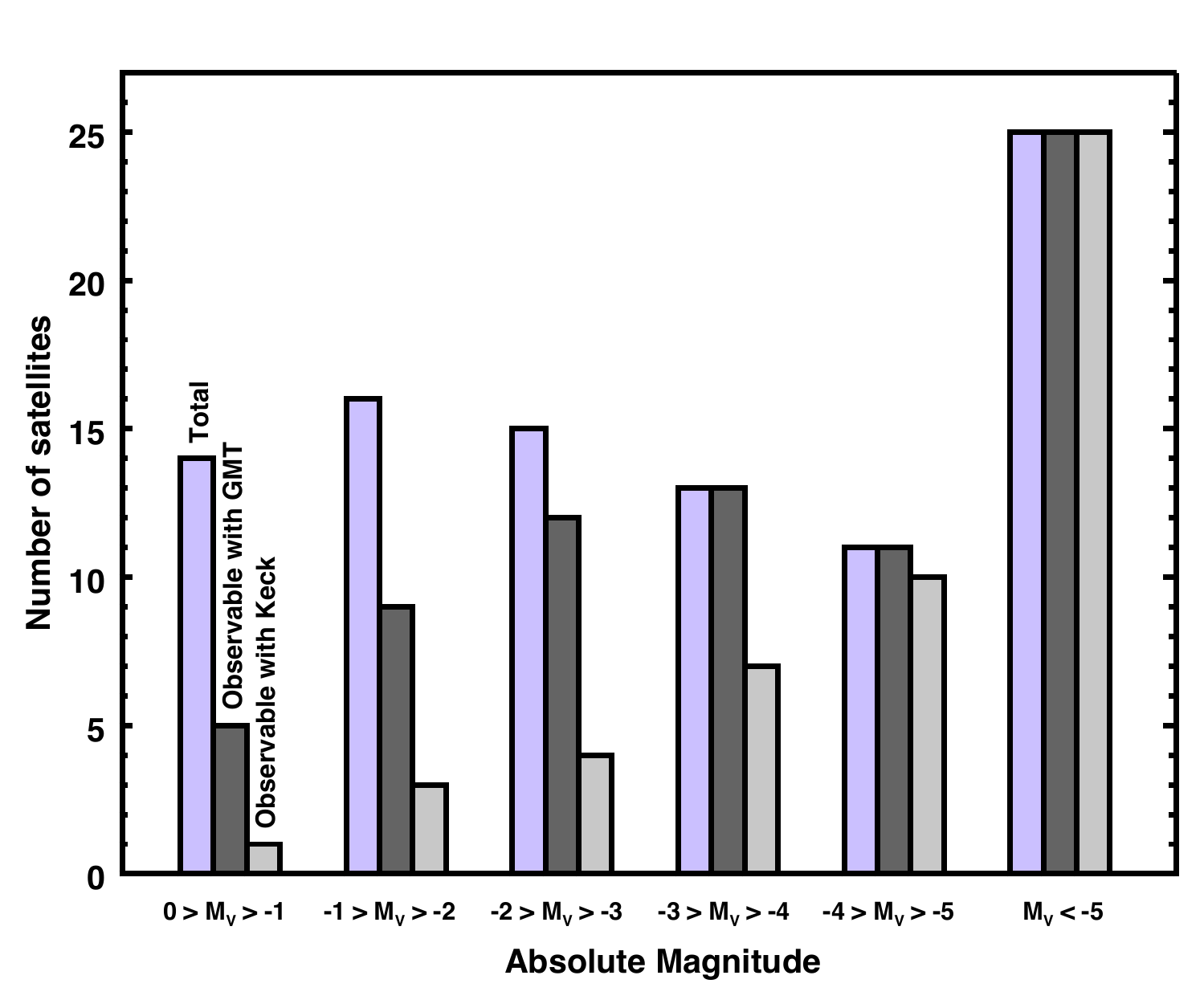}
  \caption{Possibility of spectroscopic follow-up for the LSST satellite population as a function of distance (left) and magnitude (right). Current telescopes will be able to measure velocity dispersions for $\roughly50\%$ of the expected satellites, while a GSMT can measure velocity dispersions for $\roughly80\%$.}
  \label{fig:specfollowup_distance}
\end{figure}

Spectroscopy of individual stars in the faint Milky Way satellites that will be identified with LSST will require deep observations with multiplexed spectrographs on large telescopes.  Measurements of the stellar velocity dispersions of these systems can be obtained either with 8--10-meter-class telescopes or with the next generation of 25--30-meter-class telescopes.  To assess the feasibility of spectroscopy of LSST satellites, we create a mock sample of dwarfs based on projections of the sensitivity of LSST satellite searches. We estimate the limiting magnitude set by targeting the brightest 20 stars in each simulated dwarf and apply exposure time calculators for Keck/DEIMOS and GMT/GMACS to determine the integration time needed for each satellite, with the additional constraint that no object would be observed for more than 30 hours ($\sim3$~nights).
As illustrated in \figref{specfollowup_distance}, spectroscopy of a nearly complete sample of satellites can be pushed $\roughly 2$~mag fainter in luminosity and a factor of $\roughly 2$ farther in distance with plausible investments of observing time on a GSMT than with existing facilities.

In addition to inferring the minimum dark matter halo mass, kinematics from stellar spectroscopy can also reveal the inner densities of the lowest-luminosity dwarf galaxies, for which baryonic effects are minimal and dark matter physics can be separated from the astrophysics of galaxy formation \citep{governato2012,read2017}. \AHGP{Problem: Justin read shows that density profile measurement requires about a thousand stars.  It's (relatively) easy to get a central density, hard to do a profile.} A direct measurement of the inner density in these dwarf galaxies will allow us to distinguish between collisionless CDM,  which predicts a cuspy NFW profile, and SIDM, which predicts lower central densities \citep[][though see \citealt{Nishikawa:2019lsc}]{2012MNRAS.423.3740V,Rocha:2012jg}. Moreover, the stellar kinematics will also reveal the integral of the dark matter density profile in dwarf galaxies (i.e., the J-factor), which is an essential input for constraints on the dark matter self-annihilation cross section for indirect dark matter searches in X-ray and $\gamma$-ray experiments \citep[\eg,][]{1108.3546}.

\subsection{Stellar Streams \Contact{Ting}}
%\Contributors{Ting, Denis ...}

As discussed in \secref{stream_gaps}, subhalo encounters with cold stellar streams will induce density perturbations that will be detectable by LSST, constraining the minimum dark matter halo mass and the mass function of dark matter halos from $\roughly10^5$--$10^9 \Msun$. In addition, these flybys cause velocity perturbations that correlate with the density variations.  The velocity signal near stream gaps can be measured either via line-of-sight velocity measurements from spectroscopy or tangential velocity measurements from astrometry, improving the precision with which the perturber mass can be determined.
The velocity variation (peak to peak) from these flybys will be small. To estimate the amplitude of the perturbation, we consider a stream orbiting the Milky Way at a distance of 14 kpc and compute the typical maximum velocity kick expected over its lifetime of 5 Gyr using the formalism from \citet{erkal2016}.  The velocity change is $\roughly \{0.6, 0.3, 0.1\} \kms$ for subhalos in the range $\{ 10^{7.5}, 10^{6.5}, 10^{5.5} \} \Msun$, respectively. %$\{ 10^7-10^8, 10^6-10^7, 10^5-10^6\} \Msun$

Due to the low density of the stream stars, a massively multiplexed, wide field-of-view spectroscopic facility such as PFS on Subaru or MSE is needed. 
Furthermore, given the expected small velocity kick amplitude, the velocity for each star determined from the spectroscopic observation need to be extremely precise.
%Furthermore, given the expected small velocity kick amplitude, the velocity accuracy for each star determined from the spectroscopic observations should be at or better than $1 \kms$ to unambiguously detect the signal with an ensemble of stream stars. 
Considering that the typical gap size for a $10^6 M_\odot$ subhalo is a few degrees, and the typical known streams has 10--100 stars/degree at $r\lesssim23$ \citep[\eg,][]{erkal2017, DeBoer:2018}, a velocity precision of $\sim1\,\mathrm{km\,s}^{-1}$ per star would allow an unambiguously detection on the signal with an ensemble of stream stars.

\subsection{Galaxy Clusters \Contact{Will}}
%\Contributors{Will, ...}

%As noted in \secref{merging_clusters}, one of largest systematics associated with merging galaxy cluster constraints on SIDM is modeling the merger. The more complex the merger, the more severe the systematics will be.
%The best means of constraining merging galaxy cluster substructure is with spectroscopic measurement of as many 
%galaxy cluster member galaxies as possible \citep[\eg,][]{Golovich:2018}.
%As noted in \citet{2016arXiv161001661N}, perhaps the best spectroscopic follow-up facilities are large telescopes with slitmask-like multi-object spectrographs, or fiber-based multiplex spectrometers with low ($\mathcal{O}(\unit{arcsec})$) fiber collision regions, due to the density of cluster members.

As noted in \secref{merging_clusters} one of largest systematics associated with merging galaxy cluster constraints on SIDM is modeling the merger dynamics, most importantly the relative speed at the time of
pericenter. This can be inferred from the observed line-of-sight speed by correcting for the viewing angle and evolving the equations of motion back to the time of pericenter. Therefore, a minimal requirement is a good measurement of the systemic line-of-sight speed of each subcluster, which implies spectroscopic redshifts for hundreds of members of each subcluster \citep[\eg,][]{Golovich:2018}. In addition, a thorough spectroscopic survey of the surrounding $\sim$10 arcmin region is required to test for the presence of additional substructures that could affect reconstruction of the merger scenario. 10-m class collecting area is sufficient for most known mergers, but because member galaxies can be packed quite closely in their respective subclusters, fiber collision avoidance is a limiting factor for many instruments designed for wide-field highly multiplexed spectroscopy.  As noted in \citet{2016arXiv161001661N}, perhaps the best spectroscopic follow-up facilities are large telescopes with slitmask-like multi-object spectrographs, or fiber-based multiplex spectrographs with low ($\mathcal{O}(\unit{arcsec})$) fiber collision regions.

% High-resolution imaging
\section{High-Resolution Imaging}
\label{sec:highres}

\Contributors{William A.\ Dawson, Christopher D.\ Fassnacht, Charles R.\ Keeton, David Wittman, Marc Moniez}

Since the LSST point spread function (PSF) is limited to a median angular resolution of $\roughly 0.7\arcsec$ by a combination of instrumental and  atmospheric effects \citep{0805.2366}, there are many dark matter science cases where higher resolution imaging from space or ground-based adaptive optics (AO) facilities, which can reach $\roughly 0.01\arcsec$ in some cases, can be highly complementary. We briefly summarize some of these cases in this subsection and relate them to the dark matter science capabilities of LSST.
\WAD{This section currently focuses on high resolution optical imaging, however it is worth considering other wavelengths, especially radio.}

% Astrometric microlensing of compact dark matter
\subsection{Astrometric Microlensing}
%\subsection{Astrometric Microlensing of Compact Dark Matter \Contact{Will}}
\label{sec:astrometric_microlens}
%\Contributors{Will, ...}

Related to photometric microlensing (\secref{microlensing}), astrometric microlensing relies on the fact that the two images generated during a compact object lensing event will be of differing brightness, and the brightness ratio of these two images will vary throughout the duration of the lensing event.
The two images will be of most similar brightness when the projected lens-source separation is at its minimum.
By precisely measuring the astrometry of these blended images as a function of time and combining with the LSST photometric microlensing measurement one can break the lens mass-distance degeneracy and precisely measure the mass and location of individual black holes \citep{2015ApJ...814L..11Y}.
Astrometric microlensing will require precise astrometry from ground-based optical/NIR systems, space-based observatories, or longer microwave or radio wavelength observations.
Astrometric microlensing may even give access to less compact dark matter substructures \citep{1804.01991}.

In addition, high cadence, around the clock monitoring of microlensing events will be possible from organized teams of small- to medium-sized telescopes (\eg, MiNDSTEp\footnote{http://www.mindstep-science.org/}, RoboNet\footnote{https://robonet.lco.global/}, MicroFUN\footnote{http://www.astronomy.ohio-state.edu/~microfun/}, and PLANET\footnote{http://planet.iap.fr/}). 
These follow-up observations will be sensitive to distorsions with respect to the point-source, point-lens rectilinear approximation.
It will be critical that these collaborations receive automated alerts when LSST detects probable microlensing, or generically non-standard, transient events. 
Such efforts are planned as part of the LSST Alert Stream \citep{0805.2366}.

%These teams are currently fed by the MOA and OGLE-IV alert systems, and could be in perfect synergy with the LSST microlensing detections, providing extra information on individual events that cannot be extracted from the LSST data alone. 
%Tsapras, Y. et al., 2009, AN, 330, 4T: RoboNet-II: Follow-up observations of microlensing events with a robotic network of telescopes
%http://www.planet-legacy.org/

% Strong-microlensing
\subsection{Strong Microlensing \Contact{Will Dawson}}
\begin{figure}
\label{fig:strong_microlensing}
\centering
\includegraphics[width=0.5\columnwidth]{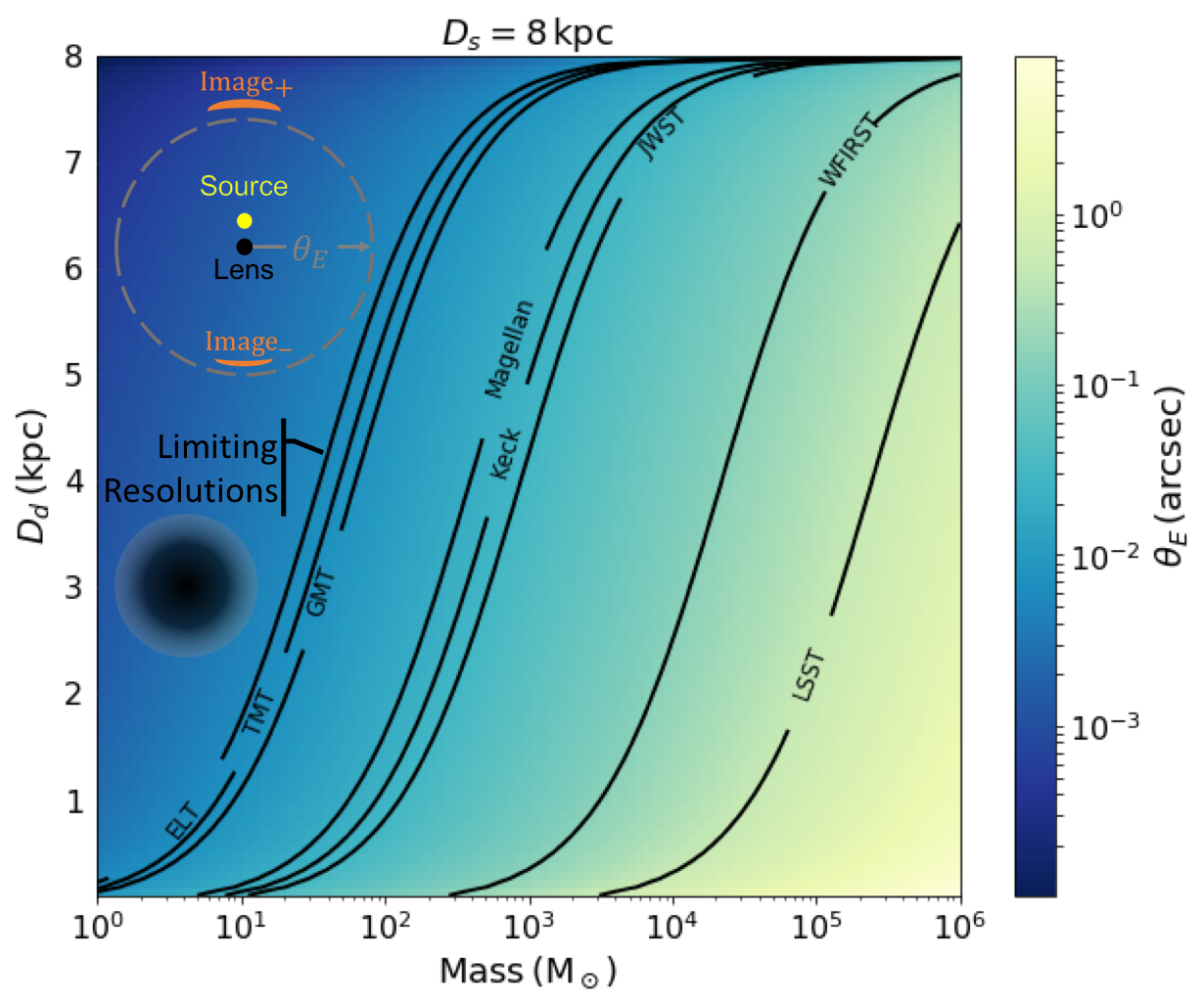}
\caption{The Einstein radius (i.e., 1/2 the separation of the multiple images) for microlensing lensing events as a function of compact object mass and distance (assuming a source distance of $8\kpc$). Parameter space below a black curve indicates that the multiple lensed source images will be resolvable by that telescope.}
\end{figure}

%\subsection{Strong-Microlensing of Compact Dark Matter \Contact{Will Dawson}}
Strong microlensing is related to astrometric microlensing.
The Einstein radius of a given lens, which is approximately the separation of the multiple images in a compact object lensing scenario, scales as $\sqrt{M_\mathrm{lens}}$.
In the intermediate mass black hole range, the separation of the two images approaches that of the resolution of various optical ground and spaced-based telescopes (\figref{strong_microlensing}).
If the multiple images can be resolved and their flux ratio measured, precise measurements of the mass and distance of the lens are achievable.

% Merging Galaxy Clusters
\subsection{Merging Galaxy Clusters and Cluster Subhalos}
%\Contributors{Will D., Dave W., ...}

Most dark matter constraints from merging galaxy clusters (\secref{halo_profile_clusters}) and cluster subhalos (\secref{halo_profile_clusters}) rely on accurately measuring the distribution of dark matter in (sub)clusters via gravitational lensing.
Strong and weak gravitational lensing both benefit from high-resolution imaging.
For strong lensing, high-resolution imaging enables better detection and characterization of strongly lensed background images in the dense cluster environment.
Similarly high-resolution imaging provides $\roughly4$ times more lensed source galaxies per unit area than ground-based imaging at similar depths, which enable higher-resolution weak gravitational lensing.
Historically, the Hubble Space Telescope (HST) has provided this higher-resolution imaging, although space-based telescopes such as JWST, Euclid, and WFIRST may take on most of this burden in the era of LSST.

% Strong gravitational lensing
\subsection{Strong Gravitational Lensing}
\label{sec:SLcomplement}
%\Contributors{Chris F.} 

All three approaches to use strong gravitational lens systems to make inferences on the nature of dark matter  described in \secref{stronglens} utilize LSST as a lens-finding facility.
Once the lenses are found, the dark matter science requires follow-up observations with other facilities.
The flux-ratio anomaly approach requires imaging that spatially resolves the lensed images from each other at a wavelength at which microlensing does not affect the image fluxes.
These observations can be in optical/near-IR wavelengths, utilizing IFU spectrographs behind the adaptive optics systems on ELTs to isolate the emission from the narrow-line regions of the lensed AGN, at mid-IR wavelengths with JWST, or at radio wavelengths for the subset of LSST lenses that are radio-loud.
The gravitational imaging and power-spectrum approaches both require milliarcsecond-scale angular resolution imaging for best results.
These observations require either ELT adaptive optics imaging or VLBI radio imaging of the targets.
ALMA can also be used in its most extended configuration, although this will not achieve as high a resolution as the ELTs and VLBI in most cases.
SKA is expected to greatly increase the number of radio-detected strong lenses by several orders of magnitude \citep[\eg][]{2015aska.confE..84M}.

% Indirect Detection
\section{Indirect Detection }
\Contributors{Esra Bulbul, Johann Cohen-Tanugi, Alessandro Cuoco, Alex Drlica-Wagner, Christopher Eckner, Shunsaku Horiuchi, Manuel Meyer, Andrew B.\ Pace, Ethan O.\ Nadler, Eric Nuss, Gabrijela Zaharijas}

In regions of high dark matter density, dark matter particles could continue to annihilate or decay through the same process that set their relic abundance.
Of specific interest are energetic photons (i.e., X-rays and $\gamma$-rays), since photons are generically produced by the annihilation/decay of dark matter in many models (either directly or as secondary products following the production of quarks or leptons). In addition, astrophysical phenomena in extreme environments could lead to conversion between Standard Model particles and the dark sector (e.g., ALPs), which could be observable through the emission of energetic photons or via alterations in astrophysical spectra.
By precisely mapping the distribution of dark matter and tracking extreme events (\eg, CCSNe) LSST will enable more sensitive searches for energetic particles originating from the dark sector.

Conventional indirect detection searches focus predominantly on WIMPs with masses between several \GeV and tens of \TeV. 
The annihilation or decay of these particles could produce energetic Standard Model particles detectable by current or future experiments.
The most sensitive and robust indirect searches for dark matter rely on a precise determination of the distribution of dark matter in the universe.
The integrated flux of energetic Standard Model particles, $\phi_s$ (${\rm particles} \cm^{-2} \second^{-1}$), expected from dark matter annihilation in a density distribution, $\rho(\vect{r})$, is given by

\begin{equation}
   \phi_s(\Delta\Omega) =
    \underbrace{ \frac{1}{4\pi} \frac{\Gamma}{m_{\chi}^{a}}\int^{E_{\max}}_{E_{\min}}\frac{\text{d}N}{\text{d}E}\text{d}E}_{\rm particle\ physics}
    \cdot
    \underbrace{\vphantom{\int_{E_{\min}}} \int_{\Delta\Omega}\int_{\rm l.o.s.}\rho^{a}(\vect{r})\text{d}l\text{d}\Omega '}_{\rm astrophysics}\,.
    \label{eqn:indirect}
\end{equation}
%\Big\{\Big\}
\noindent Here, the ``particle physics'' term is strictly dependent on the dark matter particle physics properties---i.e., the particle mass, $m_\chi$,  the interaction rate, $\Gamma$, and the differential particle yield per interaction, $\text{d}N/\text{d}E$, integrated over the experimental energy range.
The second term, denoted ``astrophysics,'' represents the line-of-sight integral through the dark matter distribution integrated over a solid angle $\Delta\Omega$. 
For cases of dark matter annihilation, the interaction rate is set by the thermally averaged self-annihilation cross section, $\Gamma = \sigmav/2$, and the astrophysical integral is performed over the square of the dark matter density ($a=2$). 
The resulting astrophysical term is referred to as the ``\Jfactor'' \citep[\eg,][]{1998APh.....9..137B}. 
In cases of dark matter decay, the interaction rate is inversely proportional to the lifetime of the dark matter particle, $\Gamma = 1/\tau$, and the integral is performed over the dark matter density, $a=1$. 
The resulting term is known as the ``\Dfactor'' \citep[\eg][]{1408.0002}.
Qualitatively, the astrophysics term encapsulates the spatial distribution of the dark matter signal, while the particle physics term sets its spectral character. 
LSST will improve the sensitivity to dark matter particle physics by improving our understanding of the astrophysics term.
While these improvements will influence a wide range of indirect detection experiments, in this section we focus predominantly on $\gamma$-ray measurements.

\subsection{Milky Way Satellite Galaxies \Contact{Andrew}}
\label{sec:dwarfs_id}
\Contributors{Andrew B.\ Pace, Alex Drlica-Wagner, Ethan O.\ Nadler, Manuel Meyer, Christopher Eckner, Gabrijela Zaharijas}

\begin{figure}[t]
\centering
\includegraphics[width=0.75\columnwidth]{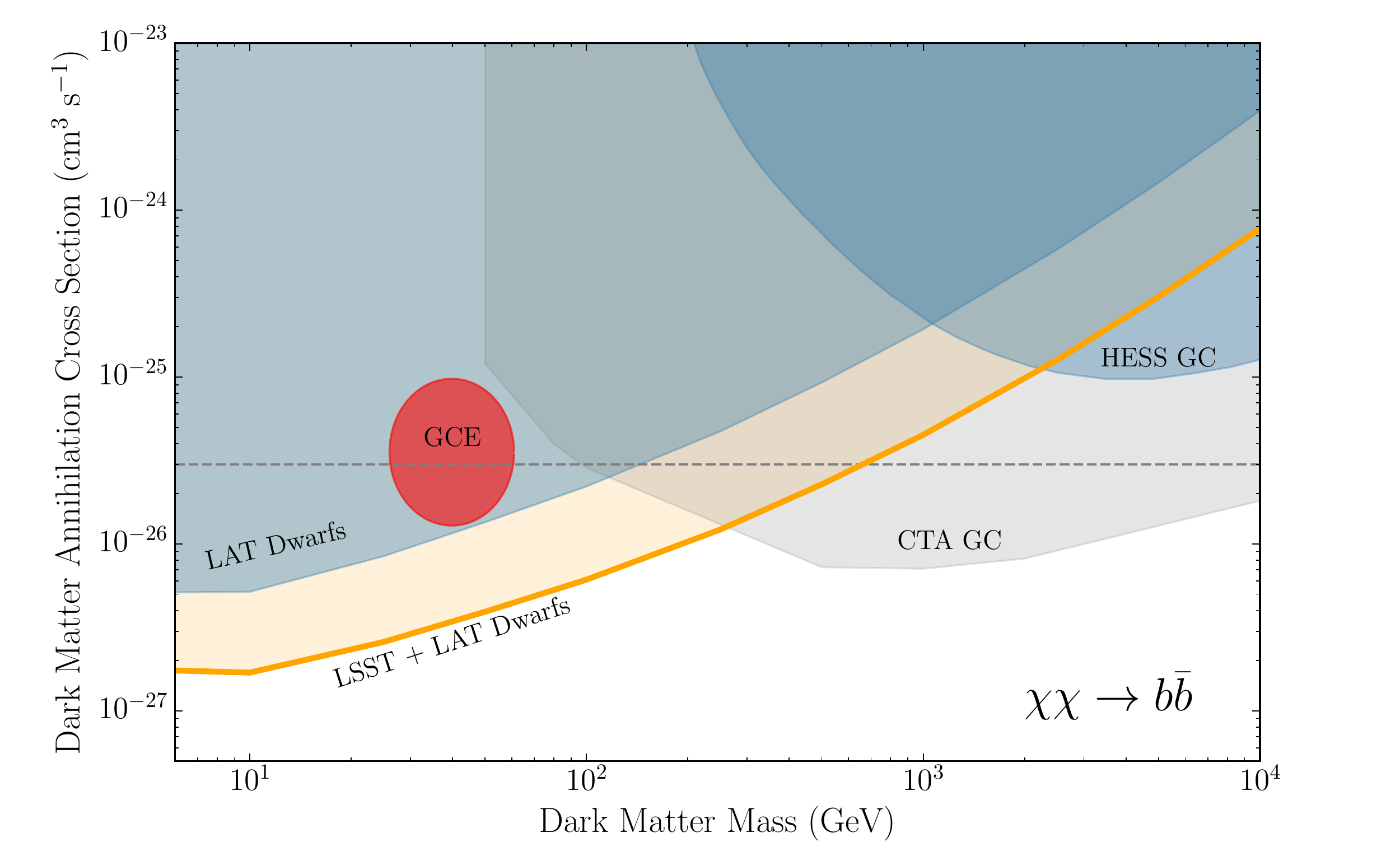}
\caption{Constraints on dark matter annihilation to $b\bar{b}$ from {\it Fermi-LAT} observations of Milky Way satellite galaxies \citep[LAT Dwarfs;][]{Ackermann:2015} and HESS observations of the Galactic Center \citep[HESS GC;][]{1607.08142}. 
A bracketing range of dark matter interpretations to the  Fermi-LAT Galactic Center Excess is shown in red \citep[GCE;][]{1402.6703, Gordon:2013, Abazajian:2014}.
Projected sensitivity to dark matter annihilation combining LSST discoveries of new Milky Way satellites, improved spectroscopy of these galaxies, and continued Fermi-LAT observations is shown in gold. This projection assumes 18 years of Fermi-LAT data, a factor of 3 increase in the integrated J-factor, and a factor of 2 improvement from improved spectroscopy. 
Preliminary projected sensitivity for CTA observations of the GC (500h; no systematics) are shown in gray \citep[CTA GC;][]{Eckner:2018}.
\label{fig:indirect}
}
\end{figure}

Gamma-ray observations of Milky Way satellite galaxies currently provide the most robust and sensitive constraints on the dark matter self-annihilation cross section for GeV- to TeV-mass particles \citep[\eg][]{Ackermann:2014, Geringer-Sameth:2015, Ackermann:2015, 1812.06986}.
The sensitivity of these searches will improve by combining new Milky Way satellite galaxies discovered by LSST, more precise \Jfactor measurements from novel spectroscopic observations, and additional Fermi-LAT data. 
We estimate each of these contributions to predict the improved sensitivity of dark matter annihilation searches in dwarf galaxies in the era of LSST.

To estimate the improvement in the integrated \Jfactor of the Milky Way satellite galaxy population, we combine cosmological zoom-in simulations of Milky Way dark matter substructure with a semi-analytic model to convert subhalo density profiles to \Jfactor estimates (this approach is is similar to that of \citealt{1309.4780}). 
Our simulation-based model accounts for modulations to dark matter-only subhalo populations due to baryonic physics, and we marginalize over the dependence of subhalo populations on host halo properties by sampling subhalo populations from a large number of hosts \citep{Nadler:2018}. 
To obtain an estimate for the increase in the integrated \Jfactor, we select a host halo with the largest number of nearby subhalos, consistent with recent observations of an overabundance of nearby satellites associated with the Milky Way \citep{Kim:2017iwr, Graus:2018}. 
We exclude subhalos with heliocentric distances $< 20 \kpc$ to avoid anomalously large projections due to a single nearby satellite.
We follow the analytic formalism presented by \citet{1604.05599} and \citet{1802.06811} to convert the dark matter profiles of our simulated subhalos to \Jfactors.  
This approach estimates the \Jfactor of each subhalo based on $r_{\max}$, $V_{\max}$, and heliocentric distance. 
%We calculate the cumulative $J$-factor within 100 kpc we find for the most optimistic case 
%(i.e., selecting the host halo with the highest cumulative $J$-factor and ignoring the effects of subhalo disruption) 
We find that the cumulative \Jfactor within 100\kpc may increase by as much as a factor of 3 relative to the known dSphs with measured \Jfactors. 

Recent studies have suggested that an additional factor of 2 improvement in sensitivity may be possible through better spectroscopic measurements  of the stars in known satellite galaxies \citep{Albert:2017}, and we include this factor in our projections.
In addition, current constraints from the Fermi-LAT Collaboration used 6 years of data \citep{Ackermann:2015}; however, the Fermi-LAT has collected more than 10 years of data and could continue to collect data for another 10+ years.
These additional data will improve the statistical sensitivity of the $\gamma$-ray search most drastically for large dark matter particle masses ($>500\GeV$).
We quantitatively evaluate the improvement from continued Fermi-LAT data taking using the results of \cite{Charles:2016}.
We combine the predicted improvements from new dwarfs, better determined \Jfactors, and more Fermi-LAT data into a projected sensitivity for future searches for dark matter annihilation in dwarf galaxies in \figref{indirect}.

\subsection{Cross-Correlation with Gamma Rays \Contact{Horiuchi,Alessandro C}}
\label{sec:igrb_id}
\Contributors{Shunsaku Horiuchi, Alessandro Cuoco}

Wide-area weak-lensing measurements from LSST will help extract  potential dark matter contributions to the isotropic gamma-ray background \citep[IGRB;][]{1410.3696}. The IGRB is defined as the residual all-sky $\gamma$-ray emission after subtracting individually detected sources and the Galactic diffuse emission, and provides the distance frontier of indirect dark matter searches with $\gamma$-rays. Contributions to the IGRB include unresolved sources that are individually too faint to be detected---e.g., blazars \citep{1110.3787,1310.0006}, star-forming galaxies \citep{1206.1346}, and misaligned AGNs \citep{1304.0908}---as well as a potential contribution from dark matter annihilation \citep{1312.0608,1501.05464,1501.05301,1608.07289}. Analyses of the IGRB intensity spectrum, auto-correlation angular power spectrum, and photon count statistics show that a linear combination of astrophysical sources can explain the observed IGRB, but the uncertainties are still large \citep[e.g.,][]{1502.02866}.

LSST will prove invaluable by mapping the distribution of matter on large scales via measurements of galaxy clustering and of cosmic shear from weak gravitational lensing. 
Since  cosmological $\gamma$-ray emission from dark matter annihilation follows the same underlying dark matter distribution traced by cosmic shear and galaxies, cross correlating them yields novel information on the composition of the IGRB \citep{1212.5018,1312.4403,1411.4651,1506.01030,Lisanti:2018}. 
Compared to the IGRB intensity or auto-correlation, the cross correlation will yield more than $\roughly 10$ times higher sensitivity to dark matter \citep{1411.4651,1503.05922}.
Cross-correlations with galaxy catalogs have been derived in \cite{1103.4861,1503.05918,1709.01940} up to $z \sim 0.6$, 
which is the largest redshift where current catalogs have enough sky-coverage and galaxy density to robustly
detect the correlation.  On the other hand, the IGRB is expected to extend up to $z \sim 2$--$3$ \citep{1502.02866}. 
LSST, with its large sky-coverage and galaxy density and broad redshift range, thus fills this gap to map the IGRB-LSS cross-correlation up to high redshift. 
A complete mapping of the IGRB up to $z \sim 3$ will constitute a crucial tool to robustly separate the different 
astrophysical contributions, as well as to isolate the DM annihilation signal, breaking the degeneracies which
are present when only low-redshift results are used \citep{1506.01030}.    

In contrast to galaxy catalogs, cosmic shear has the advantage of being an unbiased tracer of the dark matter distribution, which mitigates many of the systematics from using galaxies to trace dark matter---i.e., assumptions about the relationship between galaxy luminosity and halo mass, reliance on assumptions of hydrostatic equilibrium, and strong correlations with astrophysical $\gamma$-ray emission. At present, weak lensing surveys of several hundred square degrees allow studies of the IGRB to probe slightly above the thermal annihilation cross section \citep{1404.5503,1607.02187,1611.03554}. A simple forecast for LSST can be made by scaling the covariance matrix of the correlation estimator by the sky coverage. This shows that a combination of LSST lensing maps and all-sky Fermi-LAT data will reach a sensitivity where it is possible to \textit{detect} at $3\sigma$ WIMP annihilation to $b\bar{b}$ at the thermal cross cross section for dark matter particle masses up to 100 GeV \citep{1404.5503}.

From the $\gamma$-ray side, improvements in the mapping of the cross-correlation with galaxies and cosmic shear are expected with the next generation of $\gamma$-ray instruments: AMEGO\footnote{\url{https://asd.gsfc.nasa.gov/amego/index.html}} and eASTROGAM\footnote{\url{http://eastrogam.iaps.inaf.it/}} \citep{1711.01265}.
At the present, the main limitation in detecting the cross-correlation at GeV and sub-GeV energies is instrumental angular resolution.  
eASTROGAM will have an angular resolution 5--6 times better than the Fermi-LAT in the energy range from $100 \MeV$ to $1 \GeV$~\citep{1711.01265}. This will translate in harmonic space into a multipole reach 5-6 times larger than presently achievable, leading to stronger constraints from cross correlations.
Precise measurements of the cross-correlation at sub-GeV energies will further improve the ability to separate the astrophysical IGRB sources from the DM signal, increasing the sensitivity to the latter. 

\subsection{Axion-like Particles from Supernovae \Contact{Manuel}}
\label{sec:alp_id}
\Contributors{Manuel Meyer}

Axion-like particles (ALPs) might be produced during CCSNe explosions through the conversion of thermal photons in the electro-static fields of protons and ions, \ie, through the Primakoff effect \citep{1996slfp.book.....R}.  
Similar to neutrinos, ALPs would quickly escape the core and, if they are sufficiently light ($m_\phi \lesssim 10^{-9} \eV$), they could convert into $\gamma$-rays in the magnetic field of the Milky Way and/or the host galaxy of the CCSN. 
The resulting $\gamma$-rays would arrive in temporal coincidence with the neutrinos in a burst lasting tens of seconds with a 
thermal spectrum peaking at 60\MeV, depending on the mass of the progenitor \citep{2015JCAP...02..006P}.
The non-observation of a $\gamma$-ray burst from SN1987A, which occurred in the LMC, has been used to derive stringent constraints on the photon-ALP coupling $g_{\phi\gamma}<5.3\times10^{-12}\GeV^{-1}$ for $m_\phi < 4.4\times10^{-10}\eV$ \citep{1996PhLB..383..439B, 1996PhRvL..77.2372G,2015JCAP...02..006P}.
In the case of a CCSN within the Milky Way, the \textit{Fermi} LAT could improve these limits by more than an order of magnitude \citep{2017PhRvL.118a1103M}. 
However, with a Galactic supernova rate of $\roughly 3$ per century \citep[\eg,][]{2013ApJ...778..164A}, and the LAT field of view of 20\% of the sky, the chance to observe at least one such event in the next five years is $\sim 0.03 \times 0.2 \times 5 = 0.03$ (assuming that the occurrence of SNe is a Poisson process). This estimate is still optimistic since the supernova rate is calculated for the entire Galaxy, which is not inside the field of view at any given moment.\footnote{If the supernova is sufficiently nearby or the photon-ALP coupling is close to current limits, a signal could be detected with the BGO detectors (senisitive up to 40\MeV) of the \emph{Fermi} Gamma-ray Burst Monitor, which observes the entire unocculted sky.}
Increasing the search volume to extragalactic SNe is the obvious way to overcome this low rate. 
However, for CCSNe beyond the LMC and SMC, current-generation neutrino detectors lack the sensitivity to detect a signal \citep[\eg,][]{2011PhRvD..83l3008K}, and hence no precise time stamp will be provided for ALP-induced $\gamma$-ray emission; however, well-sampled optical light curves can be used to estimate the explosion time on the time scale of hours \citep{2010APh....33...19C}. 
LSST will detect a plethora of CCSNe light curves \citep{Lien:2009}. 
Estimates for the delay between the core collapse and the shock breakout range from minutes for massive Wolf-Rayet stars (type Ib/c) to days for red supergiants (type II) \citep{2013ApJ...778...81K}. 
Thus, SN type Ib/c caught early after their shock breakout and with subsequently well-sampled light curves are a prime target for the search of an ALP-induced $\gamma$-ray burst. 

Since the $\gamma$-ray flux scales as $g_{\phi\gamma}^4 / d^2$, where $d$ is the luminosity distance, the sensitivity for $g_{\phi\gamma}$ scales as $\sqrt{d}$. 
Limits of the order of $g_{\phi\gamma} \lesssim 2\times10^{-12}\,\mathrm{GeV}^{-1}$ should be possible for a single supernova in M31 ($d=778\kpc$) \citep{2017PhRvL.118a1103M}. 
If one allows these limits to degrade by a factor of 10, constraints better than those from CAST \citep{Anastassopoulos:2017} should still be possible for $d\lesssim 80\Mpc$ ($z \lesssim 0.02$) for a single SN assuming that the time of the core collapse is precisely known. 
LSST is expected to detect tens of type Ib/c CCSNe each year with redshifts $z \lesssim 0.02$ \citep{Goldstein:2018} and could conduct such searches in conjunction with the \textit{Fermi} satellite or future $\gamma$-ray satellites like AMEGO, eASTROGAM, or Gamma-400 \citep[\eg,][]{2017ICRC...35..910C,1502.02976} 
A stacking analysis of the $\gamma$-ray data with explosions times estimated from LSST light curves provides the exciting possibility to probe photon-ALP couplings in the regime where ALPs make up the entirety of the dark matter.

% Direct Detection
\section{Direct Detection }
\label{sec:direct}
\Contributors{Kimberly K.\ Boddy, Alex Drlica-Wagner, Cora Dvorkin, Vera Gluscevic, Ethan O.\ Nadler, Lina Necib, Justin I.\ Read}

Direct detection experiments seek to directly detect interactions between particles in the Galactic dark matter halo and an experimental apparatus. In the case of WIMP dark matter, these experiments are sensitive to the scattering between dark matter particles and atomic nuclei \citep[\eg,][]{1509.08767}.
Interpreting the results of direct detection experiments in the context of dark matter necessarily relies on astrophysical measurements of the distribution of dark matter.
In addition, limitations from the energy thresholds and shielding of direct detection experiments can limit the ranges of dark matter particle masses and cross sections that can be probed.
In this section, we describe how LSST will complement direct detection experiments by improving measurements of the local phase-space density of dark matter, and how cosmological measurements with LSST can help probe dark matter  masses and cross sections outside the range accessible to direct detection experiments.

\subsection{Local Dark Matter Distribution \Contact{Lina}}
\Contributors{Justin I.\ Read, Lina Necib}

\def\rhodm{\rho_\mathrm{\chi}}
\def\rhodmlab{\tilde{\rho}_\mathrm{\chi}}
\def\rhodmext{\rho_\mathrm{\chi,ext}}

The signal strength of dark matter scattering in direct detection experiments depends on the local dark matter density and its velocity distribution. 
For spin-independent scattering of dark matter particles off atomic nuclei, the recoil rate (per unit mass, nuclear recoil energy $E$, and time) in such experiments is given by \citep[\eg,][]{1996APh.....6...87L}: 

\begin{equation}
\frac{dR}{dE} = \frac{\rhodmlab \sigma_\chi |F(E)|^2}{2 m_\chi \mu^2} \int_{v>\sqrt{m_N E/2\mu^2}}^{v_\mathrm{max}} \frac{f({\bf v},t)}{v}d^3 {\bf v} 
\label{eqn:recoilrate} 
\end{equation} 
where $\sigma_\chi$ and $m_\chi$ are the interaction cross section and mass of the dark matter particle, $|F(E)|$ is a nuclear form factor that depends on the choice of detector material, $m_N$ is the mass of the target nucleus, $\mu$ is the reduced mass of the dark matter-nucleus system, $v = |{\bf v}|$ is the speed of the dark matter particles, $f({\bf v},t)$ is the velocity distribution function of the dark matter particles, $v_\mathrm{max} = 533^{+54}_{-41}\kms$ (at 90\% confidence) is the Galactic escape speed \citep{2014AA...562A..91P}; and $\rhodmlab$ is the dark matter density within the detector. 
 
From equation \ref{eqn:recoilrate}, we can see that $\rhodmlab$ is trivially degenerate with the properties of the dark matter particle, $\sigma_\chi/m_\chi$. For this reason, significant effort has gone into estimating the amount of dark matter within a few hundred parsecs of the Sun, $\rhodm$, from which we can extrapolate $\rhodmlab$ \citep[see][for a review]{1404.1938}. The latest values favor $\rhodm \sim 0.5\GeV \cm^{-3}$, with an uncertainty of order $20-30$\% \citep[\eg,][]{2014A&A...571A..92B,2018MNRAS.478.1677S}. With the advent of unprecedented data from the \Gaia satellite, the systematic and random errors on $\rhodm$ will continue to fall \citep{1404.1938}. However, equally important in equation \ref{eqn:recoilrate} is the velocity distribution function of dark matter, $f({\bf v},t)$, which is much more challenging to measure.\footnote{The time dependence of $f({\bf v},t)$ owes primarily to the motion of the Earth around the Sun and is, therefore, straightforward to calculate \citep[\eg,][]{1986PhRvD..33.3495D}.}

The shape of $f({\bf v})$ has been constrained primarily by numerical simulations of structure formation in the standard cosmological model \citep[\eg,][]{2009MNRAS.395..797V,1210.2721}. Such simulations include treatments to model the effects of unresolved substructure and debris, and the impact of dark matter particles scattering within the solar system \citep[\eg,][]{2009PhRvD..79j3531P}. However, such effects can only be treated statistically and might not apply to the real $f({\bf v})$ in our Galaxy.

With the advent of LSST, we will be able to {\it empirically} probe $f({\bf v})$ with unprecedented precision. The key idea is to use the oldest and most metal poor stars, which were accreted onto the Milky Way as it formed, as luminous tracers of the underlying dark matter halo \citep{Lisanti:2011as,Kuhlen:2012fz,2014MNRAS.445L..21T,Lisanti:2014dva,2018PhRvL.120d1102H,Necib:2018b}. Such accreted stars also trace the presence of a `dark disk' formed from the late accretion of massive and more metal rich satellites \citep{1989AJ.....98.1554L,2008MNRAS.389.1041R,2009MNRAS.397...44R,2014MNRAS.444..515R,2015MNRAS.450.2874R}, and structures that are not yet fully phase mixed like `debris flows' \citep[\eg,][]{Lisanti:2011as,2018MNRAS.477.1472B,2018Natur.563...85H,necib2018} and tidal streams \citep[\eg,][]{2005PhRvD..71d3516F,1807.09004}. All of these structures imprint features on $f({\bf v})$ that alter the expected flux at a given recoil energy in dark matter detection experiments, and the expected annual modulation signal \citep[\eg,][]{2005PhRvD..71d3516F,2009ApJ...696..920B,2018arXiv181011468E}.

The wide sky coverage and depth of LSST will allow us to select metal poor halo star candidates in statistically significant quantities \citep[\eg,][]{2017MNRAS.471.2587S}, with proper motion data available for the brighter stars. Combined with follow-up spectroscopy, this will provide a direct probe of the velocity distribution of the Milky Way's smooth phase-mixed component \citep{2018PhRvL.120d1102H}. In addition, LSST will find a slew of new structures and streams, allowing us to probe also the non-phased mixed component \citep{2005PhRvD..71d3516F,2018arXiv181011468E}. Finally, combining LSST with spectroscopic surveys of the disk will allow us to place ever tighter constraints on the possible presence of a dark disk \citep{2015MNRAS.450.2874R}.

\subsection{Cosmic Baryon Scattering \Contact{Vera}}
\Contributors{Vera Gluscevic, Kimberly K.\ Boddy, Ethan O.\ Nadler, Alex Drlica-Wagner, Cora Dvorkin}

The most sensitive direct searches for dark matter seek to detect the scattering of dark matter particles from the local Galactic halo in underground detectors \citep[\eg,][]{1509.08767}. 
They have unprecedented sensitivity to WIMPs with masses above a GeV, but are limited by kinematics when searching for lighter particles. 
New experimental techniques are being explored to directly search for sub-GeV models of dark matter \citep{Battaglieri:2017aum}. 
However, due to atmospheric and terrestrial shielding, most direct dark matter experiments are largely insensitive to dark matter particles with large scattering cross sections. 
Current null results from conventional direct detection experiments motivate broad searches in regions of parameter space that are largely inaccessible to underground experiments. 

%This has motivated a number of experimental tests using astrophysical and cosmological measurements including the X-ray Quantum Calorimeter experiment \citep[XQC;][]{0704.0794} and studies of dark matter-cosmic-ray interactions \citep{Cappiello:2018hsu,1810.10543}.

\begin{figure}
\centering
\includegraphics[width=0.75\columnwidth]{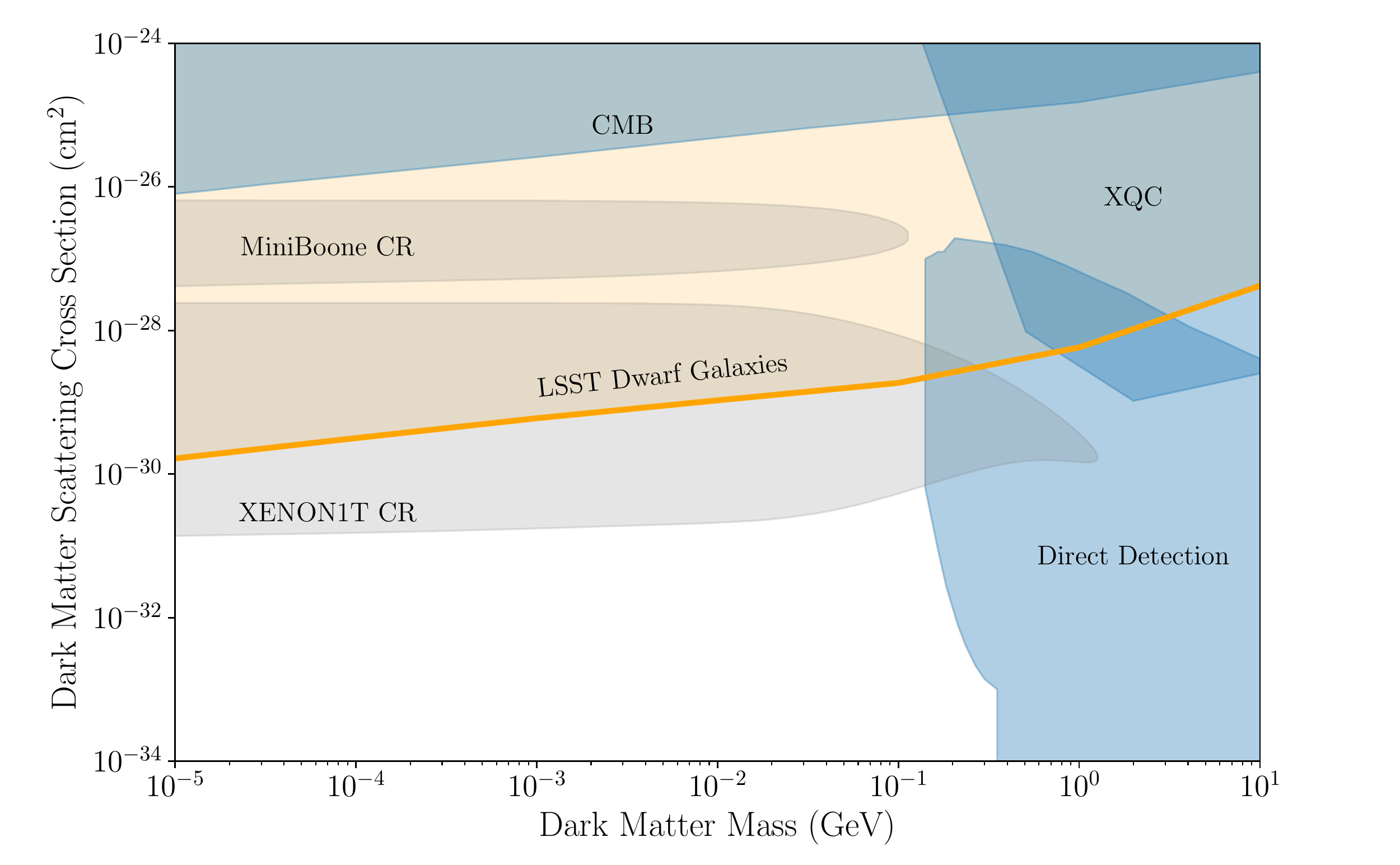}
\caption{
Constraints on dark matter-baryon scattering through a velocity-independent, spin-independent contact interaction with protons. 
Existing constraints (shown in blue) include measurements of the CMB power spectrum \citep[CMB;][]{Gluscevic:2017ywp} and constraints from the X-ray Quantum Calorimeter experiment \citep[XQC;][]{0704.0794}. Direct detection constraints include results from CRESST-III \citep{1711.07692}, the CRESST 2017 surface run \citep{1707.06749}, and XENON1T \citep{1705.06655}, as interpreted by \citet[][]{1802.04764}. %\citep{2018PhRvD..97l3013K}.
Additional constraints that include the effects of cosmic-ray heating of dark matter are shown in gray \citep[][]{1810.10543}.
The gold line shows the projected sensitivity of LSST to dark matter-baryon scattering through observations of Milky Way satellite dwarf galaxies following the prescription of \citet{Nadler:2019}.
}
\label{fig:dd}
\end{figure}

Cosmological and astrophysical observables are sensitive to the scattering of sub-GeV particles with baryons at any point in cosmic history. 
These observations can constrain the interaction cross section to arbitrarily high values and are not subject to uncertainties in the local astrophysical properties of dark matter particles \citep[\eg,][]{1210.2721,1404.1938}. 
If dark matter particles scatter with baryons, they will transfer momentum between the two cosmological fluids, affecting density fluctuations and suppressing power at small scales. 
This power suppression can be captured by a variety of observables including measurements of the CMB \citep{Dvorkin:2013cea,Gluscevic:2017ywp} and the Lyman-$\alpha$ forest \citep{Xu:2018efh}.
Assuming a velocity-independent, spin-independent contact interaction, cosmological constraints can be directly compared against those from direct detection experiments \citep[\eg,][]{Boddy:2018kfv}.
In \figref{dd}, we compare existing constraints on dark matter-baryon scattering from analyses of the CMB and direct-detection searches.\footnote{We caution the reader that this figure does not include a comprehensive list of current constraints, but rather serves to illustrate the complementarity of cosmological and direct detection probes.} 
To estimate the future sensitivity of LSST, we map the projected WDM constraints presented in \secref{smallest_galaxies} to dark matter-baryon scattering constraints by matching the characteristic cutoff scale in the matter power spectrum probed by the lowest-mass subhalos LSST can detect via observations of Milky Way satellite galaxies \citep{Nadler:2019}. 
LSST will deliver measurements of observables that trace matter fluctuations on even smaller scales (\eg, stellar stream gaps), which will potentially extend the sensitivity of these astrophysical and cosmological searches even farther beyond the reach of Planck.

%%%%%%%%%%%%%%%%%%%%%%%%%%%%%%%%%%%%%%%%%%%%%%%%%%%%%%%%%%%%%%%%%%%%%%%%%%%%%%%%
% Discovery
%%%%%%%%%%%%%%%%%%%%%%%%%%%%%%%%%%%%%%%%%%%%%%%%%%%%%%%%%%%%%%%%%%%%%%%%%%%%%%%%
\chapter{Discovery Potential \Contact{Francis-Yan}}
\label{sec:discovery}
\bigskip
\Contributors{Keith Bechtol, Francis-Yan Cyr-Racine, William A.\ Dawson, Alex Drlica-Wagner, Cora Dvorkin, Vera Gluscevic, Manoj Kaplinghat, Casey Lam, Jessica Lu, Michael Medford, Ethan O.\ Nadler, J.\ Anthony Tyson}

Cosmology has a long history of testing particle models of dark matter.
For instance, neutrinos were long considered a viable dark matter candidate \citep[\eg,][]{Kolb:1988}, before precise cosmological measurements made it clear that the universe contains multiple invisible components.
The 30\eV neutrino dark matter candidate is an especially interesting case study of the interplay between particle physics experiments and astrophysical observations.
\citet{Lyubimov:1980un} reported the discovery of a non-zero neutrino rest mass in the range $14\eV < m_{\nu} < 46\eV$ which was subsequently tested by several other tritium $\beta$-decay experiments over the next decade.
Neutrinos with this mass would provide a significant fraction of the critical energy density needed to close the universe, but would be relativistic at the time of decoupling (\ie, hot dark matter).
During the same period, the first stellar velocity dispersion results for dwarf spheroidal galaxies showed that these galaxies are highly dark matter dominated.
The inferred dark matter density within the central regions of the dwarfs was used to place lower limits on the neutrino rest mass that were incompatible with the 30\eV neutrino dark matter candidate \citep{Aaronson:1983,Gerhard:1992}.
Similar stories can be told of heavy leptons \citep[\eg,][]{Gunn:1978}, \CHECK{and other dark matter candidates}, which have been excluded by cosmological and astrophysical measurements.
Cosmology has continually shown that it is impossible to separate the \emph{macroscopic distribution} of dark matter from the \emph{microscopic physics} governing dark matter.

Through much of this work, we have expressed sensitivity to dark matter microphysics in terms of upper limits in the case of non-detection of deviations from the baseline CDM paradigm.
In this section, we consider two potential astrophysical discovery scenarios for non-minimal dark matter properties that could be realized in the LSST era.
In each scenario, a critical question is whether the systematic uncertainties associated with conventional astrophysical processes can be controlled at a level that would be sufficiently compelling to guide non-gravitational dark matter searches with collider, direct, and indirect detection experiments.

\section{Compact Object Discovery}
\label{sec:pbh_discovery}
\Contributors{William A.\ Dawson, Jessica Lu, Casey Lam, Michael Medford, Alex Drlica-Wagner}

\begin{figure}[t]
\centering
\includegraphics[width=0.75\columnwidth]{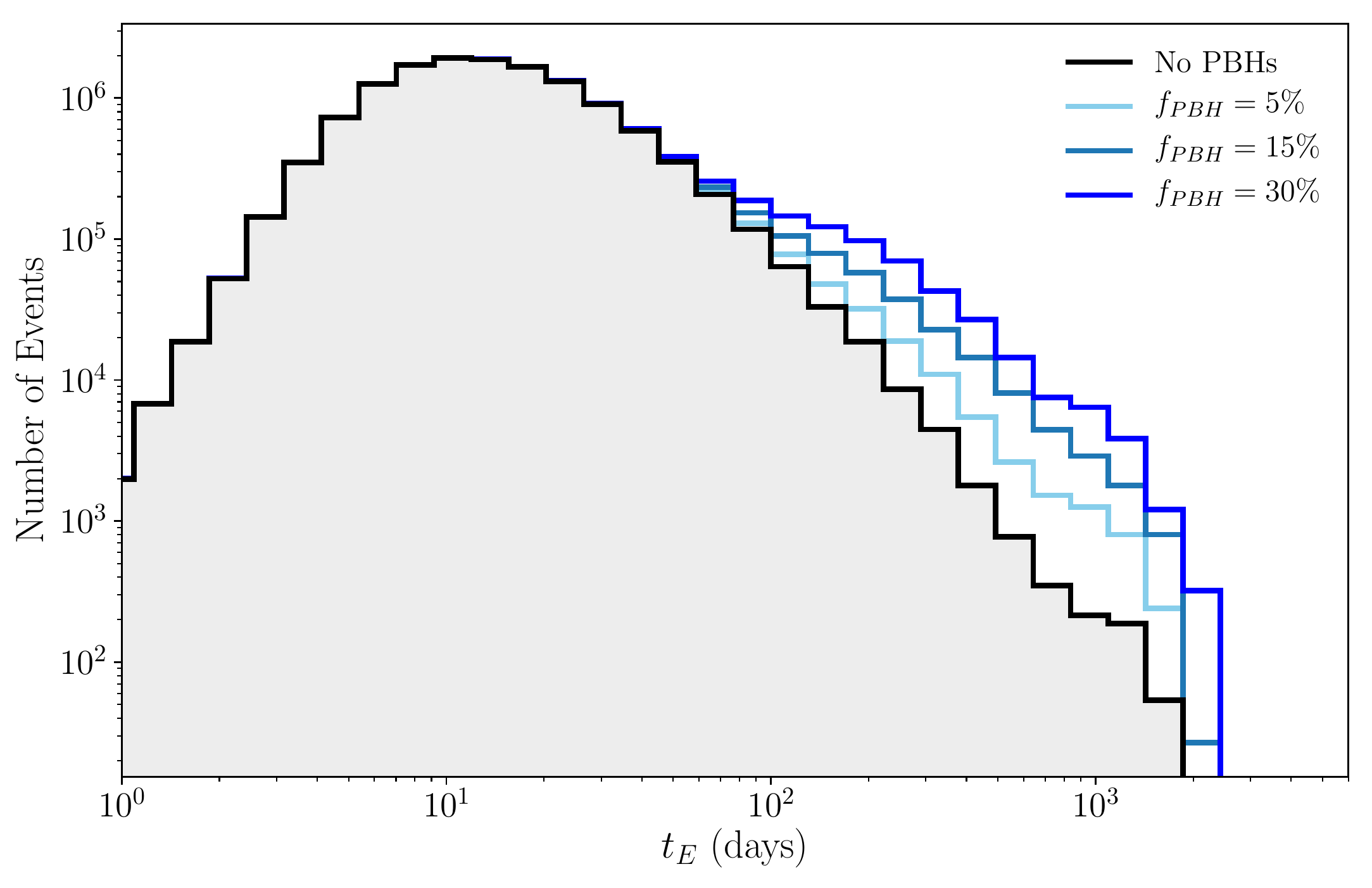}
\caption{\label{fig:macho_discovery}
  The expected number of $<\,2\theta_\mathrm{E}$ microlensing events scaled to a $10 \deg \times 10 \deg$ bulge field.
  The microlensing rate from astrophysical source (stars and compact objects resulting from stellar evolution) is shown in black.
  Blue histograms show the expected microlensing rate assuming different fractions of dark matter composed of LIGO-mass black holes (see text for details).
 The LSST detection efficiency is expected to be $\gtrsim 1\%$ for $t_{\rm E} > 100$\,days \citep[see][for details]{Lu:2019}.
}
\end{figure}

While current constraints make it unlikely that all of dark matter is composed of compact objects with a monochromatic mass function and a uniform spatial distribution, it is nearly certain that LSST will measure the mass spectrum of Galactic black holes (\figref{macho_discovery}).
In this regime, it will be necessary to test whether the observed black hole population statistics can be explained through stellar evolution, or if a novel black hole production mechanism is required (\ie, PBHs).
The discovery of an excess component to the black hole population will necessarily require a fit of the underlying population of stellar remnants and the associated astrophysical systematics. 
However, an excess of high-mass black holes ($M \gtrsim 30\Msun$) or the discovery of black hole clusters \citep[\eg,][]{Clesse:2016} could provide a smoking gun for PBH detection.
If such a PBH population is discovered, it will be possible to measure not only the fraction of dark matter in compact objects, but the compact object mass spectrum, which will in turn set constraints on the spectrum of perturbations during and after inflation \citep[\eg,][]{1702.03901}.
Knowing that some fraction of the dark matter exists as PBHs will force a re-interpretation of particle physics limits from direct and indirect searches.
Preferred regions of WIMP parameter space that are excluded under the assumption that WIMPs comprise all the dark matter will be reopened.
Somewhat counter-intuitively, the outlook for the WIMP may be stronger in a universe where PBHs make up some fraction of the dark matter density.

In \figref{macho_discovery}, we follow the analysis of \citet{Lu:2019} to present an illustration of the LSST discovery potential for a population of high-mass black holes.
\citet{Lu:2019} calculated the expected microlensing event rate from stars and stellar-mass compact objects in a $3.74 \deg^2$ field in the direction of the Galactic bulge.
In addition, a population of high-mass black holes was injected with a Gaussian mass distribution centered at $30\Msun$ with a standard deviation of $20\Msun$.
For simplicity, the injected population was assumed to have the same spatial and velocity distribution as Milky Way halo stars.
The mass density (and hence number density) of injected black holes was scaled to match 5\%, 15\%, and 30\% of the Milky Way dark matter halo.
We scale the microlensing rates from \citet{Lu:2019} to a $10\deg \times 10\deg$ Galactic bulge field, which could be observed as part of a devoted LSST mini-survey.
The resulting total microlensing event rates are shown in \figref{macho_discovery}.
The number of detectable microlensing events will subject to the LSST microlensing detection efficiency, which is expected to be $\gtrsim 1\%$ for events with $t_{\rm E} > 100$\,days \citep{Lu:2019}.
This analysis does not include the added sensitivity long-duration microlensing events from the parallax effects described in \secref{microlensing}. 
Astrometric microlensing described in \secref{astrometric_microlens} will be important for breaking the mass-distance degeneracy of the lens.

\section{WDM/SIDM Discovery}
\label{sec:wdm_sidm_discovery}
\Contributors{Francis-Yan Cyr-Racine, Ethan O.\ Nadler, Keith Bechtol, Vera Gluscevic, Alex Drlica-Wagner, Manoj Kaplinghat}

We now turn to the scenario in which dark matter possesses a particle mass or self-interaction cross section that would partially account for observed small-scale structure anomalies.
There are currently several hints of non-minimal dark matter particle properties arising from comparisons between theoretical predictions and observed galaxy populations at the dwarf galaxy scale, \ie, distances below $1 \Mpc$ and mass scales below $10^{11} \Msun$ \citep[reviewed by][]{BuckleyPeter:2017,Bullock:2017}.
However, the interpretation of these discrepancies in terms of dark matter microphysics has been hindered by uncertainties in the mapping between visible stellar populations and dark matter halos, which involves both the physics of galaxy formation as well as the connection between observable and intrinsic galaxy properties (see \secref{smallest_galaxies} and \secref{halo_profile_group}).
In a regime where systematic uncertainties are already important, it is reasonable to ask how the increased statistical power of LSST will help to resolve our current small-scale structure quandary.

We argue here that the decisive advantage of LSST is the opportunity to combine an ensemble of astrophysical dark matter probes that offer complementary perspectives on dark matter halo abundances and profile shapes, and which are affected by different sources of systematic uncertainty.
For the purpose of illustration, we outline a possible ``roadmap to discovery'' for a dark matter model that produces a cutoff in the matter power spectrum and a suppression of the central dark matter profile just below the current sensitivity limit---\ie, $\Mhm = 10^{8.5}\Msun$.
As a concrete example, we assume that these astrophysical features result from a dark matter particle model with a self-interaction cross section of $\sigmam = 2 \cmg$ and a thermal particle mass of $\mWDM = 6 \keV$.

The first indication of a discrepancy with CDM might come shortly after the first public data release of LSST survey data when automated searches for Milky Way satellites reveal only a handful of new candidate ultra-faint galaxies. 
Using the framework described in \secref{smallest_galaxies}, these observations could be combined to derive bounds on the WDM mass and the minimum halo mass for galaxy formation, $\mathcal{M}_{\rm min}$. These constraints would deviate significantly from expectations derived from CDM and the observational sensitivity of LSST, thereby hinting at a preference for new physics.

The combined depth and sky coverage of LSST will also enable the study of dwarf galaxy satellite populations around other hosts out to several Mpc, as well as the ``field'' population of isolated dwarf galaxies.
By detecting a statistical sample of low-luminosity galaxies in a wide variety of environments, LSST will provide a wealth of input data to theoreticians developing galaxy formation simulations.
In our hypothetical scenario, numerical simulations will show that it is challenging to solve the dearth of observed satellite galaxies by tuning baryonic physics models (\eg, reionization physics, supernova feedback, galaxy formation threshold).

The same LSST data set is expected to reveal many new stellar streams and gravitational lens systems, which would provide access to dark matter halos below the mass threshold of galaxy formation (\secref{stream_gaps} and \secref{stronglens}).
The search for stream gaps and lensing anomalies would be particularly well motivated since these systems would probe a halo mass regime where the discrepancy with CDM would be even more severe.
In addition, these halos are largely devoid of baryons and would be subject to different astrophysical and observational systematics.
We could expect a period of several years necessary to collect and analyze follow-up observations (both spectroscopy and high-resolution imaging) of the most favorable stream and lens systems.
The absence of lower-mass dark matter halos would greatly increase the tension between observations and the predictions of CDM.
In addition, it will be difficult to explain a dearth of dwarf galaxies and lower-mass halos with the same astrophysical systematics, strengthening the case for a fundamental physics explanation.

In parallel, spectroscopic follow-up of LSST-discovered Milky Way satellites with other telescopes (\secref{MW_sats_spec}) will probe their inner density profiles and provide further information about a possible matter power spectrum cutoff or dark matter self-interaction. These dynamical measurements might reveal an unexpected diversity in central densities measurements that might be difficult to explain within the CDM framework. In particular, the discovery of exceptionally dense or exceptionally diffuse ultra-faint satellites with properties that correlate with their orbital parameters could provide a measurement of the SIDM cross section at low velocities \citep{Nishikawa:2019lsc}. Furthermore, density profile measurements of larger dwarf galaxies outside the Milky Way using weak lensing data from LSST (\secref{halo_profile_group}) could probe the SIDM cross section in a different velocity regime. Similarly, the (non-)observation of deviations in halo profiles of galaxy clusters will be able to inform the velocity dependence of the SIDM cross section.

\begin{figure}
\centering
\includegraphics[width=0.75\columnwidth]{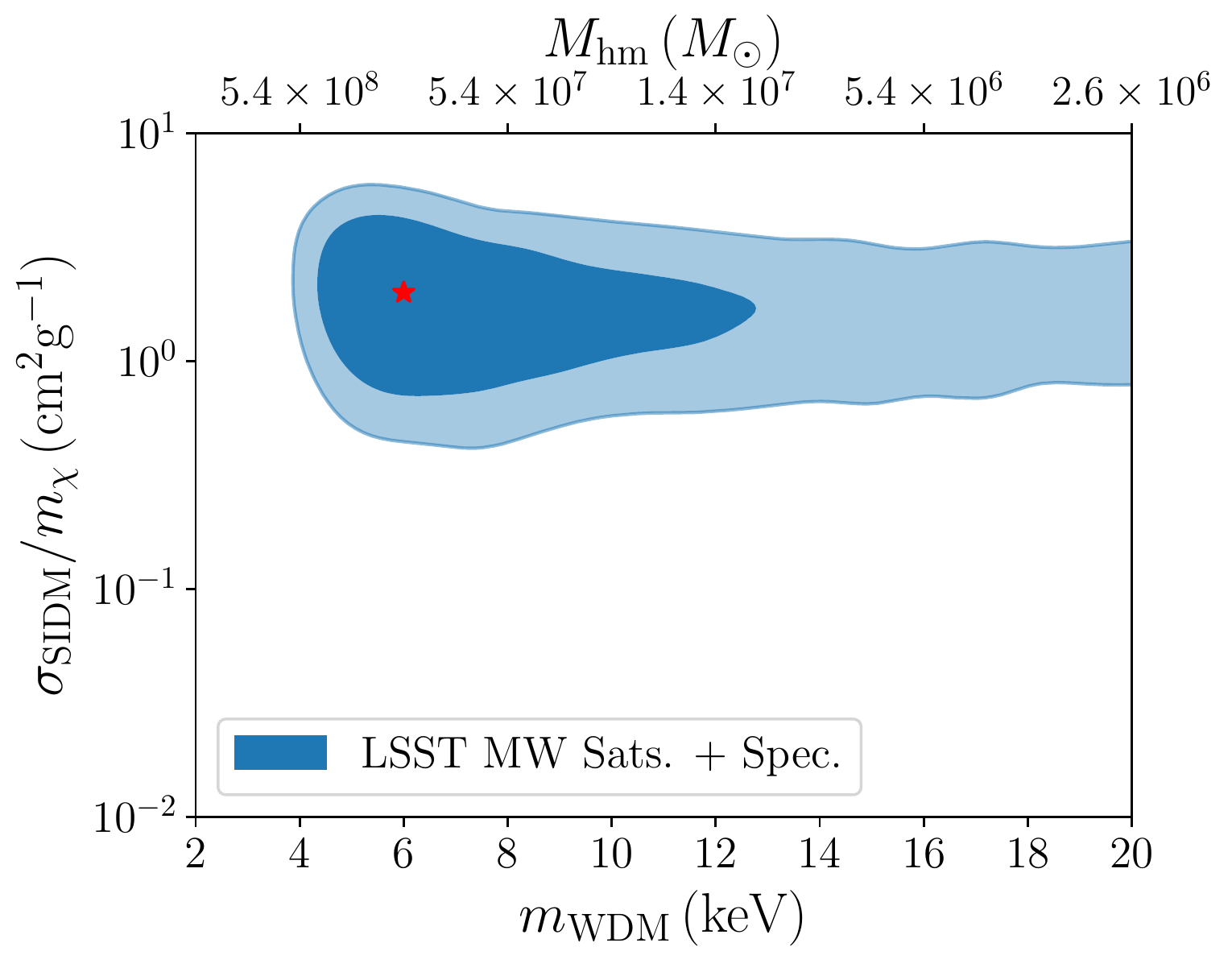}
\caption{\label{fig:sidm_wdm_disc} Example of a measurement of particle properties for a dark matter model with a self-interaction cross section and matter power spectrum cut-off just beyond current constraints ($\sigmam = 2 \cmg$ and $\mWDM = 6\keV$, indicated by the red star). Contours are created by following a procedure similar to \citet{Nadler:2018}, but augmented with the model outlined in \secref{combine_probes} to capture the effect of a power spectrum cut-off and a nonzero self-interaction cross section on the central densities of LSST-discovered Milky Way satellites with spectroscopic follow-up. We take $M_V=0$ mag and $\mu=32$ mag/arcsec$^2$ as our detection threshold for LSST. We assume a prior on the WDM mass $\propto 1/\mWDM$, and a prior on the Milky Way mass from \citet{Callingham:2018vcf}. This figure should be interpreted as a suggestive illustration of the dark matter science that will be enabled by LSST, rather than a precise forecast.  
}
\end{figure}

While much work is still needed to combine all these different probes of dark matter properties (see next section), it is informative to consider what a discovery of our fiducial non-CDM model ($\sigmam = 2 \cmg$, $\mWDM = 6 \keV$) might look like quantitatively with LSST-discovered Milky Way satellite systems. To do so, we use the galaxy-halo connection model outlined in \citet{Nadler:2018} to generate mock LSST observations of faint satellites, and use a similar procedure to that described in \secref{combine_probes} to capture the impact of self-interaction and a mass function cut-off on the central densities of these objects. We construct a binned likelihood in the space of stellar dispersion and luminosity and jointly fit for $\sigmam$ and $\mWDM$, marginalizing over several galaxy-halo connection and Milky Way host halo nuisance parameters. Using a LSST detection threshold for Milky Way satellites of $M_V = 0$ mag and $\mu=32$ mag/arcsec$^2$, we obtain the simultaneous measurement of the SIDM cross section and WDM particle mass presented in \figref{sidm_wdm_disc}. Here, the sensitivity to $\mWDM$ stems primarily from the lower number of faint satellites that the mock LSST observations contain compared the the CDM prediction, while the SIDM cross section sensitivity is driven by the diversity of central densities in the mock observations for $\sigmam = 2 \cmg$. Degeneracy with the halo mass threshold for galaxy formation causes the long tail towards large dark matter mass. Importantly, such degeneracy could be broken by combining these satellite measurements with a probe that is independent of subhalo luminosity such as stellar stream gaps or strong lensing, ultimately resulting in closed contours at high statistical significance. This would signify the start of an era of precision measurement of dark matter particle properties using astrophysical observations.

\vspace{1em} \noindent {\bf Roadmap to measurement}

The hypothetical discovery scenario described above would transform the field of astrophysical dark matter research from one of constraint to one of measurement.
In the measurement paradigm, complementary dark matter probes would be combined to break degeneracies in dark matter models and to constrain astrophysical systematics.
Such analyses would necessitate a probabilistic inference framework to self-consistently analyze multiple measurements probing the same underlying dark-matter physics. 
Such a likelihood-based framework is commonplace for modern cosmological parameter estimation with data from the CMB and current galaxy surveys, and is already under development for dark energy studies with LSST \citep{DESC:CCL}\footnote{\url{https://github.com/LSSTDESC/CCL}}. 
The extension of such a framework to the strongly non-linear regime of dark matter physics in small-scale structures will necessarily rely upon the development of physically accurate and numerically efficient procedures to simulate or emulate the observable effects resulting from varying fundamental properties of dark matter.
Such investigations have already begun through rigorous cosmological simulation of structure formation in WDM, SIDM, and FDM scenarios \citep[\eg,][]{Lovell:2013ola,Dooley:2016ajo,1807.06018,1811.11791}; however, to incorporate these results in a likelihood fit, it will be necessary to evolve effective theories of structure formation \citep[\eg,][]{Cyr-Racine:2015ihg} or quick emulation techniques.
From the data analysis side, recent studies of Milky Way satellite galaxies \citep[\eg,][]{Jethwa:2018,Nadler:2018} have begun to pave the way toward probabilistic analyses of small-scale structure, with the aim of robustly probing both galaxy physics and fundamental physics.
As these and other analyses continue to advance, it will become possible to combine likelihood functions and parameter chains from multiple different observables into a self-consistent likelihood framework. 
This likelihood space can then be scanned to produce joint constraints on dark matter properties.

The self-consistent inclusion of all available information in a joint-likelihood framework will boost the statistical significance of a combined measurement, robustly include astrophysical systematics, and break degeneracies between astrophysics and dark matter models.
The development of fast techniques to predict changes in astrophysical observables from changes to fundamental dark matter properties will allow the production of an end-to-end forward modeling framework for statistical inference.
The ability to simultaneously fit \textit{all} astrophysical observables with \emph{the same} non-minimal dark matter model, while rigorously marginalizing over relevant astrophysical systematics, will produce a compelling argument for the discovery of new dark matter physics.
The same statistical parameter estimation framework will quantify model degeneracies and enable rigorous statements about dark matter particle properties.
These results will critically guide the experimental particle physics program in a post-discovery era.

\section{Outlook}
\label{sec:outlook}
\Contributors{Alex Drlica-Wagner, Keith Bechtol, Vera Gluscevic, J. Anthony Tyson}

More than 80 years after its astrophysical discovery, the fundamental nature of dark matter remains one of the foremost open questions in physics.
Over the last several decades, an extensive experimental program has sought to determine the cosmological origin, fundamental constituents, and interaction mechanisms of dark matter. 
While the existing experimental program has largely focused on weakly-interacting massive particles, there is strong theoretical motivation to explore a broader set of dark matter candidates.
LSST provides a unique, powerful, and complementary platform to study the fundamental physics of dark matter.

LSST will detect and study the smallest dark matter halos, thereby probing the minimum mass of ultra-light dark matter and thermal warm dark matter.
Precise measurements of the density and shapes of dark matter halos will probe dark matter self-interactions, thereby accessing hidden sector and dark photon models.
Microlensing measurement have the potential to detect primordial black holes and to probe the physics of inflation at ultra-high energy scales.
Anomalous energy loss from axion-like particles could reveal itself through precise measurements of stellar populations.
In addition, LSST can uniquely test for correlations between dark matter and dark energy.

Finally, and perhaps most critically, the multi-faceted LSST data will allow novel probes of dark matter physics that have yet to be considered.
These new ideas are especially important as the absence of evidence for the most popular dark matter candidates continues to grow.
As the particle physics community seeks to diversify the experimental effort to search for dark matter, it is important to remember that astrophysical observations provide robust, empirical measurement of fundamental dark matter properties.
In the coming decade, astrophysical observations will guide other experimental efforts, while simultaneously probing unique regions of dark matter parameter space.

% ----------------------------------------------------------------------

%%%%%%%%%%%%%%%%%%%%%%%%%%%%%%%%%%%%%%%%%%%%%%%%%%%%%%%%%%%%%%%%%%%%%%%%%%%%%%%%
% Acknowledgements
%%%%%%%%%%%%%%%%%%%%%%%%%%%%%%%%%%%%%%%%%%%%%%%%%%%%%%%%%%%%%%%%%%%%%%%%%%%%%%%%
\chapter*{Acknowledgments}

We gratefully acknowledge support from the LSST Corporation Enabling Science program (grant \#2017-11).
We thank the Pittsburgh Particle Physics Astrophysics and Cosmology Center (PITT PACC) and the Space Science and Security Program at Lawrence Livermore National Laboratory for generous support of LSST Dark Matter workshops. 
%KITP
The authors would like to recognize the support of the Kavli Institute for Theoretical Physics, specifically through the 2018 program ``The Small-Scale Structure of Cold(?) Dark Matter,'' which was supported in part by the National Science Foundation under Grant No.\ NSF PHY17-48958.
This manuscript has been authored by Fermi Research Alliance, LLC under Contract No. DE-AC02-07CH11359 with the U.S. Department of Energy, Office of Science, Office of High Energy Physics. 
Part of this work performed under the auspices of the U.S.\ DOE by LLNL under Contract DE-AC52-07NA27344 and was supported by the LLNL-LDRD Program under Project No.\ 17-ERD-120.

%%% Here is where you should add your specific acknowledgments, remembering that some standard thanks will be added via the \code{desc-tex/ack/*.tex} and \code{contributions.tex} files.

%This paper has undergone internal review in the LSST Dark Energy Science Collaboration. % REQUIRED if true

%\input{contributions} % Standard papers only: author contribution statements. For examples, see http://blogs.nature.com/nautilus/2007/11/post_12.html

% This work used TBD kindly provided by Not-A-DESC Member and benefitted from comments by Another Non-DESC person.

% Standard papers only: A.B.C. acknowledges support from grant 1234 from ...

%\input{desc-tex/ack/standard} % also available: key standard_short

% This work used some telescope which is operated/funded by some agency or consortium or foundation ...

% We acknowledge the use of An-External-Tool-like-NED-or-ADS.

%{\it Facilities:} \facility{LSST}

%%%%%%%%%%%%%%%%%%%%%%%%%%%%%%%%%%%%%%%%%%%%%%%%%%%%%%%%%%%%%%%%%%%%%%%%%%%%%%%%
% Bibliography
%%%%%%%%%%%%%%%%%%%%%%%%%%%%%%%%%%%%%%%%%%%%%%%%%%%%%%%%%%%%%%%%%%%%%%%%%%%%%%%%

\clearpage

\def\bibname{References}
\chaptermark{References}

\begingroup
  \small
  \setlength{\bibsep}{0pt plus 0.5ex}
  \bibliographystyle{yahapj}
  \bibliography{main}
\endgroup

\end{document}